\documentclass[article,a4paper,11pt]{elsarticle}
	
\journal{Mechanical Systems and Signal Processing}
	
\usepackage[top=2cm,bottom=2cm,left=2cm,right=2cm]{geometry}  
\usepackage{lineno,hyperref}
\usepackage{subfigure}
\usepackage{xcolor}
\usepackage{float}
\usepackage{subfigure}
\usepackage{amsmath}
\usepackage[normalem]{ulem}
\usepackage{soul}
\usepackage{color}

\graphicspath{{Figures/}}

\begin{document}
	
\begin{frontmatter}
	
\title{A noise reduction method for force measurements
in water entry experiments based on the 
Ensemble Empirical Mode Decomposition}
	
\author{Emanuele Spinosa\corref{cor1}}
\cortext[mycorrespondingauthor]{Corresponding author}
\address{CNR-INM Via di Vallerano 139, 00128 Roma (RM), Italy -- 
Telephone: +39 06 50299296 -- Fax: +39 06 5070619}
\ead{emanuele.spinosa@inm.cnr.it}   
	
\author{Alessandro Iafrati \fnref{myfootnote}}
\address{CNR-INM Via di Vallerano 139, 00128 Roma (RM), Italy}


\begin{abstract}
	
In this paper a denoising strategy based on the EEMD (Ensemble 
Empirical Mode Decomposition) is used to reduce the background noise in 
non-stationary signals, which represent the forces measured
in scaled model testing of the emergency water landing 
of aircraft, generally referred to as ditching. 
Ditching tests are performed at a constant horizontal 
speed of 12~m/s with a controlled vertical motion, resulting in
a vertical velocity at the beginning of the impact of 0.45~m/s. 
The measured data are affected by a large amplitude broadband noise,
which has both mechanical and electronic origin.
Noise sources cannot be easily avoided or 
removed, since they are
associated with the vibrations of the structure of the towing carriage
and to the interaction between the measurement chain and the electromagnetic fields. 
The EEMD noise reduction method is based on the 
decomposition of the signal into modes and on its partial reconstruction using 
the residue, the signal-dominant modes and some further modes 
treated with a thresholding technique, which helps to retain
some of the sharp features of the signal.
The strategy is developed and tested first on a synthetic signal 
with a superimposed and known background noise. The method is then 
verified on the measurement of the inertial force acting on the fuselage
when it is moving in air, as in this case the added mass is negligible 
and the denoised force should equal the product of the mass by the
acceleration, both of them being known. 
Finally, the procedure is applied to denoise the forces
measured during the actual ditching experiment.
The  results are superior to those obtained by other 
classical filtering methods, such as a moving average 
filter and a low-pass FIR filter, particularly due to the
enhanced capabilities of the EEMD-denoising strategy here developed to
preserve the sharp features of the signals and to reduce the residual
low-frequency oscillations of spurious origin.

$\copyright$ 2021. This manuscript version is made available under the CC-BY-NC-ND 4.0 license

\url{http://creativecommons.org/licenses/by-nc-nd/4.0/}

The final published journal article can be found at \url{https://doi.org/10.1016/j.ymssp.2021.108659}

\end{abstract}
	
	
\end{frontmatter}
	

\section{Introduction}
\label{introduction}
There are cases in which the experimental setup introduces large noise
components that make it difficult even to identify the trends of the
physical signals.
This is the case of some laboratory tests performed at CNR-INM (Institute
of Marine Engineering) to investigate
the aircraft emergency landing on water, generally referred to as ditching.
The ditching experiments are performed by towing a scaled fuselage model via
a moving carriage and, once a constant horizontal speed of the carriage is
reached, by imposing a computer-controlled guided motion in the vertical plane 
that assigns both the position and the pitch angle.
During the ditching experiments, the loads acting on the fuselage 
are measured.
The nature of the ditching test is, of course, intrinsically non-stationary 
and transient, and the induced loads are characterised by sharp variations
within the observation time window. The measured loads are 
composed of different contributions, as discussed below.
In order to accelerate and to tow the model together with the actuator
systems at the desired horizontal speeds, a quite heavy carriage 
of large dimensions is used. 
Inherently, this introduces large vibrations that, in turn, are 
transferred to the fuselage model, the actuators and the measuring systems.
Furthermore, the presence of high power electric motors, together with
several electronic devices for the control of the carriage and of the
linear actuators, create an intense electromagnetic 
field, which interferes with the acquisition setup.
Consequently, despite the efforts to keep the noise level as low as
possible during the tests, the recorded signals are highly contaminated 
with noise of different nature. 

The relevant part of the signal is commonly referred to as just 
\emph{signal}, and in the present work it refers to 
the inertial and the hydrodynamic loads acting of the fuselage model during
the water entry.
The recorded time series contain both the signal and the 
\emph{background noise}, or simply the \emph{noise},
which, just like the signal, is also non-stationary.
In order to enable a correct interpretation of the experimental measurements,
a strategy allowing to distinguish between the signal and
the noise components is essential.
It is worth specifying that, due to the peculiarities of the experimental setup and 
of the specific test conditions, the noise cannot be completely removed 
from the physical data. As such, in the following the expression 
"denoise" has to be interpreted as a noise reduction rather than 
a noise suppression. 

Classical filtering techniques, such as the moving average or low-pass 
or band-stop filters, are based on the Fourier decomposition of the 
recorded signal and to the attenuation either of the highest frequency 
components or those in a predetermined frequency band.
However, in water entry problems the force is generally
characterized by steep rises and short-duration peaks (impulsive loads) 
\cite{vincent2018dynamics,tveitnes2008experimental,iafrati2015high,
iafrati2016experimental}.
Being the signal broad-banded in 
frequency, the sharp variations might be removed or significantly smoothed out
by the application of a frequency-based approach. 
This is particularly evident when the noise is induced by structural 
vibrations, since it is generally characterized by a low frequency
content and the use of a low-pass or a frequency band filter 
would strongly affect the frequency content of the signal.

In order to develop a proper denoising strategy for such non-stationary 
signals (with non-stationary noise), one option is to recur to 
time-frequency approaches, such as the Discrete Wavelet Transform 
\cite{mallat1999wavelet,percival2000wavelet} or the Empirical Mode 
Decomposition, introduced by Huang et. al 
\cite{huang2014hilbert,huang1998empirical,huang1999new}. 
Both these methods yield a so-called multi-resolution analysis, through which 
the raw signal is decomposed into a series of components or 
non-monochromatic modes, either orthogonal or quasi-orthogonal, 
with a limited frequency band, therefore suitable for a successive 
time-frequency analysis using the Hilbert transform \cite{huang2005hilbert}.
The EMD, often used in combination with the Hilbert transform,
is applied in different engineering contexts. 
It is worth mentioning
its application in naval engineering for the detection of slamming
events \cite{alsalah2021identification} and in machine
health monitoring and fault detection \cite{lei2013review,jiang2013improved,yu2005application}.
The greatest advantage of the Empirical Mode Decomposition, compared to other 
time-frequency methods, lies in its capability to conform to the signal 
itself. Therefore, the EMD does not require a predefined basis of functions, 
as it happens for the wavelet 
transform \cite{cicone2019nonstationary} or for the classical 
Fourier transform. 
The basis of the EMD is a sifting process, in which the envelopes of maxima and minima are computed 
and their average is used to retrieve the 
different modes, which are referred to as Intrinsic Mode Functions
(IMFs).

%
%

EMD and related methods are also employed to achieve a 
noise reduction of non-stationary signals. In fact, by distinguishing the
signal-dominant modes from the noise-dominant ones, it is possible to
perform a partial reconstruction accounting for the relevant modes only, 
possibly using a mode thresholding to improve the results.
A review of the first EMD-denoising techniques, applied to synthetic 
signals corrupted with white Gaussian noise or fractional Gaussian 
noise can be found in 
\cite{flandrin2004detrending,boudraa2006denoising,boudraa2007noise}. 
Noise reduction in signals with a variable mean value is sometimes referred to as ``de-trending', and 
the main approaches to achieve it through the EMD are 
outlined in \cite{wu2007trend}. 
An early application of EMD denoising to biomedical signals (ECG - 
electrocardiogram) is presented in Weng et al. \cite{weng2006ecg}. 
Therein a time-windowing of the EMD modes is 
implemented to preserve the typical waveforms of the ECG signal. 
An application of EMD-denoising to voice speech signals is presented in Khaldi et 
al \cite{khaldi2008new}.
Kopsinis and MacLaughlin 
\cite{kopsinis2008empirical,kopsinis2009development} developed an 
enhanced strategy for signal denoising using EMD, inspired by the 
wavelet thresholding method. The hard and soft thresholding techniques 
were revisited and a new thresholding method was 
proposed, the so-called interval thresholding, 
to improve the effectiveness of the denoising. 
A basic and an advanced iterative method of 
denoising using interval thresholding were proposed and tested on 
synthetic test signals. 
Komaty and Boudraa \cite{komaty2013emd} developed a strategy of EMD 
denoising in which the IMFs relevant for the reconstruction are 
selected based on the similarity of the probability density function of 
each mode to that of the original signal. 
The effectiveness of the method 
is proven through the denoising of artificial and real 
signals, to which both white and coloured noise is added. 
Klionskiy et al. \cite{klionskiy2017signal} developed a new approach 
for the application of EMD denoising to signals with heteroscedatic 
noise, i.e. noise in which the variance is not constant in time.
Tsolis et al.\cite{tsolis2011signal} developed a hybrid signal 
denoising technique based on Empirical Mode Decomposition and Higher 
Order Statistics, to reduce fractional Gaussian noise in synthetic 
signals and in RADAR signals.
Kabir and Shahnaz \cite{kabir2012denoising} continued the development 
of the EMD-denoising method developed by Weng et al \cite{weng2006ecg} 
for ECG signals, by combining an EMD with a DWT (Discrete 
Wavelet Transform) thresholding. 
Yang et al. \cite{yang2015emd} applied interval thresholding, with some 
proper modifications, to perform EMD denoising. The relevant modes are 
selected based on the similarity between the PDF of the original signal and the PDFs of the modes.

Although the classical EMD is rather suitable for the analysis of non-linear and 
non-stationary signals, it has some drawbacks. 
In particular, it exhibits an excessive sensitivity to small perturbations, 
meaning that a small change in the data can lead to rather 
different mode shapes.
Moreover, the decomposition 
is prone to mode mixing,  i.e. the undesired
condition for which a single mode may contain widely separate scales (in the
frequency domain), and to mode splitting, i.e the presence of similar
time scales in different modes\cite{ge2018theoretical},
\cite{lee2009physics},\cite{gao2008analysis}, \cite{hu2011emd}.
Among the different approaches that can be 
adopted to reduce these issues, such as Iterative Filtering
\cite{lin2009iterative},\cite{cicone2019nonstationary},\cite{cicone2016adaptive} and 
Variational Mode Decomposition \cite{dragomiretskiy2013variational},
the EEMD, introduced in \cite{wu2009ensemble}, is employed in this 
work, being quite straightforward and particularly 
indicated for the present signals, which display some intermittency. 
The EEMD also displays a lower 
sensitivity to perturbations, which is also essential 
for the present applications, as specified in the following.

%
%
In literature only a few applications of noise reduction using EEMD can  be found. 
Su et al.\cite{su2016approach} developed and applied a noise reduction 
method based on EEMD to prototypical observations on 
dam safety.
Wang et al.\cite{wang2016ensemble} developed denoising methods 
similar to that developed by Kopsinis et 
al.\cite{kopsinis2009development} by using EEMD, instead of a pure EMD. 
A modified interval thresholding method was also implemented, to suit 
the characteristics of the EEMD modes. These methods provide a better 
performance than the EMD-based ones, especially for data with low 
signal-to-noise ratio.
Wang et al. \cite{wang2019novel} developed a novel method to identify 
the frequency components of the pressure pulsation that causes damage 
in the off-design operation of a Francis turbine; a targeted 
EEMD-based denoising strategy is devised, which includes an 
autocorrelation analysis and a wavelet thresholding.
Zhang et al.\cite{zhang2016improved}
proposed an improved filtering method by performing an EEMD
decomposition, followed by a wavelet thresholding applied
to flow-induced vibration signals, recorded for the structural
health monitoring of a hydraulic structure. Bao \cite{bao2010emd}
used EEMD to retrieve  the amplitude-modulated components, 
related to the cavitation noise from ship-radiated sound, containing
a significant amount of background noise.


The existing literature on 
denoising using the EMD or the EEMD indicates that it is 
very difficult to design a strategy that is valid 
in all contexts. Instead, ad-hoc solutions must be developed and tuned
for the specific signals under examination and for the type of
background noise superimposed to the signal. Most of the works cited above 
are focused on non-stationary signals with a more or less constant trend. 
On the other hand, as mentioned above, the signals of our interest
display trends with large signal variations, sometimes also rather sharp. 
In the existing literature, transient signals are addressed only rarely to our knowledge. 
Furthermore, most works focus on reducing a 
white Gaussian noise or noise coming from one single source, whereas the type of background noise  
in the signals considered in the present paper is
associated with multiple sources, therefore it is broadband and non-stationary.
The development and the application of an EMD/EEMD denoising 
strategy to signals recorded during ditching experiments offers an 
opportunity to address such kind of noise
sources in a more systematic way than in the previous works.

In particular, in this paper an Ensemble Empirical Mode Decomposition (EEMD) denoising 
strategy is developed and applied to the analysis of the force measured 
during the controlled water entry with horizontal speed of fuselage models. 
The denoising method exploits a proper interval thresholding, to 
improve the quality of the denoised data. 
The method is tested and validated first on a synthetic signal and on a dry test,
in which only the inertial force acts, which can be easily derived based
on the mass of the body and the imposed acceleration. Hence, the method is applied to the 
real force measurements. 
\section{Experimental Setup and Data}
\label{expsetup}
%
In order to properly understand the hydrodynamic phenomena occurring 
during aircraft ditching, experimental tests on scaled fuselage models 
entering water with a horizontal speed and a controlled vertical 
trajectory have been performed at the CNR-INM Towing Tank.
The objective of these experiments is to measure the 
forces acting on
the fuselage during a guided water landing, thus allowing to retrieve 
essential information on the aircraft dynamics in this critical phase.
An accurate reproduction of the hydrodynamic phenomena and of the
fluid-structure interaction aspects occurring in the ditching phase 
necessarily requires full scale tests on sample specimens
\cite{smiley1951experimental}
\cite{iafrati2015high} \cite{iafrati2019cavitation},
\cite{spinosa2021experimental}
,whereas the investigation of the aircraft dynamics at ditching needs
small-scale free-flight tests \cite{naca2929} \cite{climent2006}\cite{zhang2012}.
Unfortunately, in the latter kind of tests it is quite difficult to achieve a precise control of
the attitude at the impact and, due to the strong non-linearity of the impact loads,
the resulting dynamics of the aircraft is not repeatable. This makes the data not
fully exploitable for the purpose of validation of the computational models.
Generally, computational models are able to integrate accurately the
equations governing the dynamic of the body motion, as long as the forces 
acting on the body are estimated correctly. Based on the above
considerations, an alternative to the free flight tests is represented by 
guided tests on scaled models, mimicking a ditching with 
an imposed pitch motion. In this way, it is possible to measure 
the loads during the ditching with precise
information on the aircraft attitude, thus making the data useful for the validation of
the loads provided by the simulation tools.

\begin{figure}[htbp]
\centering
\includegraphics[width=0.75\textwidth]{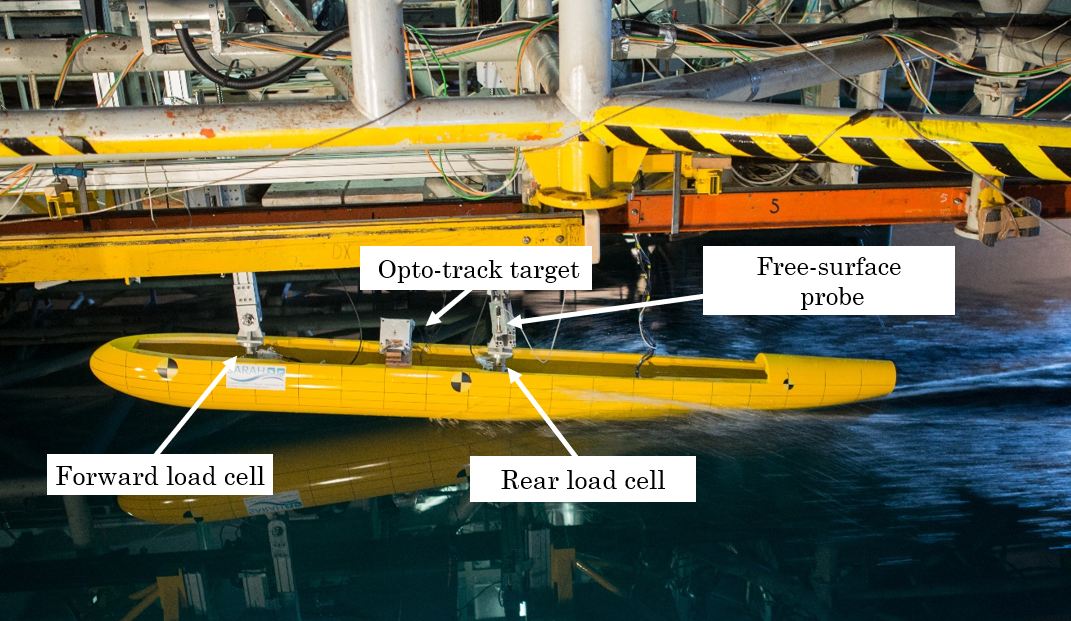}
\caption{The instrumented fuselage model installed on the towing carriage.}
\label{fig:exp_setup_S1}
\end{figure}

The experiments are performed on fuselage models, the shapes of which are
defined analytically.
The one used for the tests considered here
has a circular cross section with a diameter of 0.4~m and a length of 4~m. 
More details can be found in \cite{iafrati2020experimental}.
The tests have been performed at the CNR-INM towing tank, which is 
470~m long, 13.5~m wide and 6.5~m deep. The fuselage model is towed by a carriage at
the ditching speed, while the motion in the vertical plane is imposed by two linear
servo-actuators, computer-controlled.
The two actuators are installed on a frame, which is clamped to
the structure of the towing carriage. The experimental setup and the
instrumented fuselage are shown in Figure \ref{fig:exp_setup_S1}.
%
The fuselage models are connected to the two actuators 
by two 6-axes Kistler load cells 9306A, driven by charge amplifiers ICAM 5073A.
The cells enable the measurements of the loads in 
both the longitudinal and normal direction to the fuselage axis.
The longitudinal force components are indicated with $F_{xF}$ and $F_{xR}$ 
whereas the normal force components are denoted as $F_{zF}$ and $F_{zR}$, subscripts
$F$ and $R$ being used to distinguish the front and rear measurements, respectively.
The loads are sampled at 200~kHz and acquired by a DAQ system, which 
includes an analogue anti-aliasing filter at 78~kHz. 
For the data processing, the signals are further down-sampled to 20 kHz.

\begin{figure}[htbp]
\centering
\includegraphics[width=0.95\textwidth]{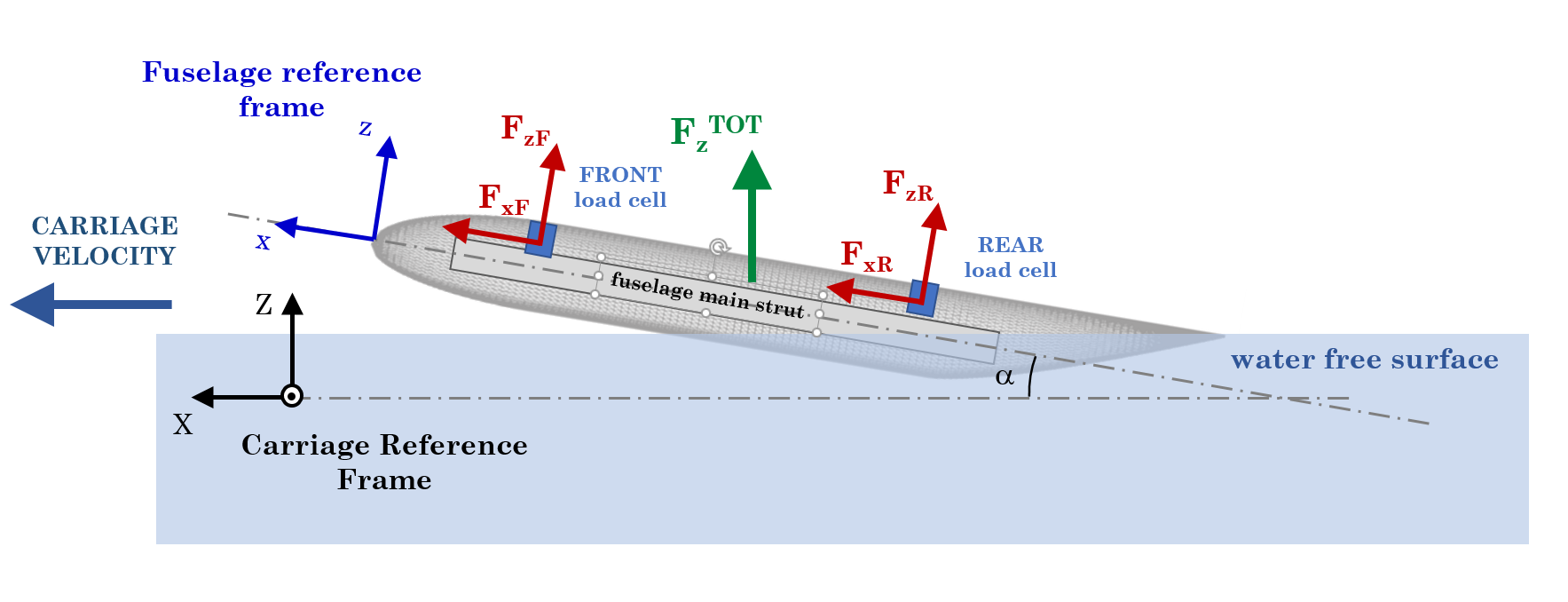}
\caption{Schematic of the reference frames used for 
the data analysis and of the force components measured by the load cells.}
\label{fig:FusRefRames}
\end{figure}

\begin{figure}[htbp]
\centering
\includegraphics[width=0.80\textwidth]{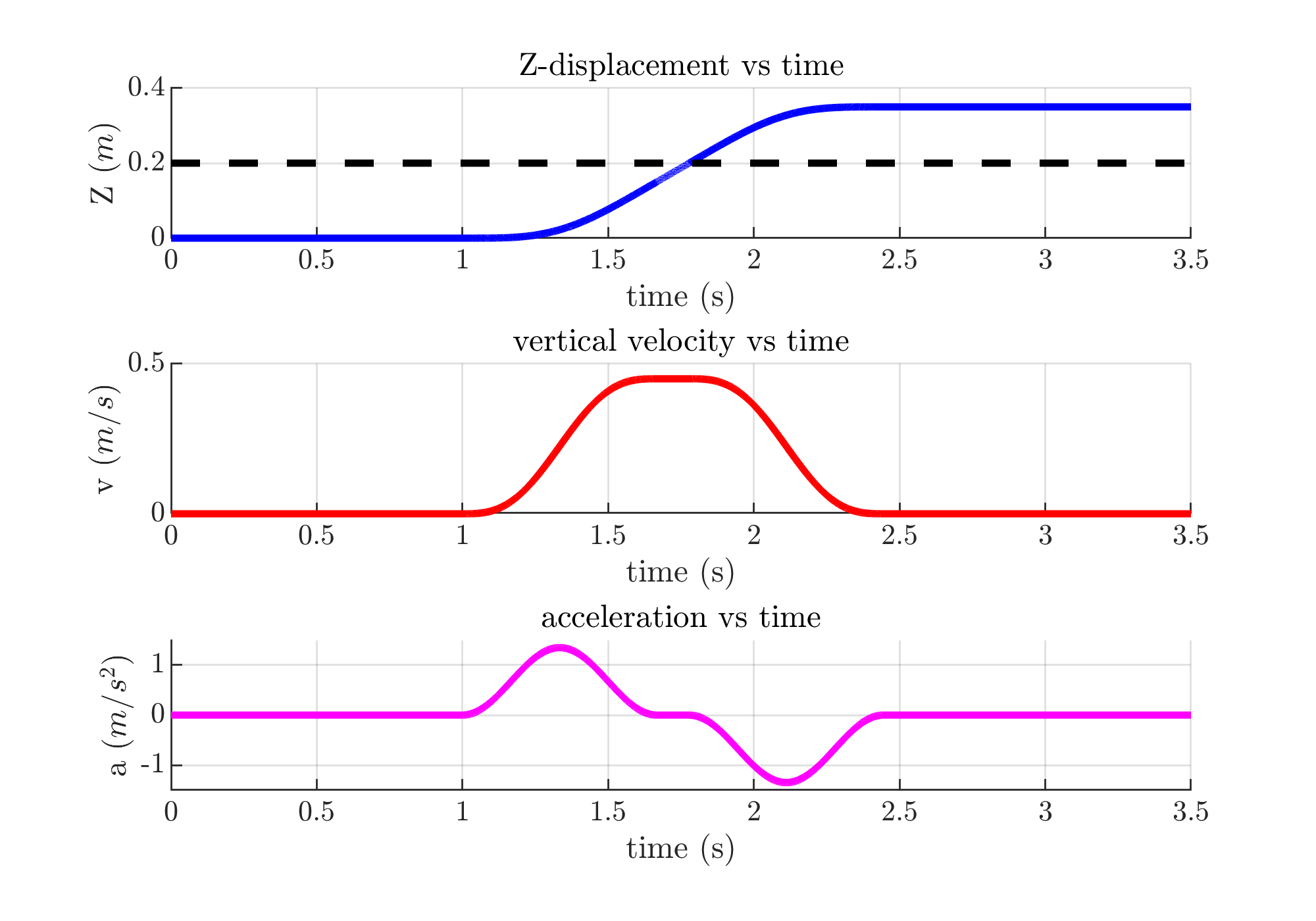}
\caption{Time histories of vertical displacement, velocity and 
acceleration of the fuselage (referred to the carriage reference 
frame, see Figure \ref{fig:FusRefRames}). The horizontal line, 
located at $Z = 0.20$~m, identifies the instant of first 
contact of the fuselage with water.}
\label{fig:trajectory}
\end{figure}

Figure \ref{fig:FusRefRames} shows a sketch of the fuselage and of the reference 
frames used for the data analysis 
as well as the force components measured by the load cells.
The vertical displacement, velocity and acceleration of the fuselage 
are relative to the reference frame $(X,Z)$ fixed to the carriage.
The forces acting
on the fuselage are expressed in the reference frame $(x,z)$ fixed to the fuselage.
At the beginning of the test, the fuselage is set at the
given attitude with the lowest point of the shape positioned 0.20~m above
the still water level.
During the run, when the carriage reaches the constant speed of 12~m/s, the
fuselage model starts its vertical motion.
The fuselage is accelerated with an analytically defined law of motion, and moves 
0.15~m vertically. At the end of the acceleration phase the velocity is 0.45~m/s, 
which corresponds to a vertical-to-horizontal velocity ratio of $V/U = 0.0375$.
Hence, the fuselage moves at a constant vertical velocity for 0.05~m 
until it touches water.
As soon as the fuselage touches the water, it starts decelerating with the same law
used in the acceleration phase, until it stops when the first contact point is 0.15~m 
below the still water level.
Therefore, the overall vertical displacement of the fuselage is 0.35~m. 
The imposed vertical displacement $Z(t)$, the velocity $v(t)$ and the 
acceleration $a(t)$ of the fuselage are shown in Figure \ref{fig:trajectory}.
The acquisition is started 1~s before the
fuselage motion (Fig. \ref{fig:trajectory}). The recorded time histories 
in that initial interval are
used both to set the reference zero values for the different transducers,
as well as to evaluate and characterize the signal noise before the fuselage descent.
As the fuselage is initially set 0.2~m above the still water level, the
intersection of the horizontal line at $Z = 0.20$~m with the fuselage position
line, allows to identify the time at which the fuselage gets in touch with the water. 
The time of the initial contact can be also estimated analytically as 
$t \simeq 1.778$~s.
It is worth remarking that the exact value may vary due to
the lowering of the free surface caused by the air-flow field induced by the
motion of the carriage and by the presence of the residual standing wave
in the tank caused by the previous runs.
In total, the two effects may be responsible for a oscillation in the water
level of $\pm$ 5~mm, which implies a variation of the impact time of $\pm$ 0.011~s.

\begin{figure}[htbp]
\centering
\includegraphics[width=0.95\textwidth]{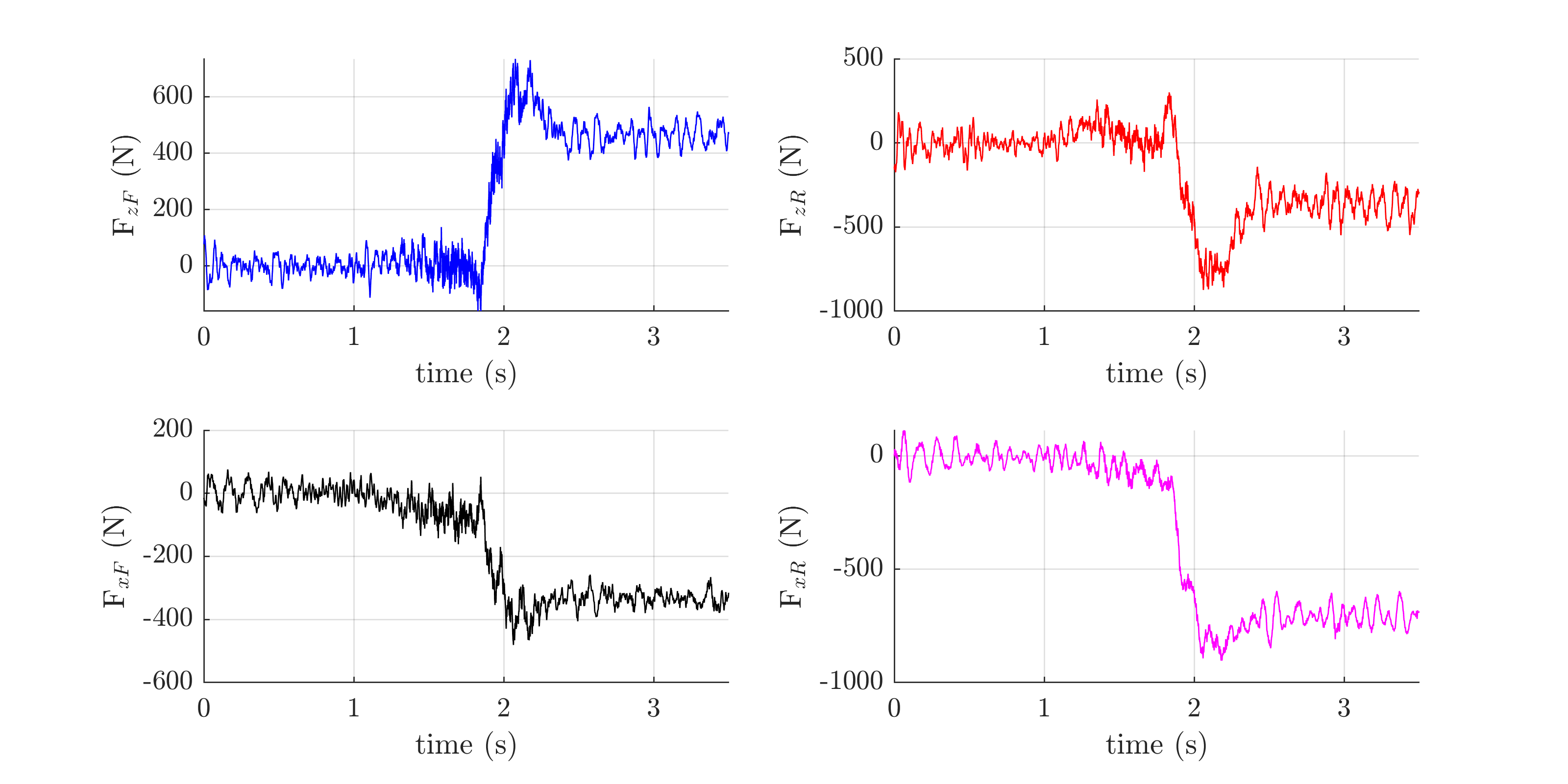}
\caption{Time histories of the forces recorded during the water 
entry with the prescribed trajectory.}
\label{fig:Denoising_Forces_4T1422N_02}
\end{figure}

A typical example of the force time histories recorded by the load cells
is provided in Figure \ref{fig:Denoising_Forces_4T1422N_02}.
By limiting the attention to the $z$-component of the loads, given the 
manoeuvre of the fuselage, two different contributions can 
be distinguished: an \emph{inertial part}, which is proportional to the 
fuselage acceleration in the various phases, and a 
\emph{hydrodynamic part}, which accounts for the force exerted by
the water on the fuselage during the entry phase.
The time histories of the measured data clearly show that the phenomena
are non-stationary, with large and rather sharp variations.
Several noise sources affect the measurements.
The frames holding the two actuator systems are firmly clamped to the
carriage structure. The rigid connection of the frames allows a precise
control of the fuselage motion, however the carriage vibrations 
are transferred to the frame without damping.
Vibrations are also associated with the control system of 
the two servo-actuators.
In addition to the mechanical vibrations, the data are affected by the 
electronic noise resulting from the 
interaction between the electromagnetic fields generated by the carriage 
motors and by the two servo-actuators with the cables and the
electronic circuitry of the data acquisition system.
Finally, other sources of noise could be ground loops and clearance adjustments.
Given the presence of noise of different nature, it is necessary to develop 
ad-hoc strategies capable to isolate and distinguish the noise from the 
physically-relevant components of the signals.
\section{Algorithm development and verification based on a synthetic signal}
\subsection{EMD algorithm}
The EMD algorithm used in the following is 
based on the original procedure 
described in \cite{huang1998empirical,huang1999new}, with some
minor modifications needed to make it applicable to the signals
considered in the present work. 
The basic principle of the EMD is to perform an iterative cycle, named \emph{sifting}, 
in which a sort of moving average of the signal is computed by averaging 
point-wise the two cubic
splines that interpolate the local maxima and minima of the 
signal, respectively. This sort of moving average, which represents the 
main trend of the signal, is subtracted from the original signal, thus yielding 
a temporary Intrinsic Mode Function (IMF). 
The temporary IMF may still be characterized by a residual trend.
To remove it, the sifting is repeated within an inner sifting cycle until the 
number of maxima/minima of the
temporary IMF is equal to the number of zero-crossing $\pm$ 1 and the 
mean value of the maxima and minima envelopes is zero \cite{huang1998empirical}.
Hence, the temporary IMF is marked as an actual IMF,
sometimes also referred to as mode in the following. 
Once the first actual IMF is identified, it is subtracted from the original signal
leading to a temporary residue. The sifting procedure is applied to the
temporary residue until a final residue, which is either 
monotone or sufficiently small, is obtained.
At the end of the procedure, the original signal results decomposed into a 
set of IMFs and a final residue, in other words the EMD is said to be complete.
	
\begin{figure}[htbp]
\centering
\subfigure[Issue due to edge mirroring.]
{\includegraphics[width=0.48\textwidth]{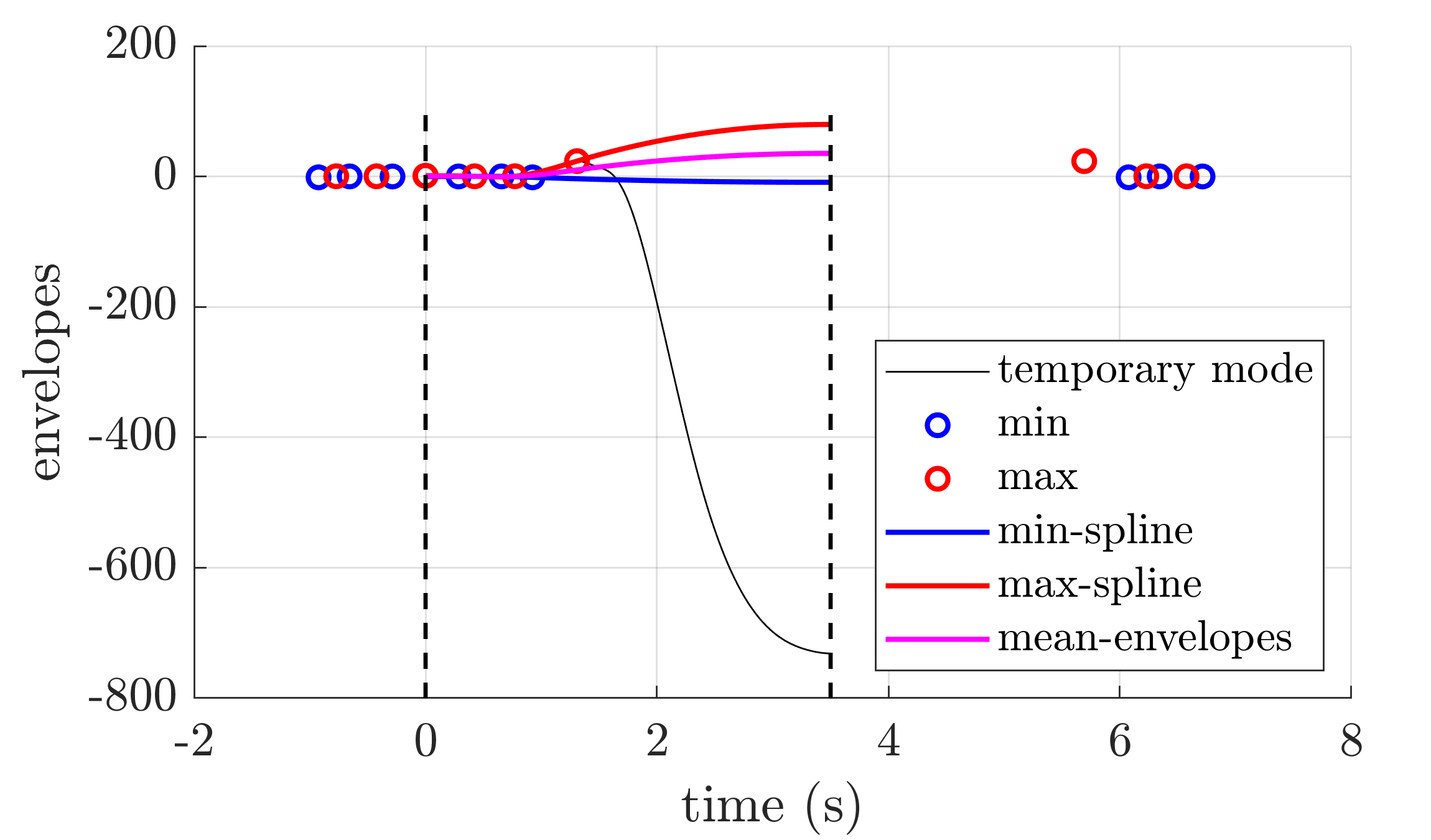}} 
\quad
\subfigure[Solution of the issue.]
{\includegraphics[width=0.48\textwidth]{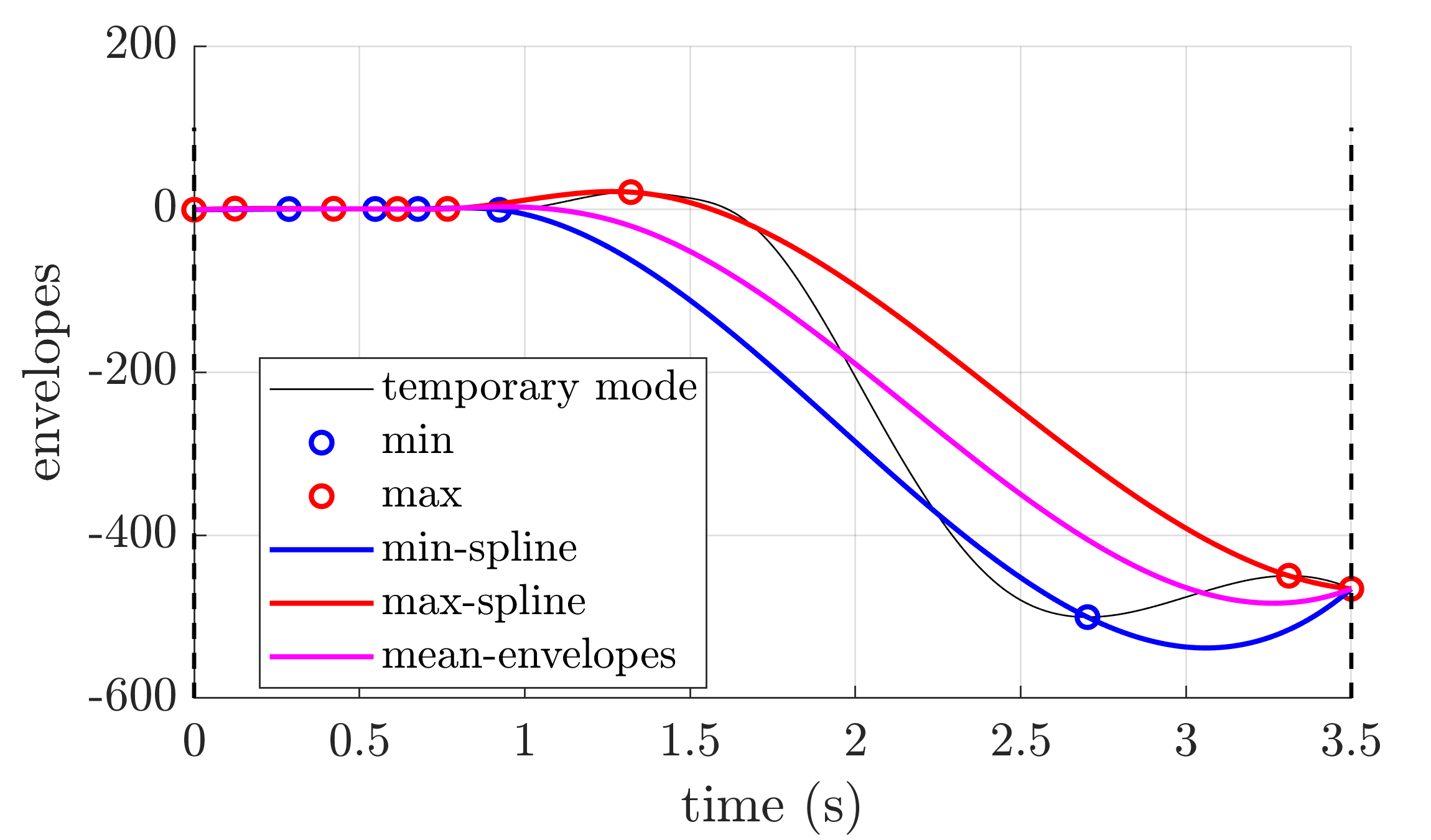}}
\caption{(a) Example of the issue encountered in the first sifting iteration 
if using edge mirroring at both ends with initial 
and final values significantly different and (b) improved prediction obtained 
by forcing the initial and final values to be both maxima and minima.}
\label{fig:edge_mirroring_issue}
\end{figure}

In the present work, the EMD algorithm is slightly modified compared to
the one first proposed in \cite{huang1998empirical}. 
First of all, the convergence of inner sifting 
cycle is based on a Cauchy criterion, i.e. the loop is terminated when the 
sum of the differences $SD$, defined as:
\begin{linenomath}
\begin{equation}
\textrm{SD}=\frac{\displaystyle{\int \left( h_{(k-1)}(t) - h_{k}(t) 
\right)^2 dt} }
{\displaystyle{\int h_{(k-1)}^2(t) dt } } \;\;,
\end{equation}
\end{linenomath}
drops below a threshold value, i.e. $SD \le SD_T$, where $h_k(t)$ is the temporary IMF 
at the $k$-th iteration. For the convergence, it is assumed $SD_T = 0.2$. 
Such a convergence criterion is proposed in \cite{huang2014hilbert},
but the definition of $SD$ suggested in \cite{wang2010intrinsic} is used,
which makes the criterion itself more stable. 
Of course, the application of such a criterion is not directly 
related with the conditions for the attainment of an actual IMF, as specified 
in \cite{huang2014hilbert}. 
However, it is considered rigorous enough for the EMD procedure
\cite{huang2014hilbert} and also provides the advantage of speeding up computation.
The attainment of an actual IMF to terminate the inner sifting loop, in fact, typically 
leads to a large number of sifting iterations.
As noticed in \cite{wang2010intrinsic}, a large number of sifting iterations 
using the cubic spline interpolation would lead to constant-amplitude IMFs, 
without any amplitude modulations, and therefore not very meaningful.
As a second difference, the loop is forced to terminate when the number of
maxima or minima is lower than 3, and for 3 points a parabolic interpolation
is used instead of cubic splines.
Finally, in order to improve the accuracy in the representation of the modes 
at the sides, according to the suggestion by Rato et al.\cite{rato2008hht},
a number of 3 maxima and minima at the two edges are mirrored, and the 
spline envelopes encompass the mirrored maxima and minima as well.
Owing to the specific behaviour of the signals considered in the present
study, the initial and final values may be quite different from each other. 
As such, the average of the spline envelopes may differ substantially from
the trend of the signal. 
This generally happens in the first sifting iteration, as shown for 
example in Figure \ref{fig:edge_mirroring_issue}.
Such a limitation may be overcome by forcing, only in the first iteration
of the inner loop, the first and last values of the signal to both maxima 
and minima. Starting from the second sifting iteration, the edge 
mirroring described above is applied.

\subsection{Ensemble Empirical Mode Decomposition (EEMD) algorithm}
\label{EEMD_algorithm}

As already mentioned in Section \ref{introduction}, the classical empirical mode 
decomposition suffers from two main limitations: sensitivity to the input perturbations 
(and to time shifts) and mode-mixing. The sensitivity to the input perturbations
is mainly introduced by the 
interpolation with cubic splines, and may also lead to an instability 
of the decomposition itself, as the spline interpolation is used repeatedly 
within the sifting loops \cite{cicone2019nonstationary,lin2009iterative}. 
An immediate consequence is that the EMD of two different repeats of the same test, 
which contain a different background noise and 
approximately the same signal, but with slight time-shifts,
may result in significantly different IMFs, particularly for the last modes.
Mode mixing (and/or mode splitting), on the other hand, 
occurs when there are oscillations at significantly different frequencies 
within the same mode and/or when oscillations at similar frequencies occur in 
different modes \cite{ge2018theoretical,lee2009physics,gao2008analysis,hu2011emd}. 
These phenomena are critical if the signal exhibits a certain degree of intermittency, 
as it happens in the present case. 
Reducing or avoiding mode mixing enables a more robust decomposition, 
leading to an easier interpretation of the physics \cite{lee2009physics}.

In order to overcome, or to partially reduce at least, the above limitations, 
the Ensemble Empirical Mode Decomposition, or EEMD, introduced in 
\cite{wu2009ensemble} is employed in this work, being rather straightforward and
very efficient in presence of signal intermittency associated to the
background noise.
The EEMD is based on the use of the classical EMD on a certain number of artificially 
perturbed signals $y_i(t)$ which are
obtained from the original signal $y(t)$ as:
\begin{linenomath}
\begin{equation}
y_i(t)=y(t)+w_i(t) \qquad i=1 \, ... \, N_e
\end{equation}
\end{linenomath}
where $w_i(t)$ is an artificial white Gaussian noise realisation
with amplitude $N_a \, \sigma$, $N_a$ being a 
parameter (typically from 0.1 to 0.2) and $\sigma$ 
the standard deviation of the original signal $y(t)$ or of a part of it.
It makes sense to compute $\sigma$ over a 
time interval in which the signal is characterized by a constant trend, so that 
it can be directly related to the level of background noise. 
Once the EMD is repeated on $N_e$ different signal perturbations, the IMFs are
retrieved from the ensemble-averages resulting from the different
EMDs. Such a procedure partially reduces the effect of random perturbations on the
final decomposition. Moreover, the addition of artificial white Gaussian 
noise to the 
original signal provides a uniformly distributed reference scale to the 
input, which contributes to reduce the mode mixing \cite{huang1998empirical},
as mentioned in Section \ref{introduction}.

\begin{figure}[htbp]
\centering
\subfigure[Pure EMD]{\includegraphics[width=0.48\textwidth]{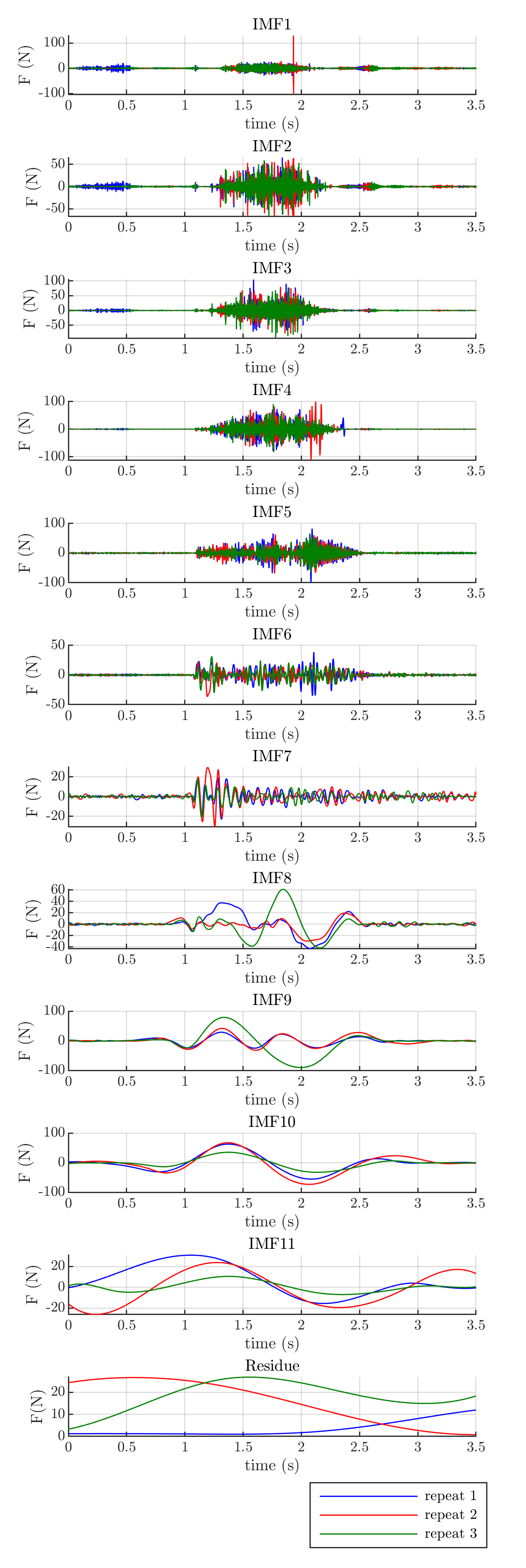}} \quad
\subfigure[EEMD]{\includegraphics[width=0.48\textwidth]{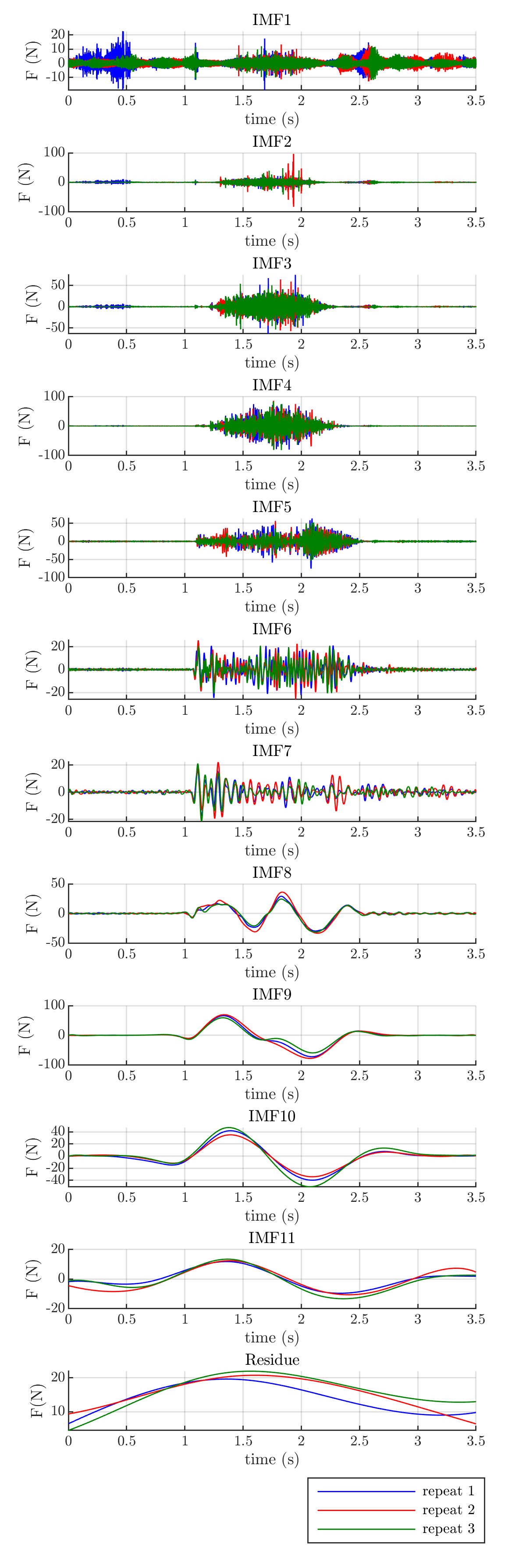}}
\caption{Comparison between the pure EMDs and the EEMDs of three 
repeats of the dry test. The EEMD is performed with $N_a = 0.1 \sigma$ 
and $N_e$=1000.}
\label{fig:IMFs_FzTOT_4T1092N_0X}
\end{figure}

The improvements that the EEMD yields compared with the
pure EMD can be seen by comparing the decompositions of different repeats of the 
experimental tests. Such a comparison can be established for 
the dry tests, i.e. the tests in which the fuselage motion is activated with 
the carriage at rest, starting from a position high enough so that the 
fuselage is always out of the water. 
This kind of test is highly repeatable and
avoids both the uncertainty in the water level as
well as the noise induced by the movement of the carriage.
The decomposition is applied to the total vertical force.
The EMD is performed on each of the three test repeats, and the
resulting IMFs and the residue are shown in Figure
\ref{fig:IMFs_FzTOT_4T1092N_0X}(a).
Hence, an EEMD with $N_a=0.1$ and $N_e$=1000 is performed
on the same data. 
The standard deviation $\sigma$ is 
computed over an interval during which the fuselage descends 
at constant speed, i.e between $t=1.70$ and $t=1.75 s$. 
The resulting IMFs and the residue are shown in Figure 
\ref{fig:IMFs_FzTOT_4T1092N_0X}(b).
The IMFs obtained from the pure EMD display significant differences 
in the shapes of the modes
extracted from the three repeats, especially for the last IMFs and
for the residue. Instead, the modes provided by the EEMD for 
the three repeats are much closer and repeatable. 
Such a result clearly testifies the reduced sensitivity of the EEMD to the
background noise, compared to the EMD. Furthermore, as suggested in \cite{wu2009ensemble}, 
the EEMD is also less 
sensitive to the end effects, which are more relevant in the last modes.
Based on the above experience and results the EEMD approach is used in the following.

\subsection{Denoising Strategy}
The principles of the denoising techniques based on the EMD and EEMD are 
described in \cite{flandrin2004detrending, boudraa2006denoising, boudraa2007noise,
wu2007trend, kopsinis2008empirical, kopsinis2009development}
and in \cite{su2016approach,wang2019novel} respectively, as briefly
reviewed in Section \ref{introduction}.
The conventional EMD denoising strategy consists first in decomposing 
the signal into its intrinsic mode functions (IMFs) and then in
distinguishing the signal-dominant from the noise-dominant modes. 
Typically, the noise-dominant modes are discarded, whereas the signal-dominant 
modes are retained. Hence, a denoised signal is obtained by partial reconstruction, i.e. 
by the sum of the residue and the signal-dominant IMFs only. 
In most cases, it is reasonable to assume that the first modes, which 
have a high frequency content, are noise-dominant, whereas the last modes, 
with a typically lower frequency content, are signal-dominant.
However, for the present applications sharp variations occur and, therefore,
some of the first modes, which are noise-dominant,
might also contain physically relevant contributions that should be preserved. 
A possibility to improve the denoising strategy is to 
include part of the noise-dominant IMFs with a thresholding.
Such a technique enables a  reduction of the background 
noise within the single modes, which can then be included in the partial 
reconstruction \cite{kopsinis2008empirical,boudraa2006denoising,kopsinis2009development}.
The most challenging aspect of such denoising methods is how to 
discern the noise-dominant modes from the signal-dominant ones, as well
as the definition of 
the thresholding strategy and the choice of the threshold value. 
All these choices strongly depend on the type of signal to be analysed and 
on the background noise to be reduced, as no universal method
is available in the existing literature. 

\begin{figure}[htbp]
\centering
\subfigure[Synthetic signal with superimposed \emph{background noise} and only \emph{signal}]
{\includegraphics[width=0.65\textwidth]{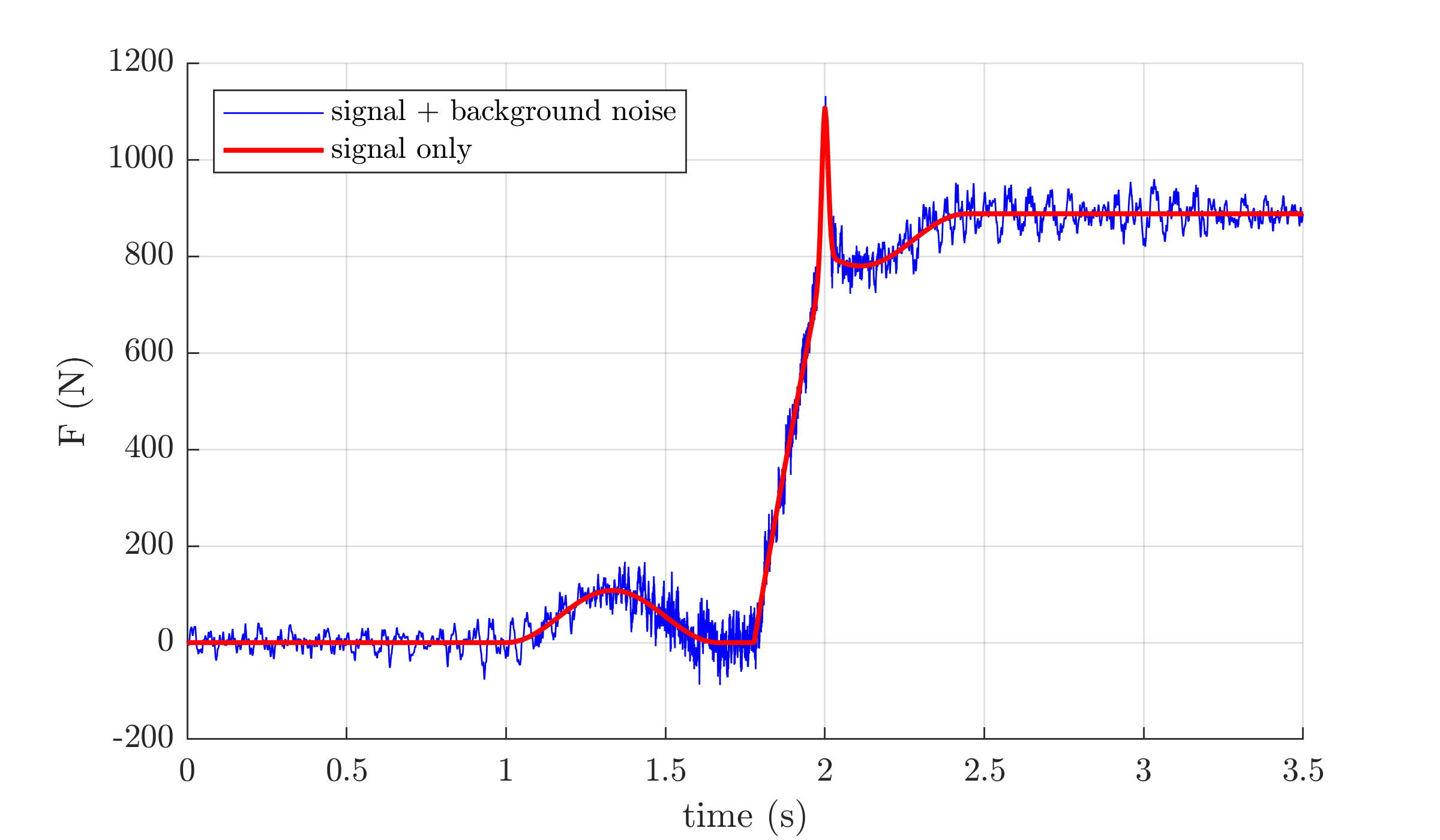}} \\ 
\subfigure[Different components of the \emph{signal}]
{\includegraphics[width=0.65\textwidth]{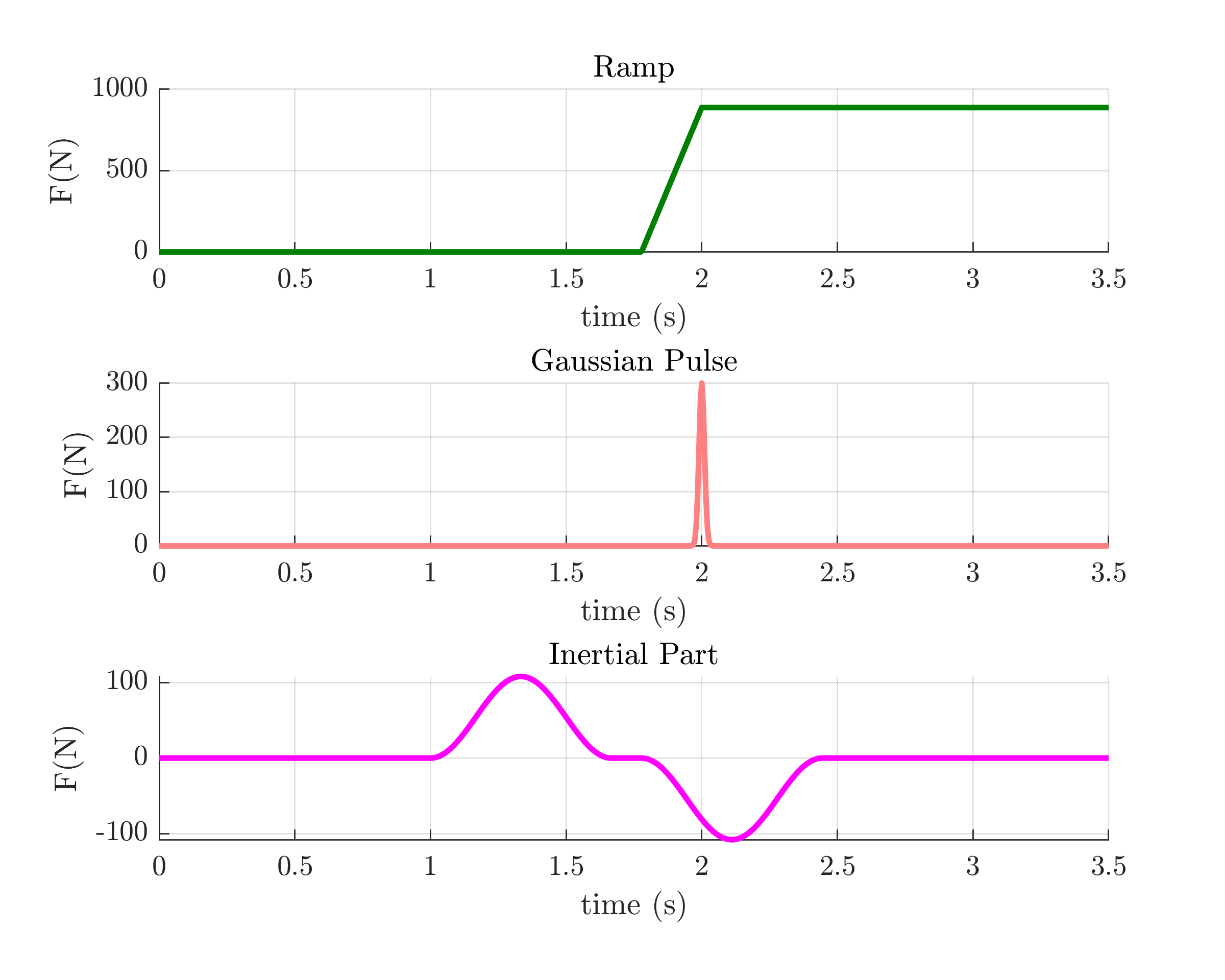}}
\caption{Synthetic signal employed to develop the denoising strategy and 
its different components.}
\label{fig:ramp_synthetic_signal}
\end{figure}

In order to devise a criterion to choose the set of parameters 
to be used in the denoising strategy, the EEMD-based method is  tested 
on a synthetic signal with a superimposed background noise. 
In this way, the effectiveness of the denoising strategy can be easily 
assessed by comparing reconstructed signal with the original one.
The synthetic signal is constructed to be representative of a typical 
force time history measured in the water entry tests. The time length of the 
synthetic signal and the sampling rate are the same as those of the 
signals recorded during the actual experiments.
The synthetic signal is constructed as as the
sum of three components, shown in Figure \ref{fig:ramp_synthetic_signal}.
The three components are a ramp, a Gaussian pulse and the inertial 
contribution, proportional to the fuselage acceleration shown in Figure 
\ref{fig:trajectory}. The ramp starts exactly at the time at
which the fuselage touches the water in the water entry 
tests, i.e. at $t=1.778$~s, as described in Section \ref{expsetup}. 
The pulse occurs at $t=2$~s. 

\begin{figure}[htbp]
\centering
\includegraphics[width=0.65\textwidth]{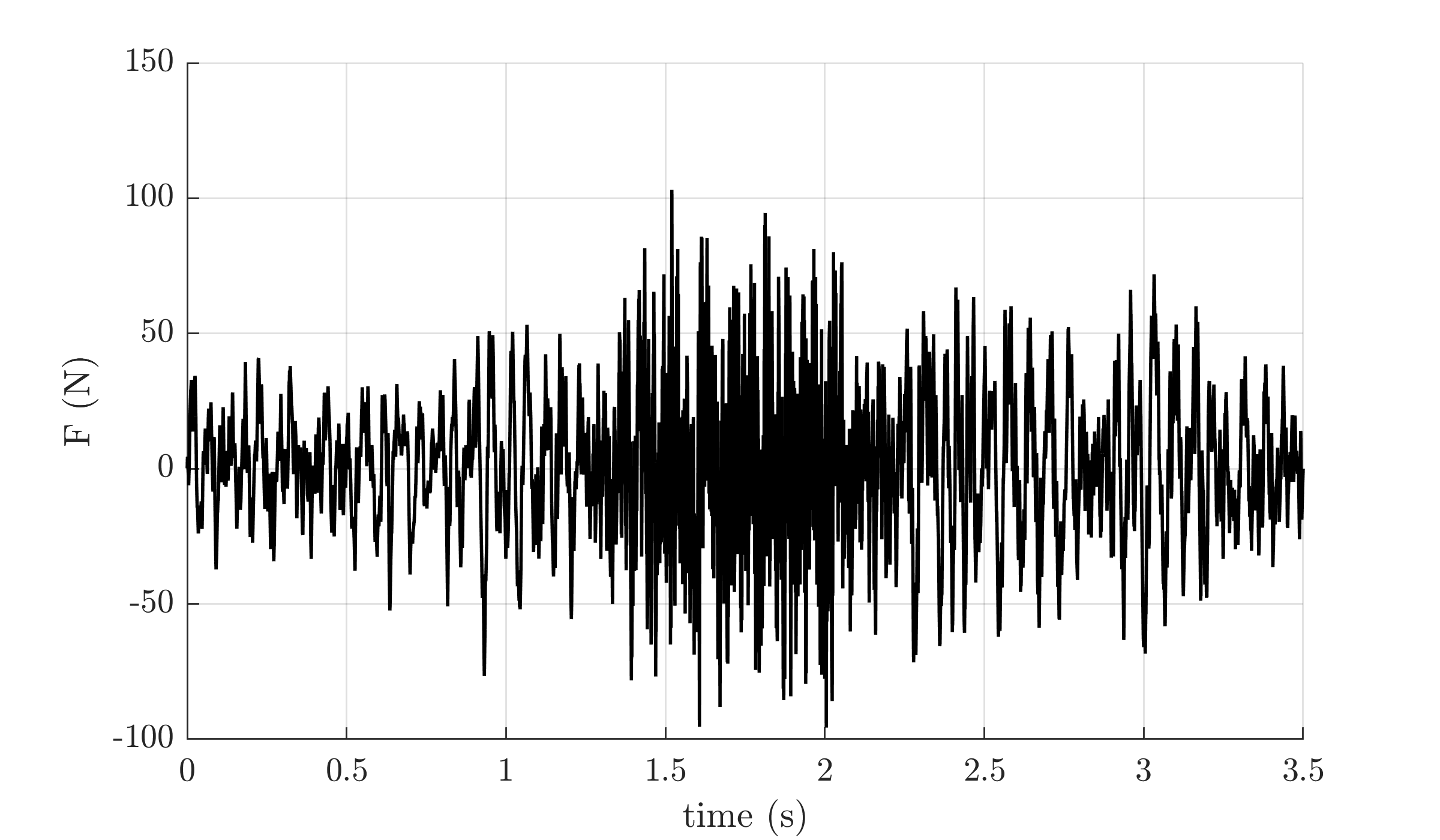}
\caption{Time history of background noise superimposed to the synthetic signal 
components.}
\label{fig:noisefromEXP}
\end{figure}

\begin{figure}[htbp]
\centering
\subfigure[Histogram and best-fitting Gaussian 
distribution]{\includegraphics[width=0.65\textwidth]{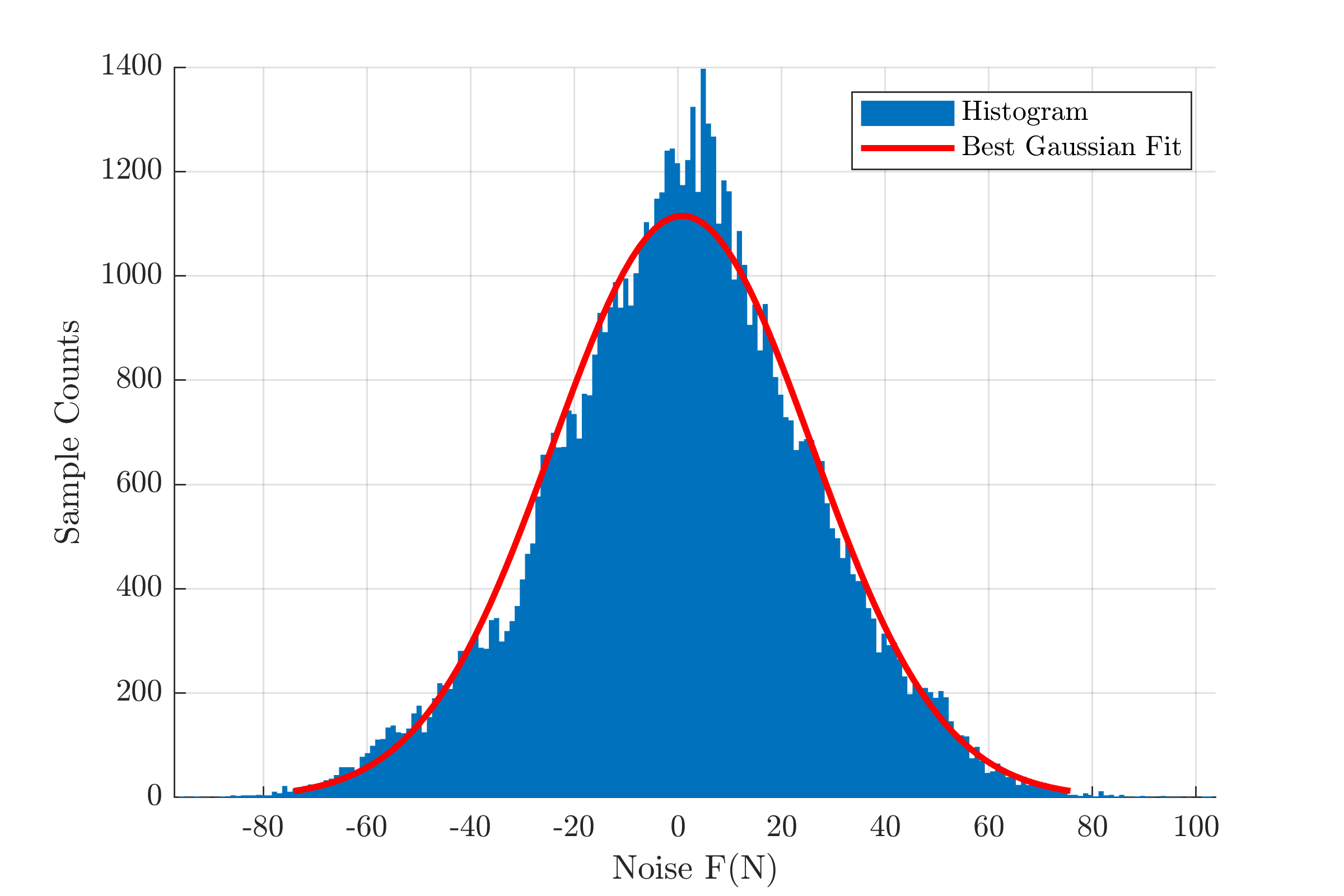}
} \quad
\subfigure[Power spectral 
density]{\includegraphics[width=0.65\textwidth]{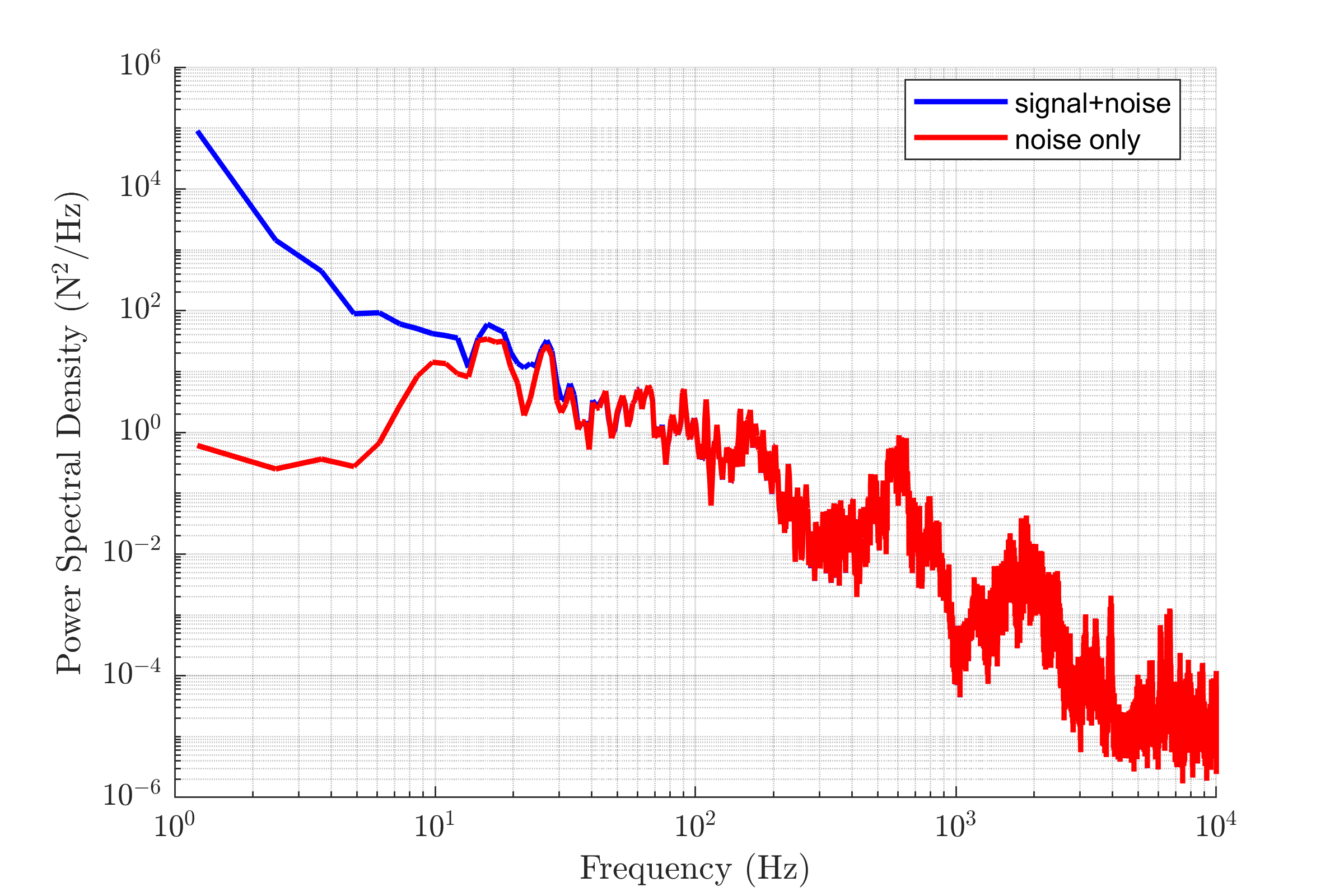}}
\caption{(a) Histogram of the background noise and best-fitted Gaussian 
distribution and (b) power spectral density of the synthetic signal 
with background noise and of the background noise only.}
\label{fig:histogram_PSD}
\end{figure}

A background noise component is added to the signal, which, to 
be more consistent with the type of noise encountered in the 
experimental tests, is obtained by subtracting the denoised
version of a typical experimental signal from the raw data. This \emph{background noise} 
time history is shown in Figure \ref{fig:noisefromEXP}.
It is observed that also the background noise 
displays significant non-stationary features, especially in the central part, 
i.e. during the vertical motion of the fuselage. 
In that time interval, the oscillations are larger in amplitude, 
and the variance of the signal is also higher.
In Figure \ref{fig:histogram_PSD}(a) the histogram of the background
noise is provided. It is worth noting that the background noise does not 
deviate significantly from the Gaussian distribution that best-fits the data, 
except for the central part, where the sample counts 
lie above the Gaussian distribution.
The power spectral density of the synthetic signal (signal+background 
noise) and of the background noise only are computed using the 
Welch's periodogram method and are shown in Figure 
\ref{fig:histogram_PSD}(b). The background noise displays a quite clear 
broad frequency content with peaks at 600~Hz, 1800~Hz and 6500~Hz. 
The power spectral density of the synthetic signal is well 
overlapped to that of the background noise for $f>$27~Hz. 
The peaks at 27~Hz and at 16~Hz occur on both the power 
spectral density of the synthetic signal and of the background noise, 
although they are slightly lower in the latter case. 

\begin{figure}[htbp]
\centering
\includegraphics[width=0.85\textwidth]{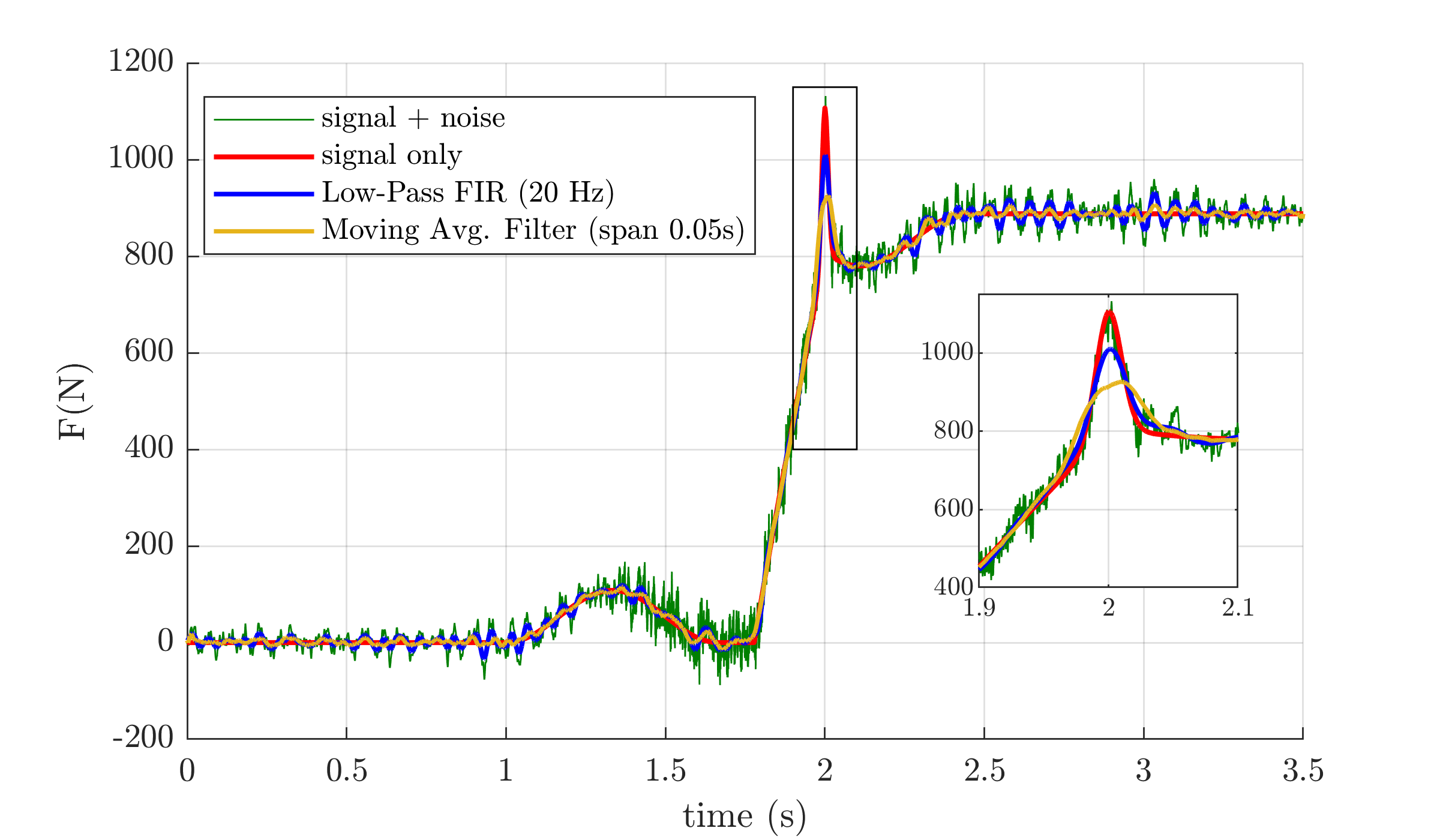}
\caption{Application to the synthetic signal 
of a FIR filter (designed with the Hanning 
window method) with a cut-off frequency 20 Hz and of a moving 
average filter with span 1000 samples or 0.05 s.}
\label{fig:classical_filters}
\end{figure}

By looking at Figure \ref{fig:histogram_PSD}(b), the simplest way to 
denoise the signals seems to be a classical low-pass FIR filter with a 
cut-off frequency of 20~Hz or a moving average filter 
spanning 1000 samples, which corresponds to a time interval of 0.05~s. 
The outputs of the two filters are shown in Figure \ref{fig:classical_filters}.
The use of the moving average filter reduces the background 
noise but smooths out significantly the Gaussian pulse at $t=2$~s. 
On the other hand, the application of the FIR filter has a lower effect 
on the peak smoothing, but it does not eliminate the oscillating behaviour. 
Given the nature of the background noise, which exhibits a 
significant low frequency content (see Figure
\ref{fig:histogram_PSD}(b)), it is very difficult to suppress the 
spurious oscillations and at the same time to preserve the sharp features of the signal. 
This is the reason why non-stationary filtering techniques are deemed essential for such cases.
\subsection{Determination of the EEMD Parameters}
%
In order to perform the EEMD, the input signal is corrupted 
with an artificial white Gaussian noise with amplitude 
 $N_a \, \sigma$, where $\sigma$ is the standard deviation of signal
computed over a time interval in which the
trend is constant, thus being representative
of the background noise. The parameters 
$N_a$ and $N_e$ have to be determined based on the characteristics of the signals.

\begin{figure}[htbp]
\centering
\includegraphics[width=0.80\textwidth]{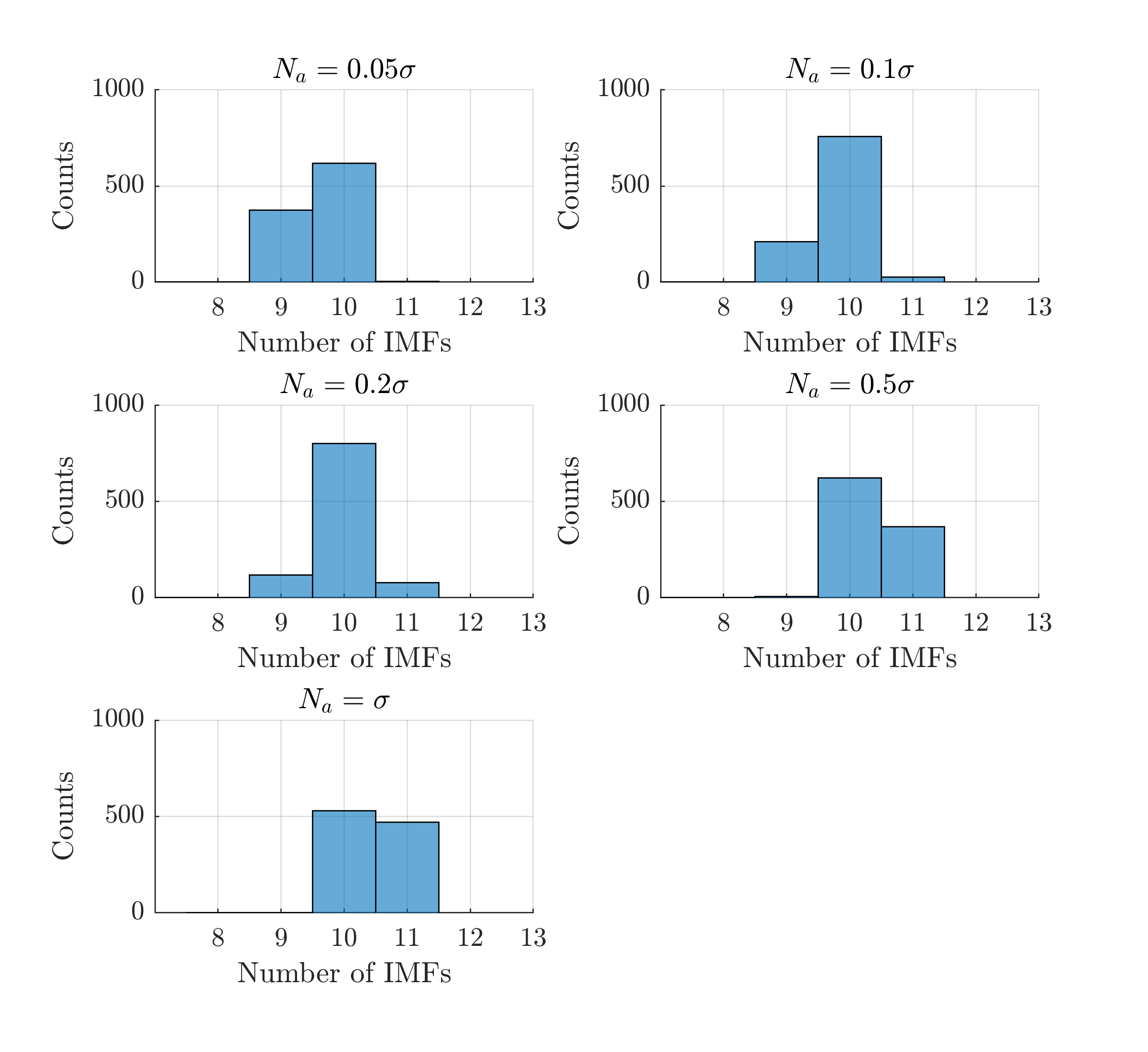}
\caption{Histograms of the number of IMFs obtained over 1000 EMDs of the
synthetic signal, with different values of the artificial noise 
amplitude $N_a \, \sigma$.}
\label{fig:histograms_IMFs_EEMD_NaXX_Ne1000_SynthMixed}
\end{figure}

The number of modes retrieved via the EEMD method is typically higher 
than what would be obtained via the standard EMD. This is because the addition of 
artificial white noise introduces more scales in the input, which
are distributed among the various IMFs and may not be eliminated completely 
through the ensemble averaging. 
As such, it is expected that if 
$N_a$ is higher, also the number of modes tends to be higher, and some 
redundant modes can also be generated \cite{zhang2010performance}.
Furthermore, due to the features of the signal, which is characterised by
sharp variations, and to the implementation of the EMD algorithm 
(in particular the stopping criterion both for the inner sifting cycle and 
for the outer while cycle), each pure EMD does not return the same number of modes.
For example, the number of IMFs obtained by performing an EEMD over 1000
realizations for different values of $N_a$ are shown in Figure
\ref{fig:histograms_IMFs_EEMD_NaXX_Ne1000_SynthMixed}.
In order to perform the ensemble average of the modes,
it is essential that the decompositions of all perturbed 
signals return the same number of modes. Unfortunately, this cannot be imposed a priori. 
Therefore, a large number of EMDs is performed and the ensemble
averaged modes are computed on the subset of the EMDs 
that return the most frequent number of modes, discarding the others.
Figure \ref{fig:histograms_IMFs_EEMD_NaXX_Ne1000_SynthMixed} shows that
 the number of modes grows when increasing the amplitude of the artificial noise.

\begin{figure}[htbp]
\centering
\subfigure[$N_e$=1, 10, 20, 50 and 100.]{\includegraphics[width=0.47\textwidth]{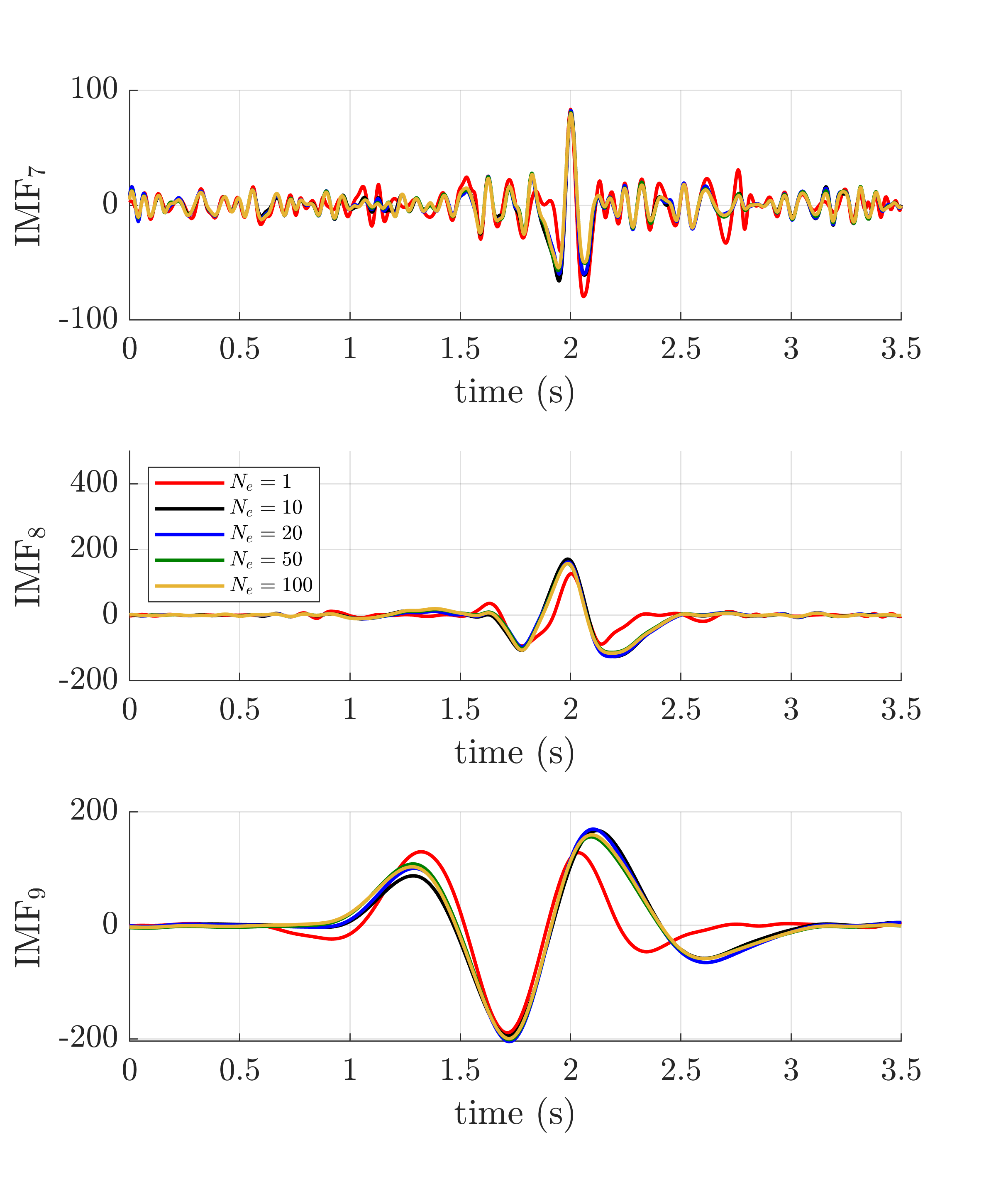}
} \quad
\subfigure[$N_e$=100, 150, 200, 500 and 1000.]{\includegraphics[width=0.47\textwidth]{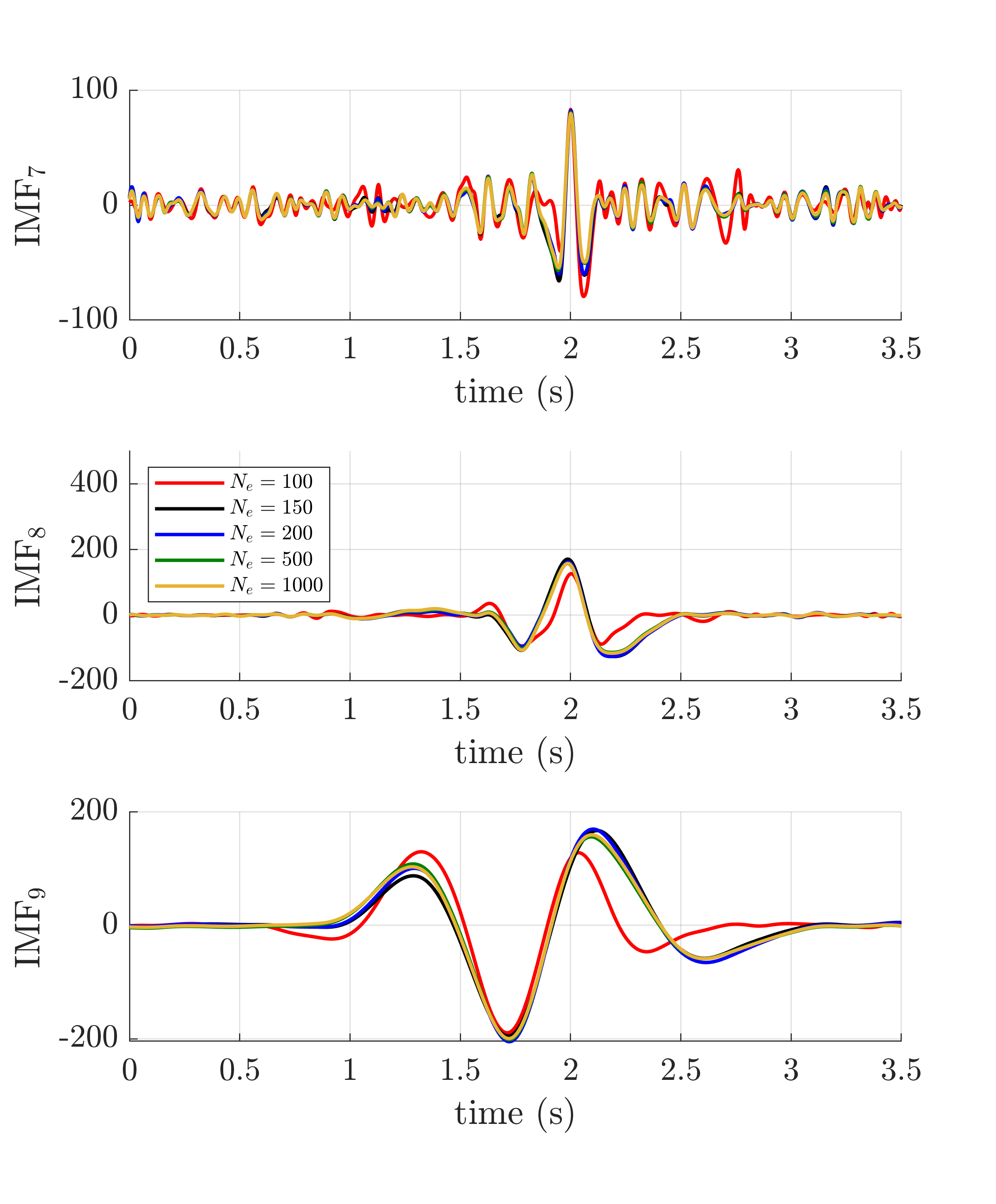}}
\caption{IMF$_7$, IMF$_8$ and IMF$_9$ obtained by increasing the 
number of artificial noise realizations using the EEMD 
algorithm on the synthetic signal: (a) $N_e$ from 1 to 100 and (b) $N_e$ from 100 to 1000}
\label{fig:EEMD_ConvergenceNe_Na01Std_SynthMixed}
\end{figure}

\begin{figure}[htbp]
\centering
\includegraphics[width=0.99\textwidth]
{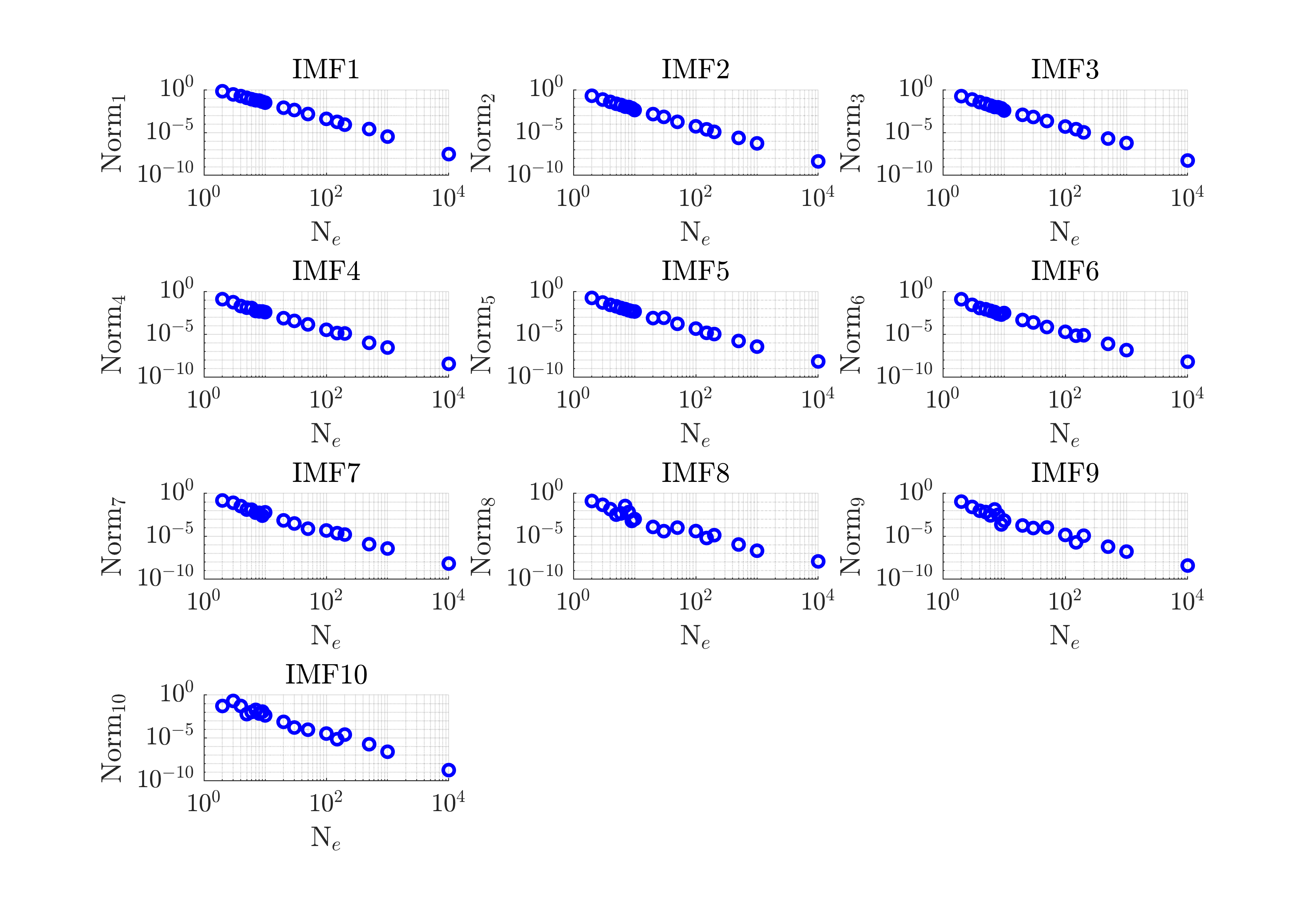}
\caption{Norm indicating the ``distance'' between a mode obtained 
with a certain value of $N_e$ and $N_e+1$, as a function of $N_e$. The EEMDs are all 
performed with $N_a$=0.1.}
\label{fig:NormsConvergence_EEMD_Na01Std_NeXX_SynthMixed}
\end{figure}

As noticed in \cite{wu2009ensemble}, the parameters $N_a$ and $N_e$ are 
not independent, because if an artificial noise realization with a 
higher amplitude is added to the original signal, a higher $N_e$ 
is required to achieve convergence in the ensemble averaging process.
An analysis of convergence is performed by assuming $N_a=0.1$, which 			 
is the value used in the following.
The EEMD modes are computed by for a different number of ensembles $N_e$. 
Figure \ref{fig:EEMD_ConvergenceNe_Na01Std_SynthMixed}(a) shows
three modes of the EEMD, namely IMF$_7$, IMF$_8$ and IMF$_9$, computed as
ensemble average of a different number of EEMDs from 10 to 100.
It can be noticed that the modes approach a certain shape when the number of realizations of the ensemble average is increased. 
In Figure \ref{fig:EEMD_ConvergenceNe_Na01Std_SynthMixed}(b)
it is shown that if the number of realizations is increased
beyond 100, the variability of the modes is much lower.
In order to introduce a more precise convergence criterion, a norm that 
measures the ``distance'' between an IMF of the EEMD computed with $N_e$ 
perturbations and an IMF obtained with $N_e +1$ perturbations is defined as:
\begin{linenomath}
\begin{equation}
\left. \text{Norm}_i \right \vert_{N_e} =\frac{ \displaystyle{\int \left( \left.
\textrm{IMF}_i(t)\right\vert_{N_e+1} -  \left.\textrm{IMF}_i(t)\right
\vert_{N_e} \right)^2 dt}}{ \displaystyle{\int \left (\left. \textrm{IMF}_i(t) 
\right \vert_{N_e+1} \right)^2 dt } } \;\;.
\end{equation}
\end{linenomath}
The tested number of ensembles varies from 2 to 10000.
The norms are plotted in Figure 
\ref{fig:NormsConvergence_EEMD_Na01Std_NeXX_SynthMixed} as a function 
of $N_e$ for all the IMFs. 
The norm for all the IMFs tends to decrease when $N_e$ 
is increased.
The last modes, in which the artificial noise 
in each perturbed signal introduces additional artificial 
low frequency components, clearly display some oscillations 
in the norm for low $N_e$.
It is assumed that a satisfactory convergence is achieved
when the norm drops below $10^{-5}$. Based on the
data shown in Figure \ref{fig:NormsConvergence_EEMD_Na01Std_NeXX_SynthMixed}
this generally happens, at least for the last modes, at about $N_e=500$.
To be even more conservative, in the following the EEMD is performed by using $N_e$=1000.

\begin{figure}[htbp]
\centering
\includegraphics[width=0.8\textwidth]{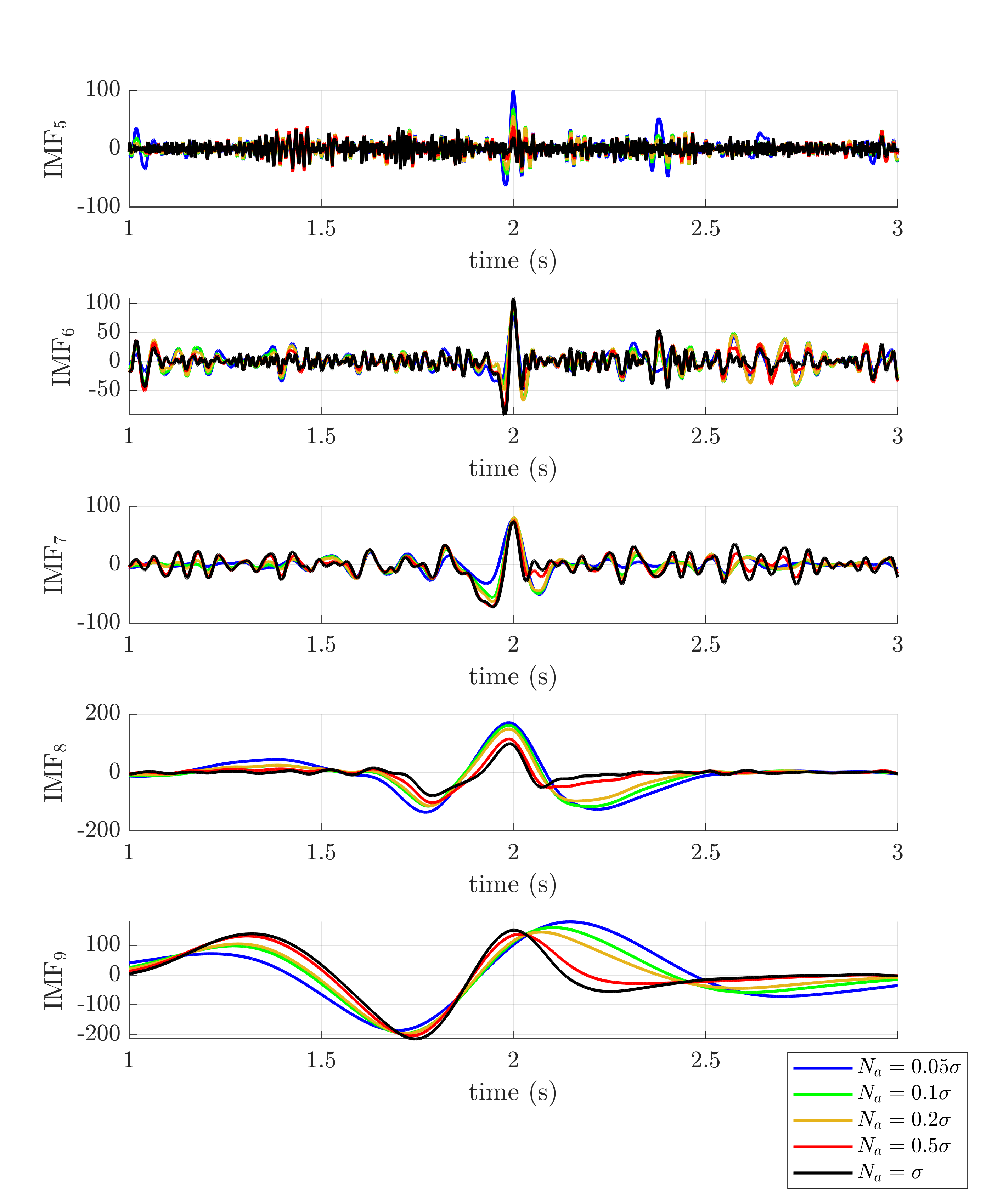}
\caption{IMF$_5$, IMF$_6$, IMF$_7$, IMF$_8$ and IMF$_9$ obtained 
from the EEMD of the synthetic signal with 
different artificial noise amplitudes and $N_e$= 1000.}
\label{fig:EEMD_Stds_NaXX_Ne1000_SynthMixed}
\end{figure}

It is less straightforward to establish the optimal value of $N_a$. 
In the paper that introduced the EEMD first
\cite{wu2009ensemble} the values of 0.1 and 0.2 for $N_a$ are proposed. It is   
also suggested choosing $N_a$ by taking into consideration the type of signal. In Figure 
\ref{fig:EEMD_Stds_NaXX_Ne1000_SynthMixed} the modes from IMF$_5$ to IMF$_9$, which result from the EEMDs with $N_e$=1000 and different values of $N_a$, are shown.
It is worth noting that in all cases the reconstruction of the 
original signal as the sum of all modes still holds,
aside from a small difference, in this case negligible, due to the fact that, differently from the EMD, the EEMD is not a complete decomposition. 
It is observed that for such 
type of signals, if $N_a$ is higher than 0.2, the peaks at t=2~s are smoothed out, 
especially in the IMF$_5$ and IMF$_8$. 
In the present case, those peaks results from the decomposition of
the Gaussian pulse component of the synthetic signal, 
see Figure \ref{fig:ramp_synthetic_signal},
hence it is necessary to preserve them.
It can be anticipated that this is achieved in this paper
by employing a thresholding strategy, see
Section \ref{thresholding_modalities}, according to which the signal values 
below a certain threshold are discarded, assuming that they are part of the background
noise of the mode. Of course, the thresholding strategy preserves better 
the peaks that are more pronounced, hence choosing $N_a < 0.2$ provides an advantage
in this sense.
Furthermore, it is also observed that an increase in the amplitude of
the artificial noise causes a corresponding growth of the amplitudes of all frequency
components, which make all modes noisier.
This is evident for instance by looking at IMF$_5$, IMF$_6$ and IMF$_7$.
The higher noise level makes the
application of the thresholding strategy more difficult, 
as some non-physical peaks and valleys may appear, which cannot be 
distinguished from the physically relevant ones.
Given these two circumstances, a value of $N_a$=0.1 is chosen. It must be 
stressed again that such a choice is made based on this specific 
type of signal and on the derived EEMD. For different type of signals, 
the optimal value of $N_a$ might be different.

\begin{figure}[htbp]
\centering
\includegraphics[width=0.90\textwidth]{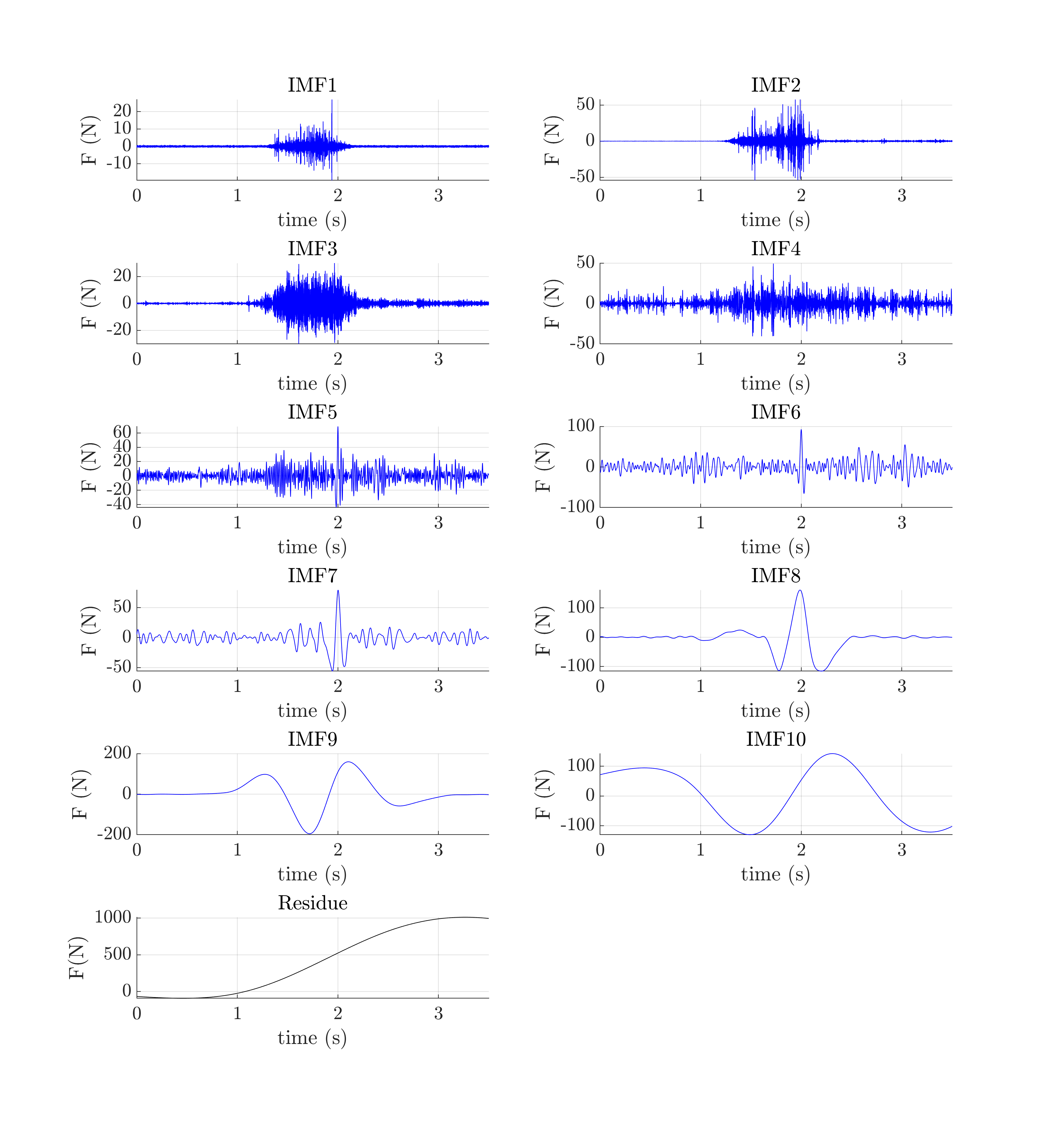}
\caption{IMFs obtained from the EEMD of the synthetic signal
with $N_e$= 1000 and $N_a$ = 0.1}
\label{fig:IMFs_EEMD_Na01_Ne1000_SynthMixed}
\end{figure}

Once the number of realizations and the artificial noise amplitude is
chosen, an EEMD with $N_a=0.1$ and $N_e=1000$ is performed, 
yielding ten IMFs and a residue. The decomposition  
is shown in Figure \ref{fig:IMFs_EEMD_Na01_Ne1000_SynthMixed}. 
The first IMFs contain mostly oscillations at 
high frequencies, 
which are presumably associated with the electronic noise. 
As a matter of fact, most of 
the oscillations with increased variance in the central time interval, 
i.e. $1.2$ s $< t < 2.1$ s, are contained in those modes. That time
interval is the one during which the fuselage is moving, and thus the 
control-system of the actuators is operating. 
On the other hand, the oscillations due to the mechanical vibrations of the 
carriage and of the actuator systems are also visible in the last 
IMFs, being characterized by a lower frequency content. 
Finally, the presence of the pulse at $t=2$~s is visible in modes from 
IMF$_5$ to IMF$_{10}$. The 
oscillation related to the pulse broadens more and more
moving towards the last modes, as expected \cite{stallone2020new}).
\subsection{Thresholding Methods}
\label{thresholding_modalities}
As anticipated, the modes are a combination of physical components and
spurious noise. In the first modes it is expected that the noise component
is dominant, but it cannot be excluded that a physical component exists as
well. Based on the above considerations, it is worth introducing a
technique which allows to preserve the relevant 
components and remove the background noise. 
One possibility to achieve such a goal is to perform a mode thresholding.
Denoising using thresholding is commonly performed in 
literature following the application of the Discrete Wavelet Transform 
\cite{donoho1995noising,donoho1994threshold,luo2012wavelet}.
In that case, the thresholding 
is applied to the wavelet coefficients, hence in 
the wavelet domain (i.e. in the time-frequency or time-scale domain). 
However, it is also possible to perform a thresholding in
the time domain, applying it directly to the IMF 
time samples, as described in \cite{kopsinis2009development}. 

Different types of thresholding can be performed: \emph{hard thresholding}, 
\emph{soft thresholding} and \emph{interval thresholding}. 
Given a signal $y(t)$, the corresponding hard 
thresholded signal $y_T(t)$ is given by:
\begin{linenomath}
\begin{equation}
y_T(t) =
\begin{cases} 
y(t) &\mbox{if } |y(t)| > T  \\
0 & \mbox{if } |y(t)| \leq T \;\;.
\end{cases}
\end{equation}
\end{linenomath}
Hard thresholding helps to preserve the highest peaks and valleys in 
the signal, but it introduces jumps at the instants where the time 
series crosses the threshold. Such jumps are quite 
evident, especially if the threshold value is relatively high. 
A soft-thresholding, or shrinkage, can be achieved by using the 
following approach:
\begin{linenomath}
\begin{equation}
y_T(t) =
\begin{cases} 
y(t)-T &\mbox{if } |y(t)| > T  \\
y(t)+T &\mbox{if } |y(t)| < -T  \\
0 & \mbox{if } |y(t)| \leq T  \;\;.
\end{cases}
\end{equation}
\end{linenomath}
With soft thresholding the modes are shrunk, so 
jumps are still present in the thresholded signal, but their amplitudes are
smaller. Even if there is a significant improvement compared to the
hard thresholding, soft thresholding is not suitable for
the present applications, in which it is required to preserve the full 
values of peaks of valleys in the time histories,
even in presence of large background noise components.

\begin{figure}[htbp]
\centering
\includegraphics[width=0.90\textwidth]{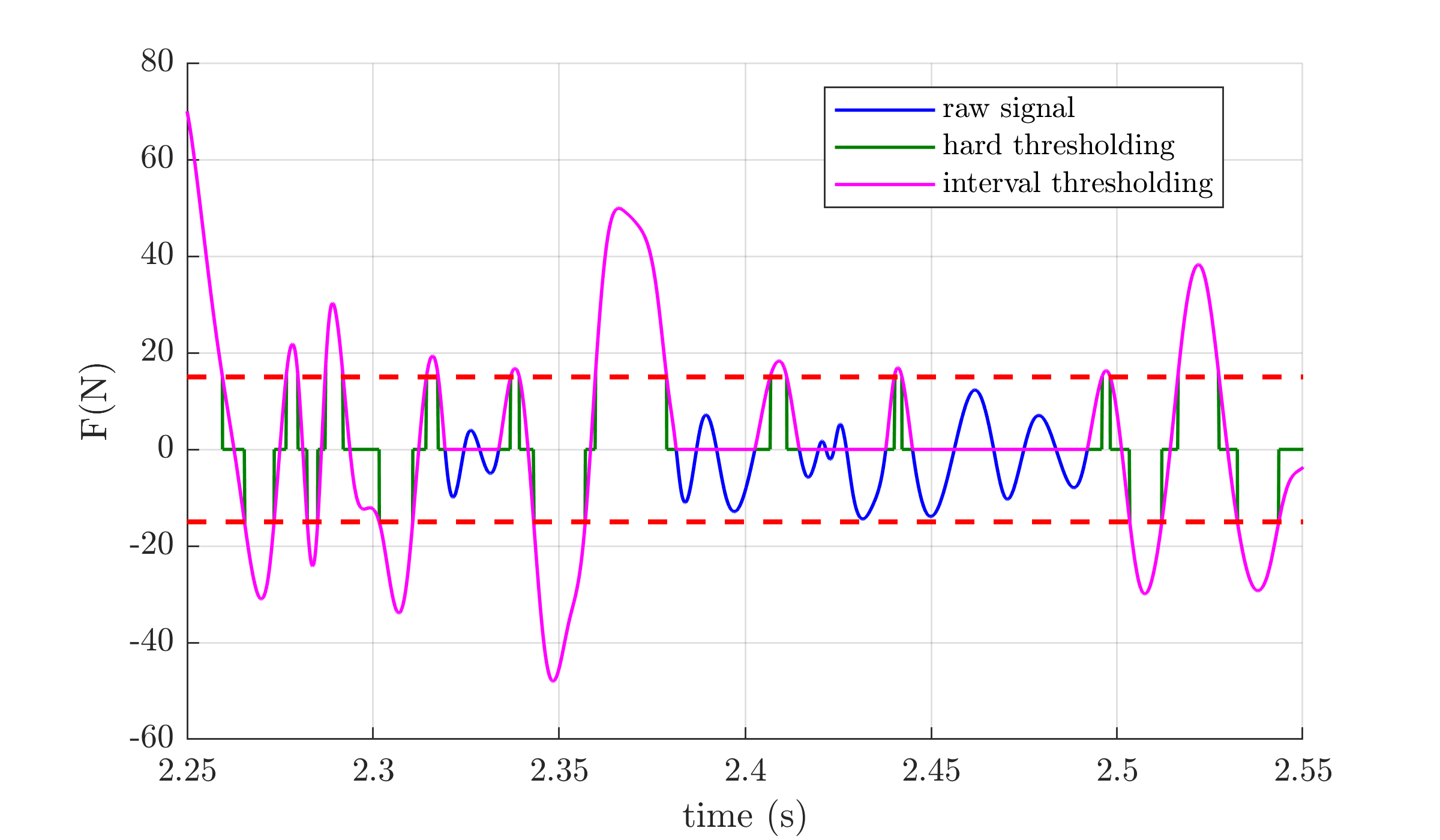}
\caption{Comparison between hard thresholding and interval 
thresholding.}
\label{fig:tresholding_hard_interval}
\end{figure}

A technique that seems more appropriate for the present applications is the hard 
\emph{interval thresholding}, introduced in 
\cite{kopsinis2009development}. The input signal $y(t)$ is initially divided 
into a series of time intervals in between two successive zero-crossings. Then 
a thresholding is performed interval-wise rather than point-wise, meaning 
that
\begin{linenomath}
\begin{equation}
y_T(t_a \leq t < t_b) =
\begin{cases} 
y(t_a \leq t < t_b) & \mbox{if } \max \left(|y(t_a \leq t < t_b)| 
\right) > T \\
0 & \mbox{if } \max \left(|y(t_a \leq t < t_b)| \right) \leq T \\
\end{cases}
\end{equation}
\end{linenomath}
where $t_a$ and $t_b$ correspond to two successive zero-crossing time 
instants. The procedure is repeated for all the time intervals in 
between two successive zero-crossings. As pointed out in 
\cite{kopsinis2009development}, the application of interval thresholding has some similarities with the
wavelet thresholding, where setting to zero a wavelet coefficient that is below a
certain value, affects a set of contiguous time samples. 
A comparison between hard thresholding and interval thresholding is shown in Figure 
\ref{fig:tresholding_hard_interval}.
Interval thresholding helps to preserve the whole semi-oscillations 
of the original signal without introducing undesired jumps
in the thresholded output. Owing to such good peculiarities of the method, the
interval thresholding is used in the following.

\subsection{Denoising procedure and results}
%
Through the EMD or EEMD, the background noise is removed 
from the original signal by discarding some of the first modes 
in the reconstruction of the signal. 
Formally, the ``denoised'' signal is reconstructed by summing 
the last $N-M+1$ modes and the residue, which is:
\begin{linenomath}
\begin{equation}
\label{EEMD_partial_reconstruction}
\hat{y}_M(t) = \sum_{i=M}^N \textrm{IMF}_i (t) + \textrm{Res}(t)
\end{equation}
\end{linenomath}
where $M$ is the first mode considered in the reconstruction and $N$ is 
the total number of modes. This approach is referred to
as \emph{conventional EMD denoising}, see \cite{kopsinis2008empirical, 
kopsinis2009development}.

\begin{figure}[htbp]
\centering
\subfigure[Synthetic signal 
only]{\includegraphics[width=0.48\textwidth]{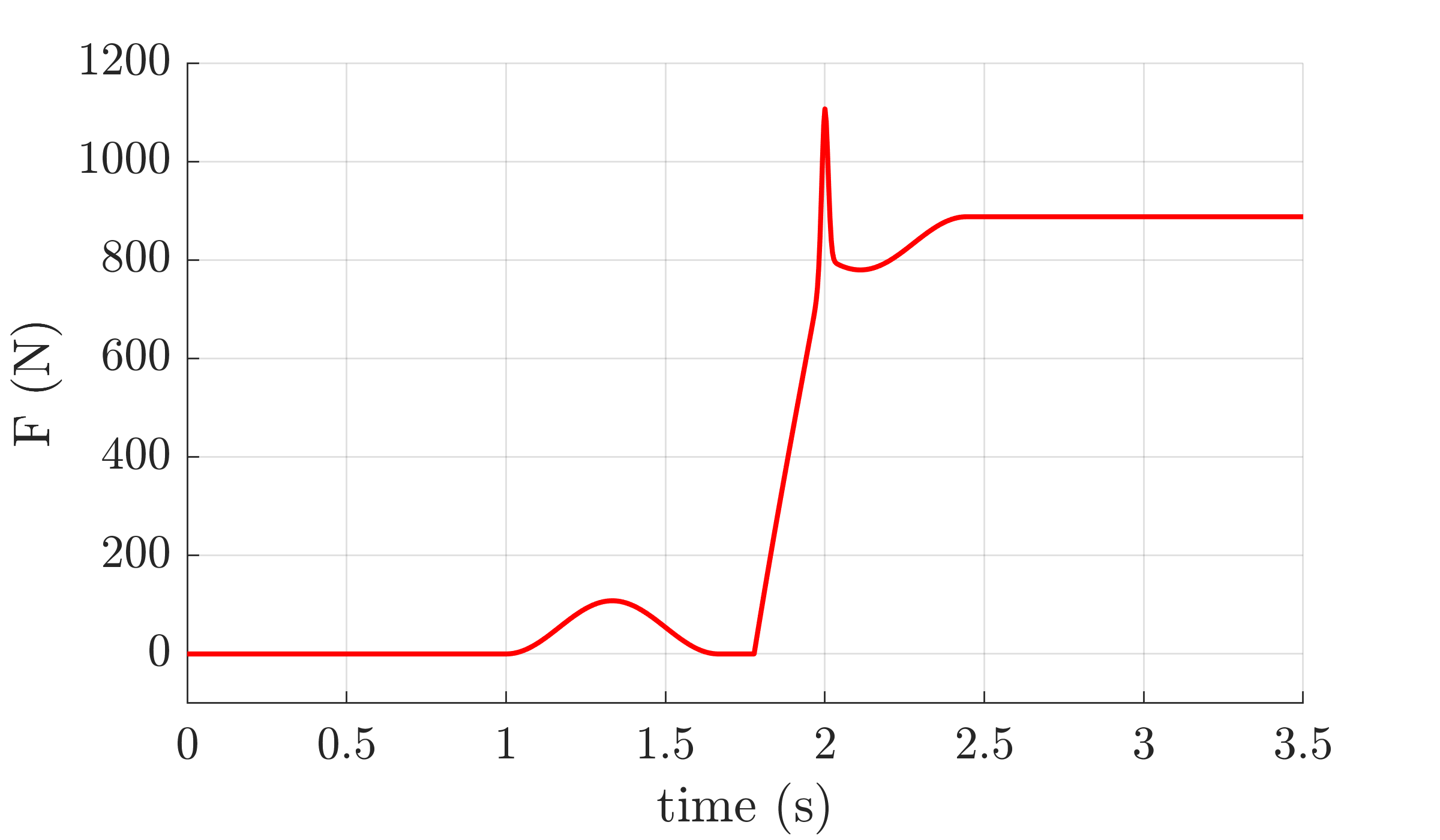}} 
\quad
\subfigure[Synthetic signal plus background 
noise]{\includegraphics[width=0.48\textwidth]{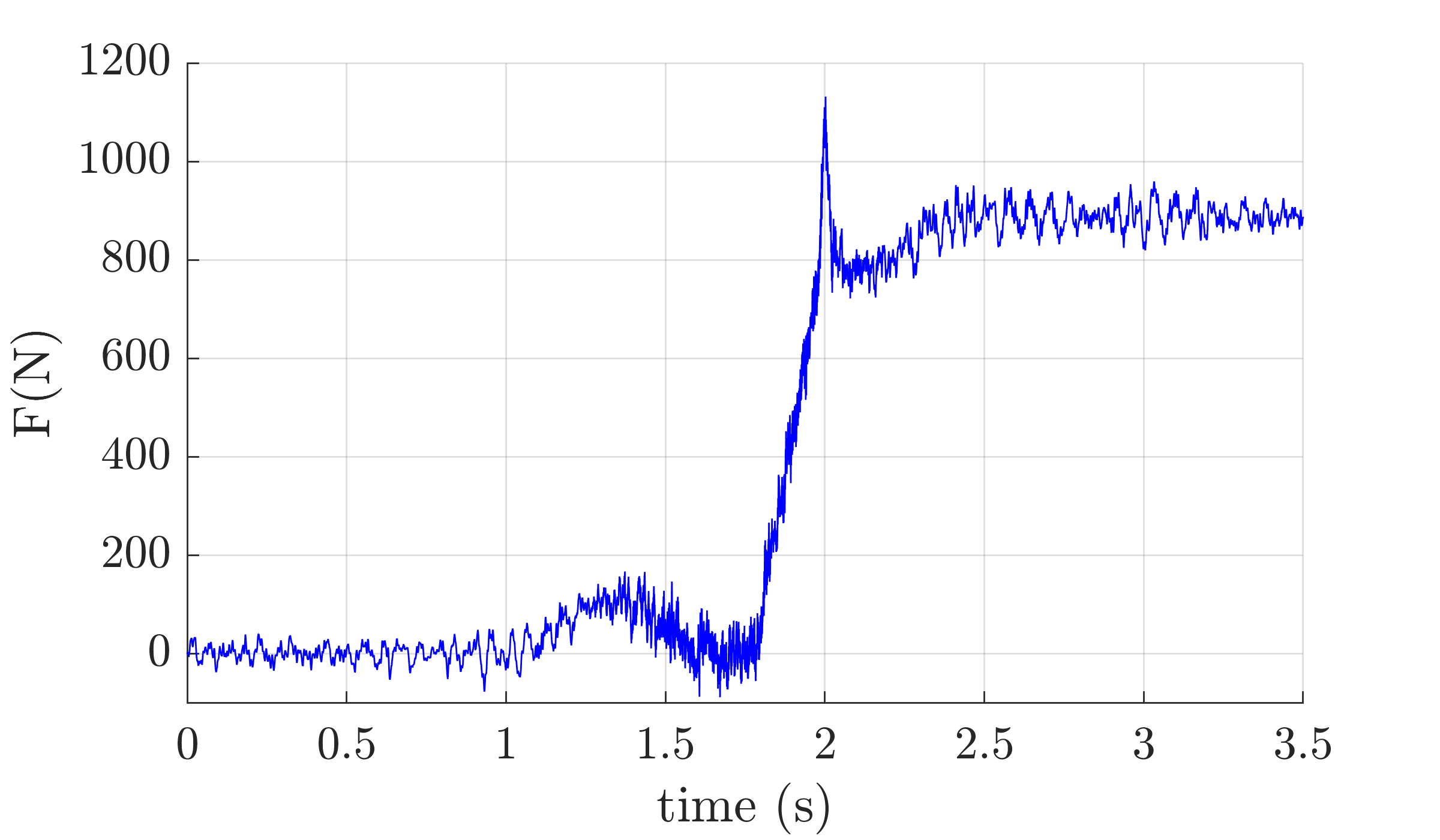}} 
\\
\subfigure[EEMD Synthetic signal 
only]{\includegraphics[width=0.48\textwidth]{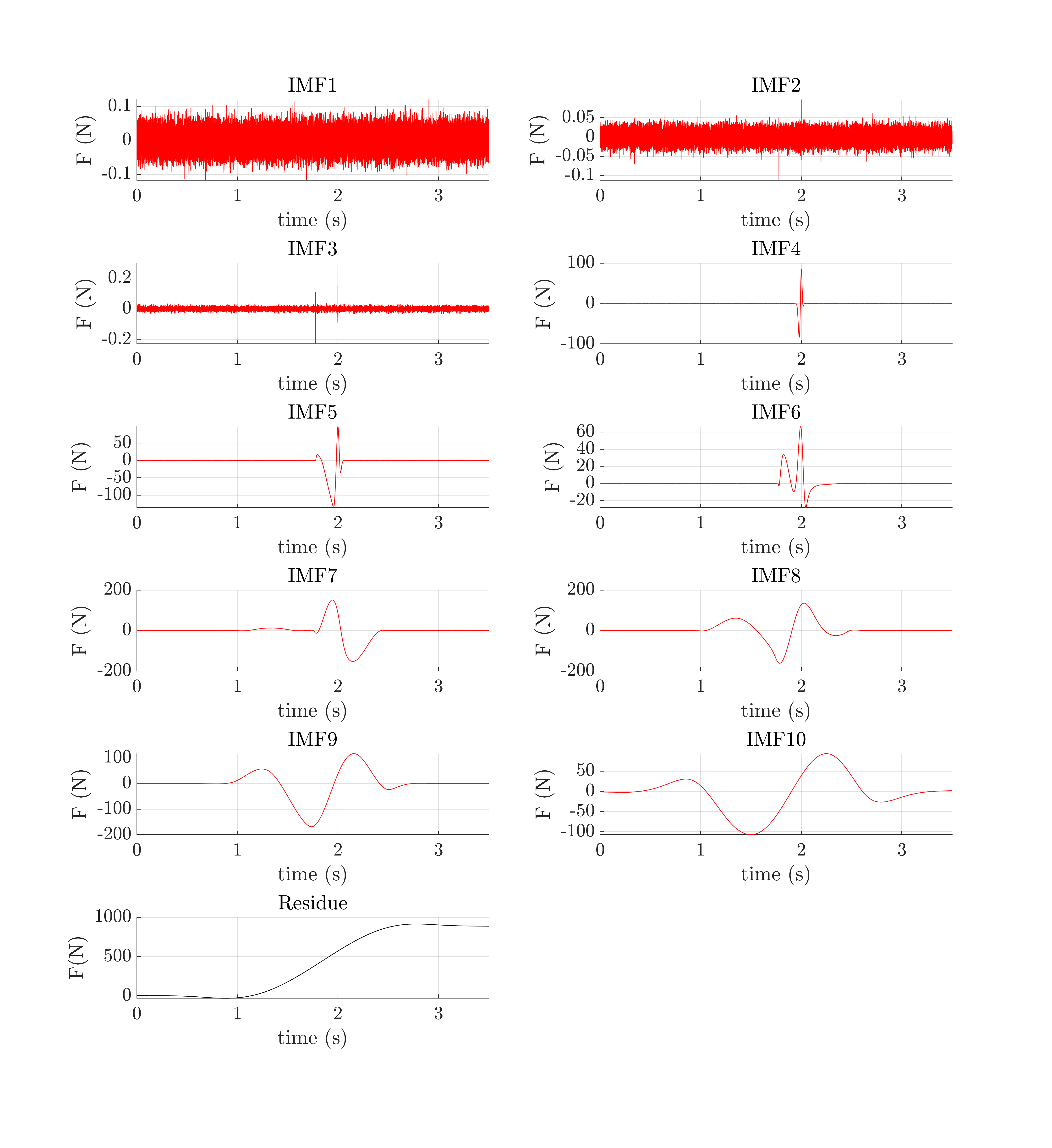}} \quad 
\subfigure[EEMD Synthetic signal plus background noise]
{\includegraphics[width=0.48\textwidth]
{IMFs_EEMD_Na01_Ne1000_SynthMixed.png}}
\caption{Synthetic signals (a) without and (b) with background noise and 
related EEMDs, (c) and (d) respectively.}
\label{fig:EEMD_Synthetic_EEMD_Comparisons}
\end{figure}

The most difficult part of the approach concerns the choice
of the number of modes to be accounted for and disregarded in 
the reconstruction, a choice that is
strongly dependent on the specific problem and on the noise
characteristics.
In this regard, some considerations can be made based on the
synthetic signal without the background noise.
It is worth noticing that this synthetic signal cannot be treated 
with the classical EMD, since it
does not contain sufficient maxima or minima to start the sifting cycles.
Nevertheless, the problem can be overcome by performing the EEMD.
For the reader's convenience, the signals without and with the background noise 
are shown in Figure \ref{fig:EEMD_Synthetic_EEMD_Comparisons}(a) and 
(b) respectively, and the corresponding EEMDs with $N_a$=0.1 and 
$N_e$=1000 are shown in Figure 
\ref{fig:EEMD_Synthetic_EEMD_Comparisons}(c) and (d) respectively.
Unfortunately, it is not possible to find a mode to mode correspondence 
between the two decompositions. 
However, the EEMD of the signal only is useful as it provides a 
visual 
indication of the shape of the modes that would be attained without 
background noise. 
By looking at Figure \ref{fig:EEMD_Synthetic_EEMD_Comparisons}(c), the first 
two IMFs of the 
synthetic signal just account for some residual noise that is not 
eliminated during the ensemble averaging; the amplitude of these modes
is rather small, though.  
The most relevant oscillations that should be accounted for in 
the signal reconstruction appear from the IMF$_3$ and beyond. 
The residue accounts mainly for the increasing trend of the signal, from 
a null value at the beginning to a positive constant value at the end. 
By looking at the EEMD of the signal with noise, shown in Figure 
\ref{fig:EEMD_Synthetic_EEMD_Comparisons}(d), it is possible to 
recognize modes with shapes quite similar to those of the
undisturbed signal. 
The background noise spreads throughout all the modes, although it 
looks negligible in the last three modes, 
i.e. IMF$_8$, IMF$_9$ and IMF$_{10}$, and it is smaller
than any prevailing mode oscillation in modes IMF$_5$, IMF$_6$ and
IMF$_7$.
The background noise is instead dominant in the first modes, 
from IMF$_1$ to IMF$_4$, where no prevailing oscillations are 
visible. Therefore, the latter modes can be disregarded in the
partial reconstruction for denoising.

\begin{figure}[htbp]
\centering
\subfigure[Original IMFs]{\includegraphics[width=0.47\textwidth]{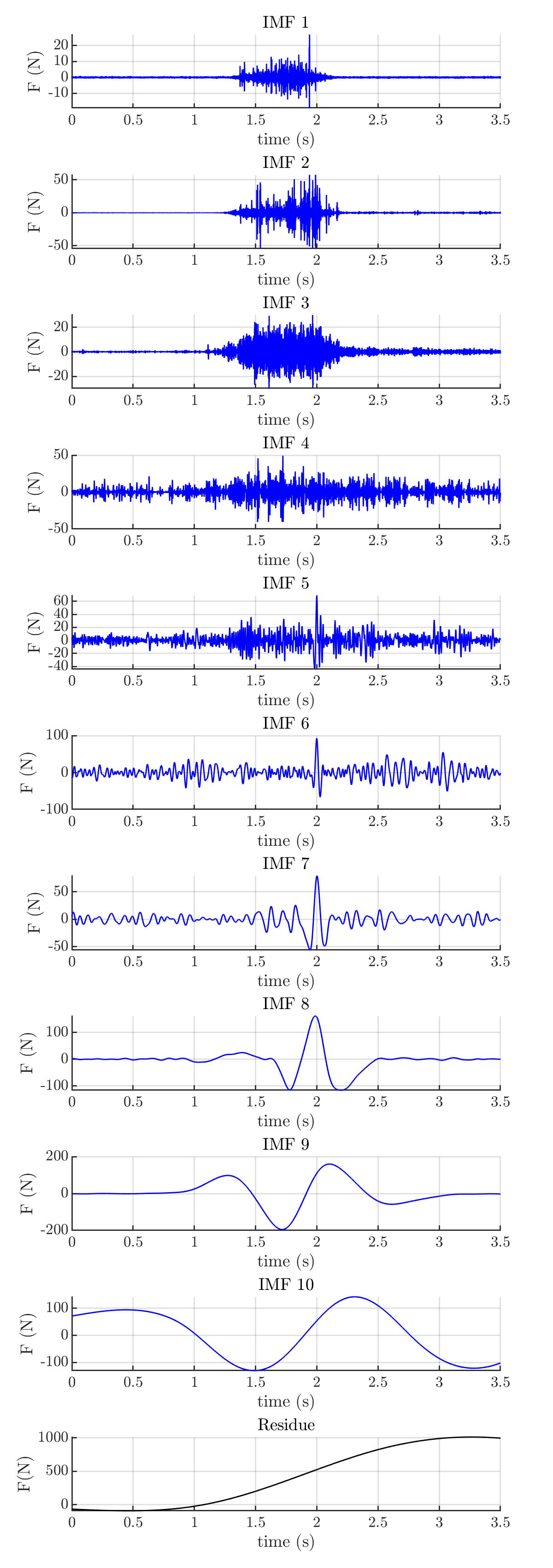}} \quad 
\subfigure[Thresholded IMFs]{\includegraphics[width=0.47\textwidth]{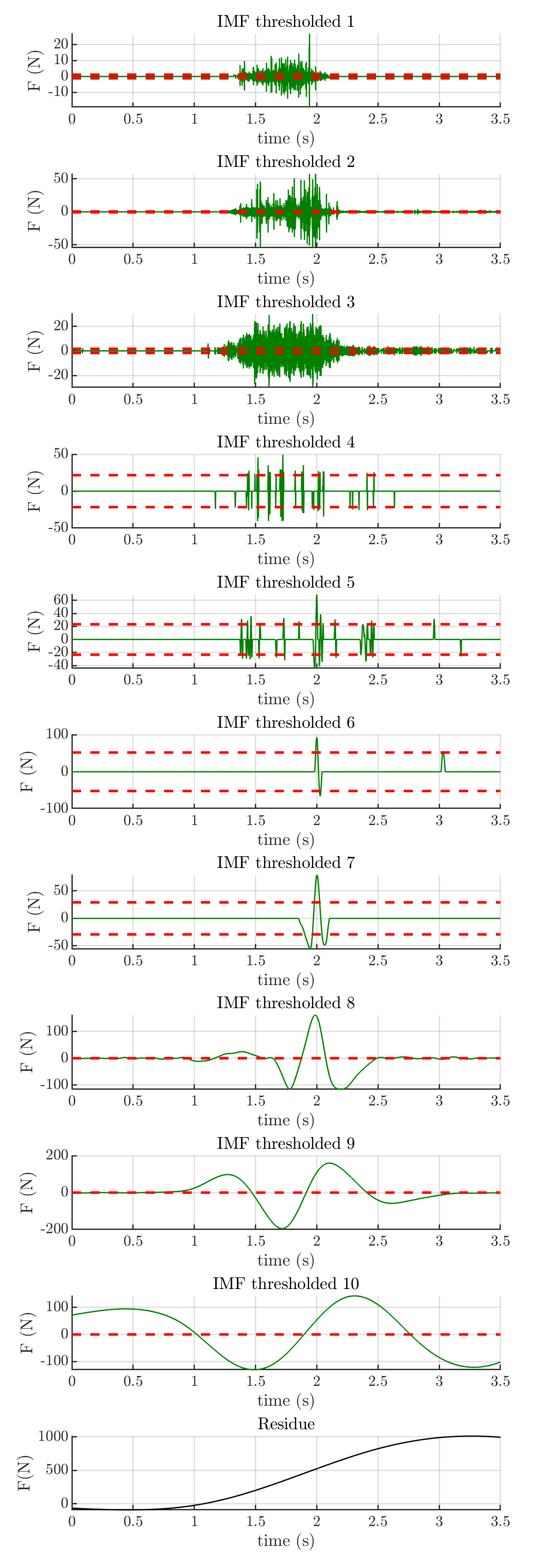}}
\caption{Comparison between the original IMFs and the thresholded IMFs of the synthetic signal using interval thresholding.}
\label{fig:ThresholdedIMFs_SyntheticMixed}
\end{figure}

Based on the above considerations, it is concluded that,
for such specific signal types, compared to the conventional EMD or EEMD 
denoising, an improved noise deduction can be obtained by performing a 
thresholding of 
IMF$_5$, IMF$_6$ and IMF$_7$, and then reconstructing the denoised signal, 
denoted with $\hat{y}_T(t)$, as:
\begin{linenomath}
\begin{equation}
\label{EEMD_reconstruction}
\hat{y}_T(t) = \sum_{i=L}^{(M-1)} \textrm{IMF}_{T,i}(t) + 
\sum_{i=M}^{N} \textrm{IMF}_{i}(t) + \textrm{Res}(t)
\end{equation}
\end{linenomath}
where $\textrm{IMF}_{T,i}(t)$ indicates the $i^{th}$ thresholded IMF, $L=5$ and $M=8$.
Given the background noise that is superimposed to the signal (see 
Figure \ref{fig:noisefromEXP}), it makes sense to define a 
mode-dependent threshold as 
\begin{linenomath}
\begin{equation}
T_i=N_{\sigma} \, \sigma_i \qquad i=1 \, ... \, N
\end{equation}
\end{linenomath}
where $N$ is, again, the number of EEMD modes, $\sigma_i$ is the 
standard deviation of each mode over the initial time interval, from 
$t=0.05$ s to $t=0.95$~s, and $N_{\sigma}$ is a parameter
to be determined. 
The interval over which $\sigma_i$ is computed for the
threshold definition is the one in which 
only the noise induced by the vibrations of the
carriage and of the actuator system is present. In fact, as
anticipated, such contributions to the background noise affect all modes.
In the central part of the acquisition the amplitude of the
background noise is higher but it is mostly associated with the 
electronic noise, which has a relatively high frequency content and is
almost completely contained in the first modes, which are discarded in the partial
reconstruction.
A comparison between the original IMFs derived from the EEMD and the 
IMFs after the application of the interval thresholding is shown in Figure 
\ref{fig:ThresholdedIMFs_SyntheticMixed}. 
The threshold for last three modes is set to 0, i.e. those modes are not 
thresholded, whereas it is $T_i=N_{\sigma} \sigma_i$ with $N_{\sigma}=4.75$ 
for the other modes. These values are chosen by trial and 
error, in order to remove the undesired oscillations in IMF$_5$, IMF$_6$ 
and IMF$_7$. 

\begin{figure}[htbp]
\centering
\includegraphics[width=0.85\textwidth]
{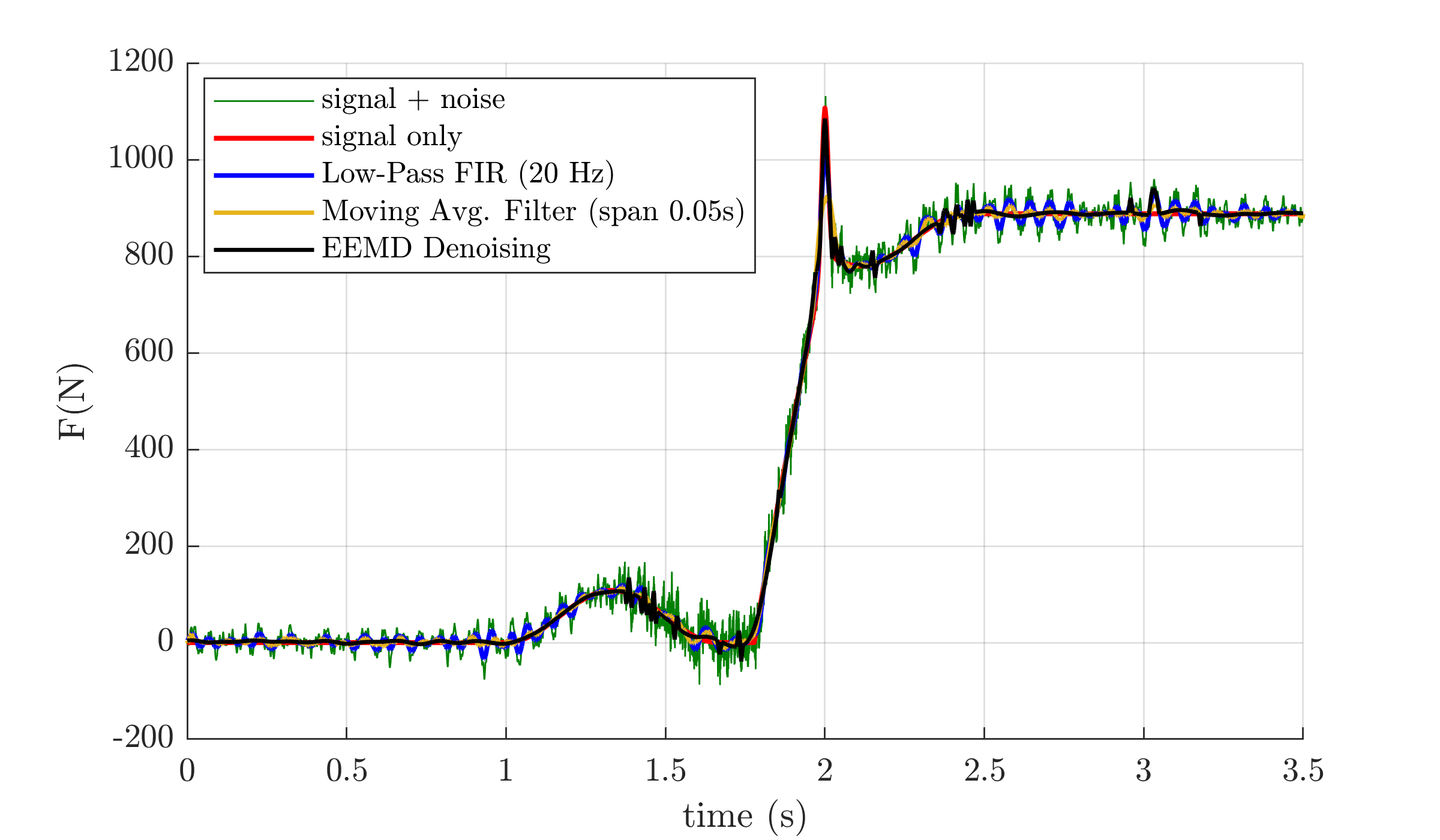}
\caption{Comparison between the different denoising techniques: moving 
average filter (span 1000 samples i.e. 0.05 s), low-pass FIR 
Hanning window filter (cut-off frequency 20 Hz) and the EEMD denoising with interval 
thresholding.}
\label{fig:EEMDDenoising_T4sigmaINIT_Nkeep3_Nmore3_EEMD_Na01_Ne1000_SynthMixed}
\end{figure}

\begin{figure}[htbp]
\centering
\subfigure[Close-up view about the Gaussian 
peak]{\includegraphics[width=0.48\textwidth]{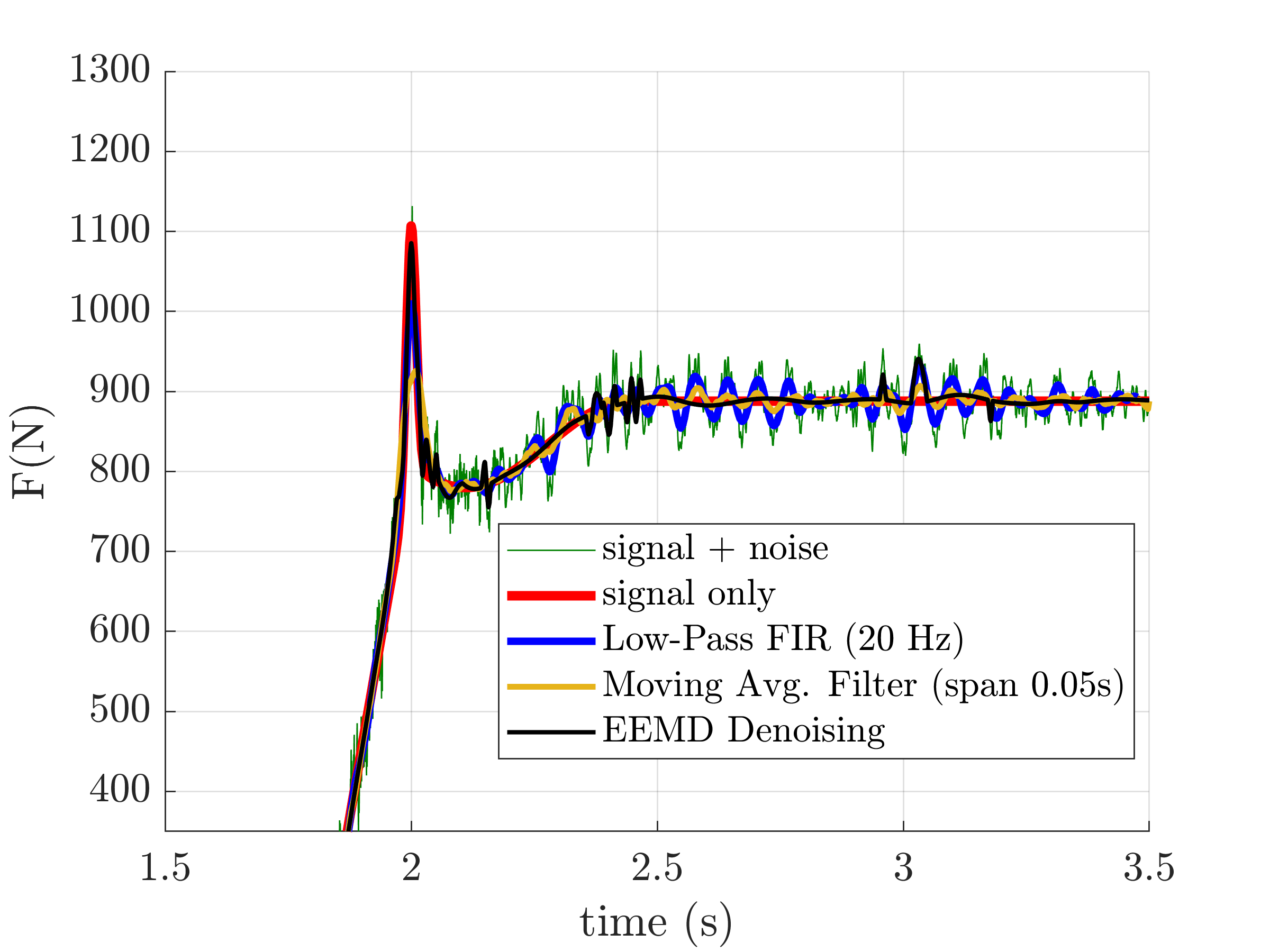}} \quad
\subfigure[Close-up view near the 
ramp]{\includegraphics[width=0.48\textwidth]{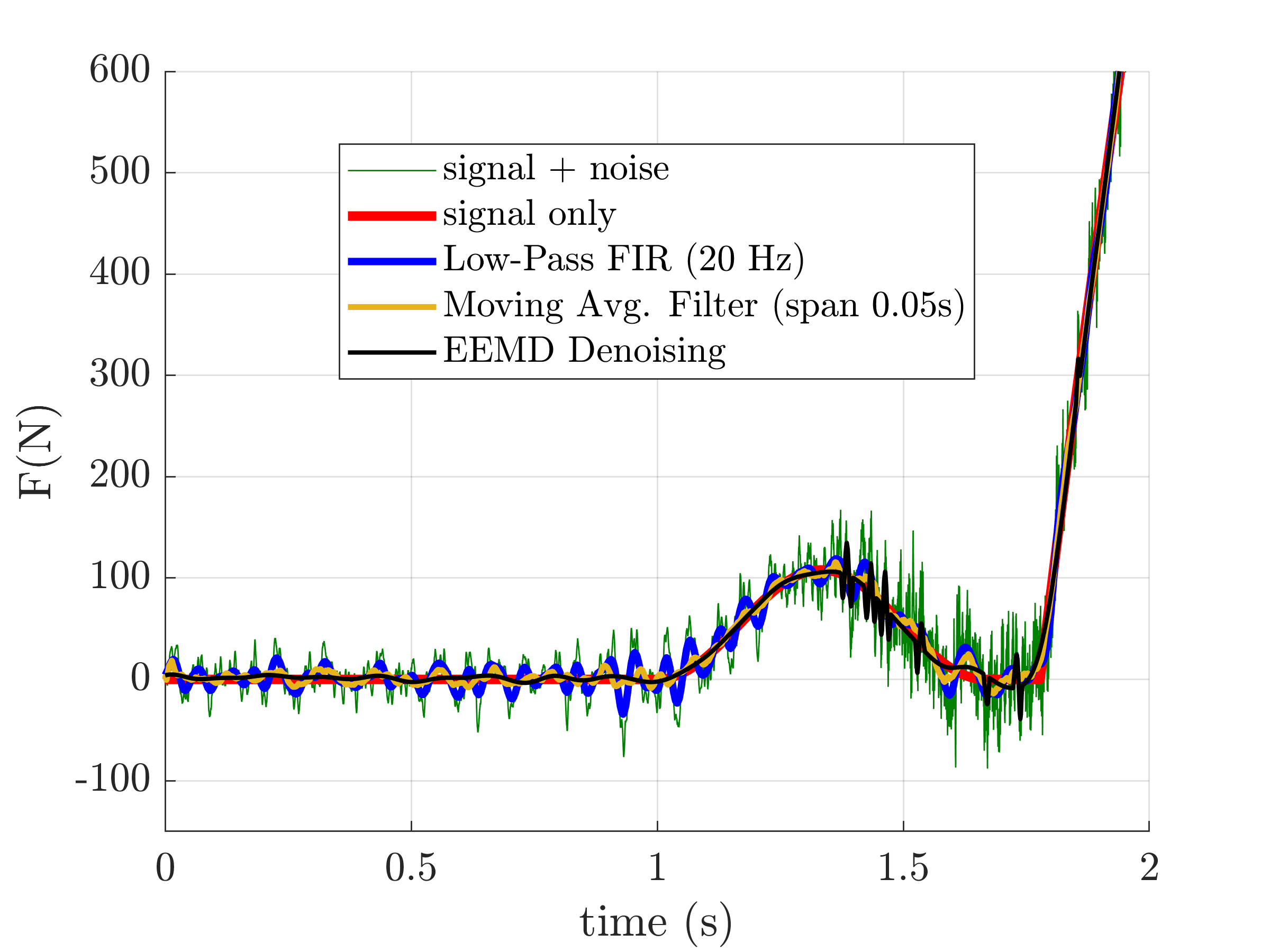}}
\caption{Detailed views
 of the comparison shown in Figure
\ref{fig:EEMDDenoising_T4sigmaINIT_Nkeep3_Nmore3_EEMD_Na01_Ne1000_SynthMixed}.}
\label{fig:zoomed_details_EEMD_synthramp}
\end{figure}

In Figure 
\ref{fig:EEMDDenoising_T4sigmaINIT_Nkeep3_Nmore3_EEMD_Na01_Ne1000_SynthMixed} 
the result of the denoising technique described above is
compared with the original synthetic signal, the signal only and the 
result of FIR low-pass and the moving average filter.
The comparison shows that the EEMD denoising, integrated with the
interval thresholding, provides superior 
results compared to other techniques and it is able to retrieve the original 
signal hidden in the background noise with just a few spurious
oscillations and with only a limited smoothing of the pulse.
The good quality of the reconstruction can also be appreciated from the
close-up views provided in Figure \ref{fig:zoomed_details_EEMD_synthramp}.

\subsection{Guidelines for the development of a EEMD denoising strategy transient
signals}
\label{guidelines_EEMD_denoising}

The application of the denoising strategy to the present
synthetic signal enables to draw some guidelines 
for the determination of the parameters of the EEMD denoising of a 
general transient signal in terms of the number of modes to be included 
in the partial reconstruction and the threshold values.

At first a visual method can 
be applied. 
If the technique is applied to a synthetic signal with superimposed 
background noise,
as done before, it is possible to perform an EEMD decomposition
of the synthetic signal only and of the signal plus noise.
By comparing the two EEMDs,
it is possible to visually establish how the background noise is distributed across
the different modes, and thus to identify the modes that should be 
preserved in full, the ones that should be 
thresholded and to determine an initial threshold value. The determined parameters
can be further tuned through a trial-and-error 
approach.

For a real measurement, the signal only
is of course not available. 
In some cases it is possible to derive an expected behaviour
of the time history of the measured quantity from theoretical 
considerations or from numerical simulations.
Such solutions provide a clean signal that can be decomposed
through the EEMD. The EEMDs of the clean and of the measured signal,
similarly to the case of the synthetic signal, can help 
to choose the denoising 
parameters either visually or with a more sophisticated methods.

If, instead, a theoretical or a numerical solution is not available, 
it is suggested to build a synthetic clean signal that is very similar 
in shape to that under examination and to examine its EEMD, together
with that of the noisy signal, in order to test different
combination of the denoising parameters.
\section{Application of the denoising method to the experimental 
measurements}
\label{EEMD_Thresholding_Exp_Data}
\subsection{Dry Tests}
\begin{figure}
\centering
\includegraphics[width=0.75\textwidth]{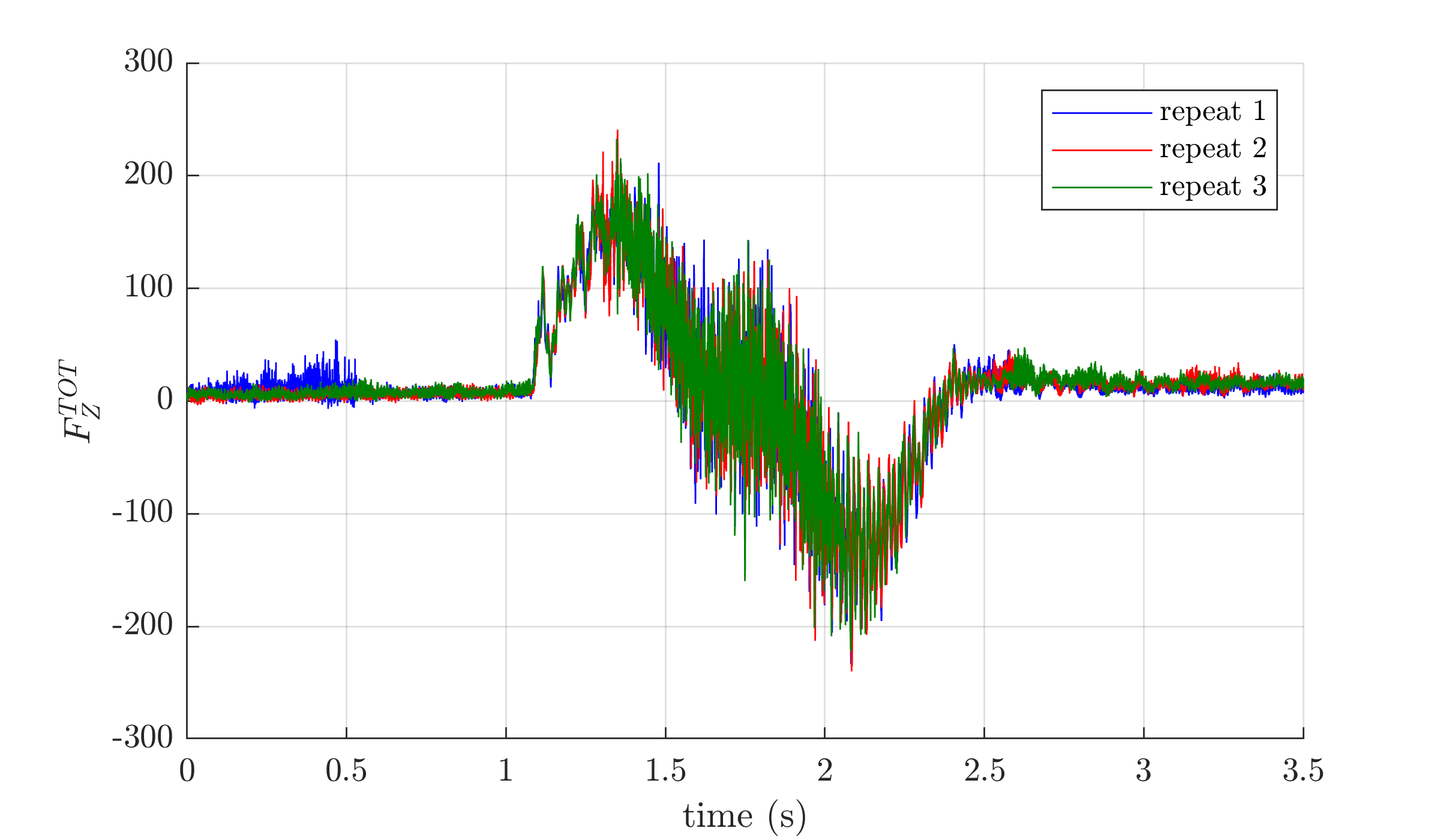}
\caption{Total force $F_{Z}^{TOT}$ measured during three repeats of the dry test 
of the fuselage.}
\label{fig:DryTest_FzTOT_S166TT_3repeats}
\end{figure}

%
The principles of the denoising technique 
developed in the previous sections are used in the following
to reduce the 
background noise of the data acquired in the experimental campaign.
As anticipated in Section \ref{EEMD_algorithm}, before starting the actual 
water entry tests, preliminary dry tests were performed, i.e. tests in which the 
fuselage is moved vertically with the same time velocity profile of Figure
\ref{fig:trajectory}, but starting from a position such that
it does not get in contact with the water at the end of the descent.
The aim of the dry test is to isolate the inertial force 
contribution that has to be subtracted from the data
acquired in water entry tests to retrieve the hydrodynamic loads. 
The total vertical force in the carriage reference frame $F_{Z}^{TOT}$, according to 
the notation provided in Figure \ref{fig:FusRefRames}, can be evaluated as:
\begin{linenomath}
\begin{equation}
\textrm{F}_{Z}^{\textrm{TOT}} \doteq \left(F_{zR}+F_{zF} \right) \, \cos(\alpha) 
+ \left( F_{xR}+F_{xF} \right) \, \sin(\alpha)
\end{equation}
\end{linenomath}
where $\alpha$ is the pitch angle. 
The inertial force is a good reference for the validation 
of the denoising strategy, since it can be
analytically computed starting from the body mass and the time history of the
acceleration (see Figure \ref{fig:trajectory}). 
The time histories of the inertial force recorded in three different repeats 
are shown in Figure \ref{fig:DryTest_FzTOT_S166TT_3repeats}.
Of course, the measured force is affected by a significant 
background noise. It is worth observing that part of the background 
noise during the descent seems to be in phase in all the three repeats, 
hence it is likely associated with the vibrations 
experienced by the actuator system.
As discussed in Section \ref{EEMD_algorithm}, the EEMD 
is more robust than the pure EMD approach.
In the dry test the EEMD parameters are chosen as $N_e$=1000 and
$N_a$=0.1. The artificial noise amplitude is  
$N_a \, \sigma$, where $\sigma$ is the standard deviation computed over the
time interval $1.70 < t <1.75 $~s, during which the fuselage descent
with a constant velocity.

\begin{figure}[htbp]
\centering
\subfigure[Inertial force 
(theory)]{\includegraphics[width=0.48\textwidth]{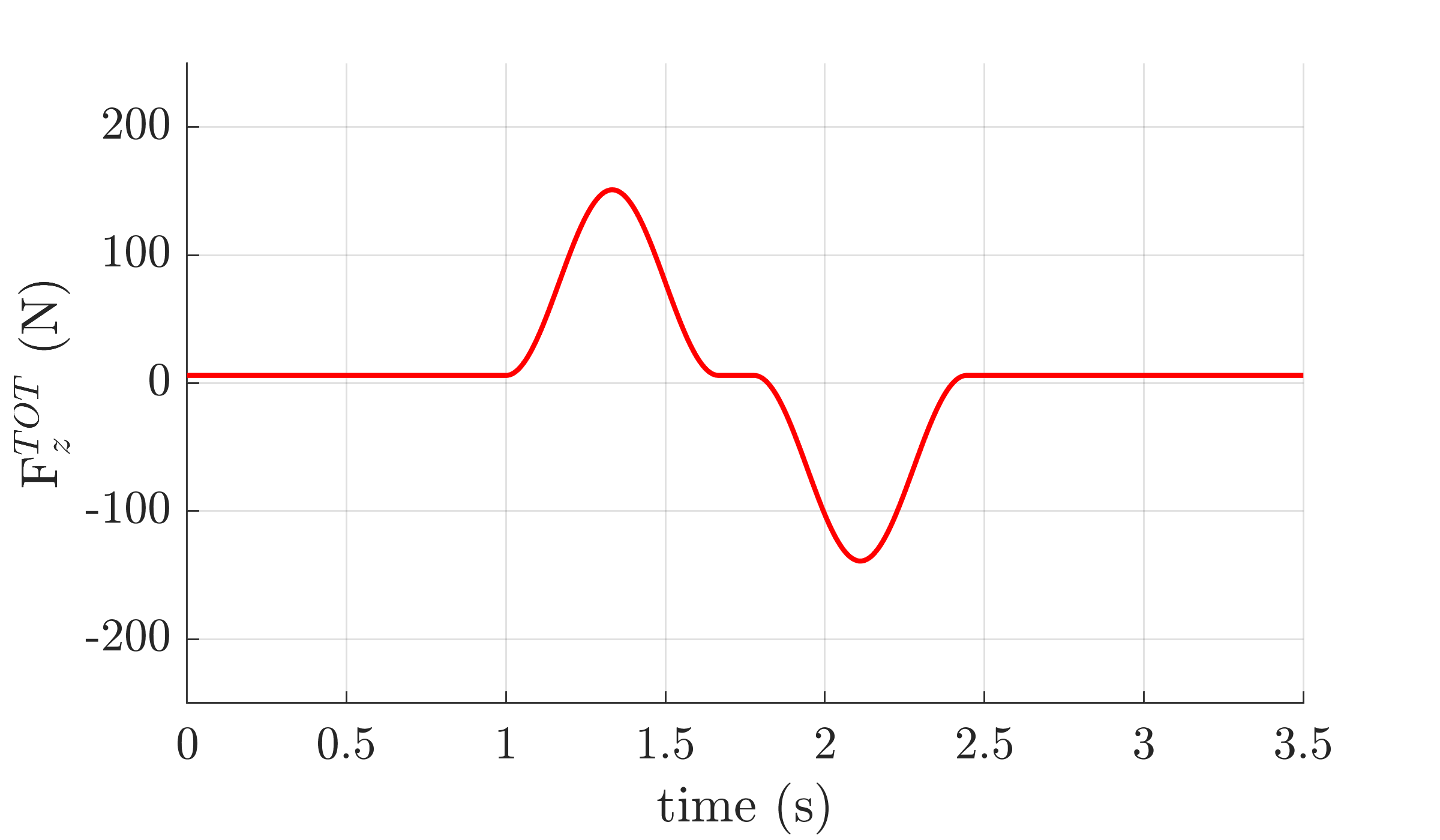}} \quad
\subfigure[Dry test $F_Z^{TOT}$ (experimental)]{\includegraphics[width=0.48\textwidth]{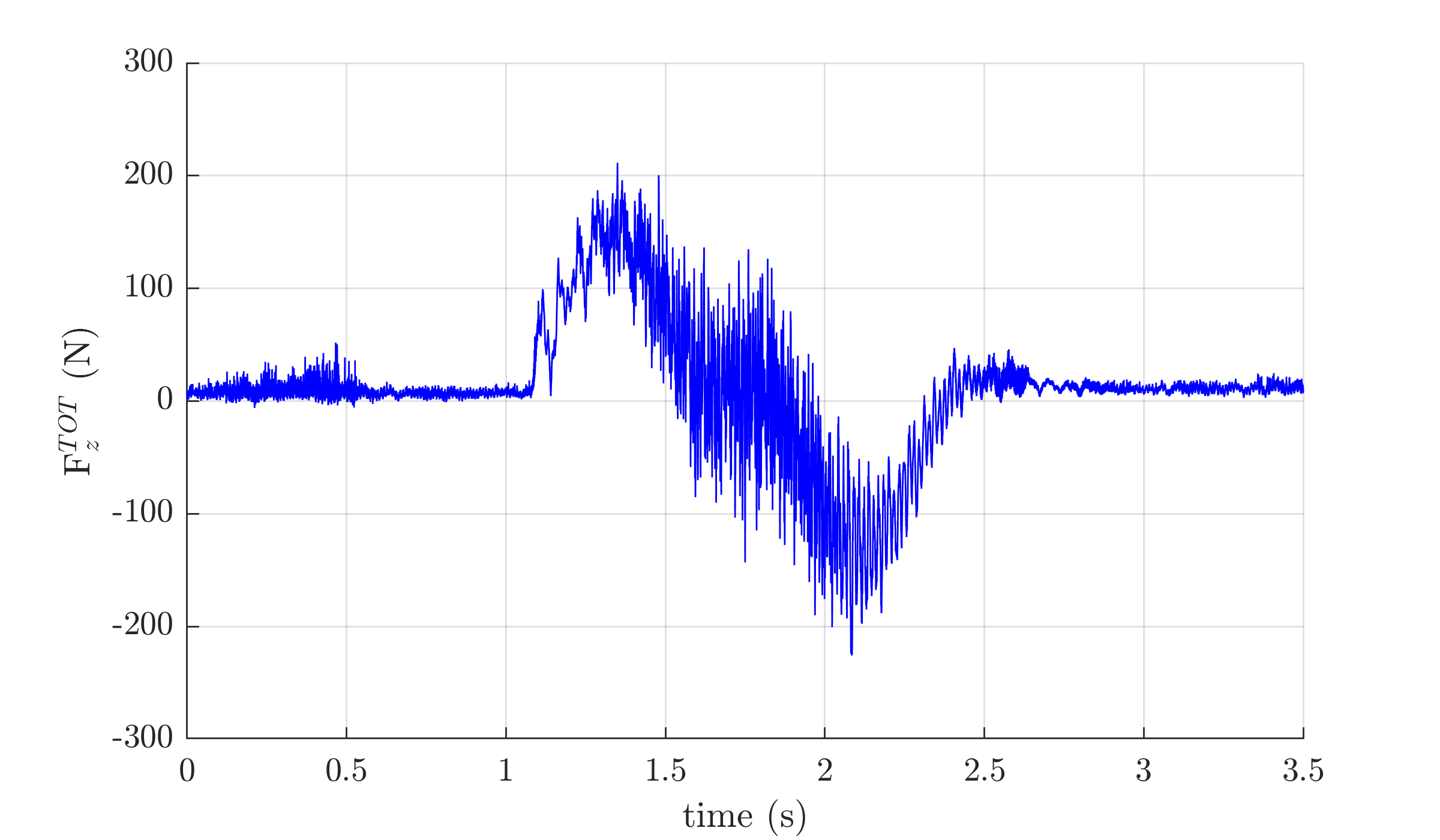}} \\
\subfigure[EEMD inertial force 
(theory)]{\includegraphics[width=0.48\textwidth]{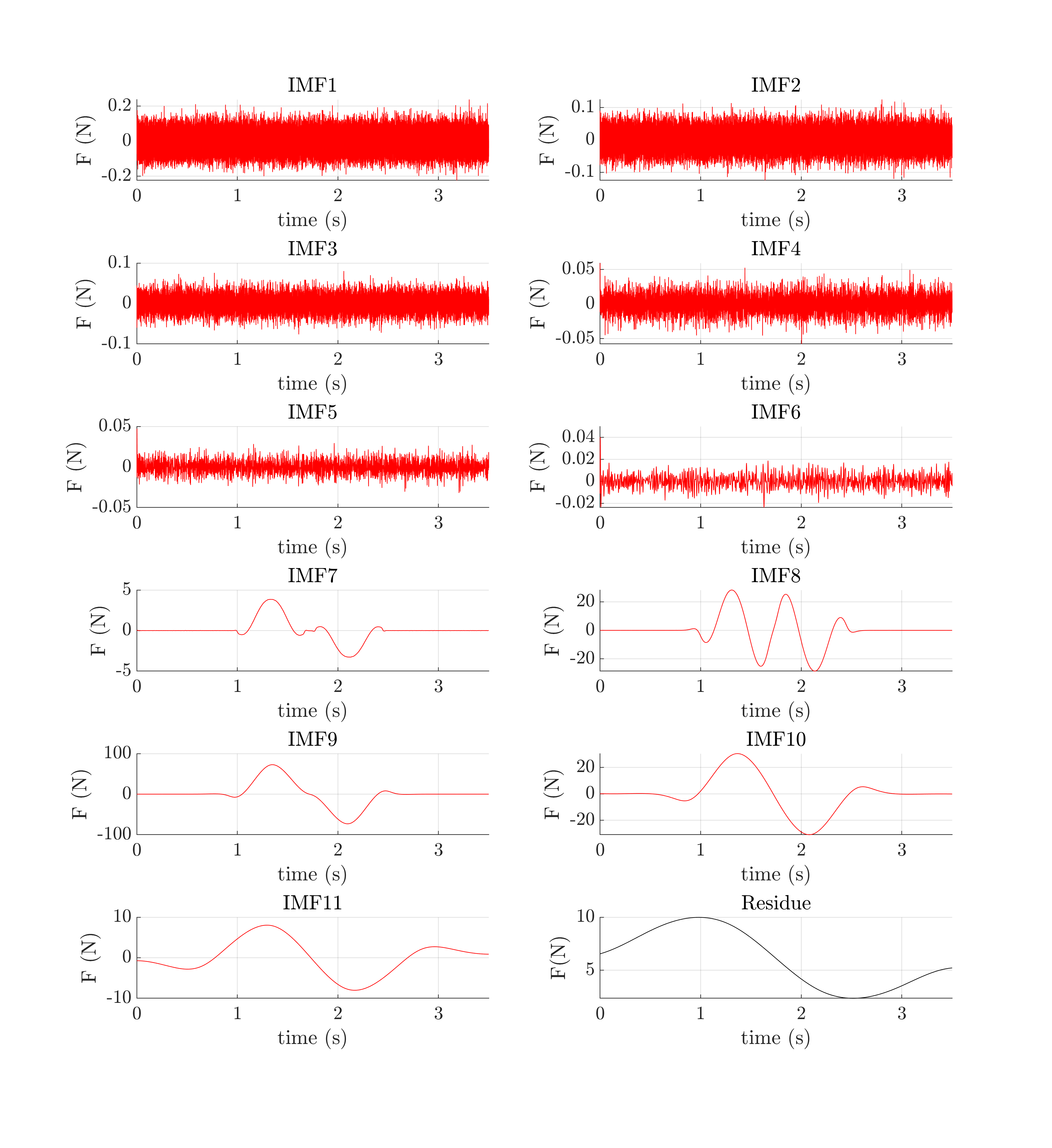}} \quad 
\subfigure[EEMD dry test $F_Z^{TOT}$ 
(experimental)]{\includegraphics[width=0.48\textwidth]{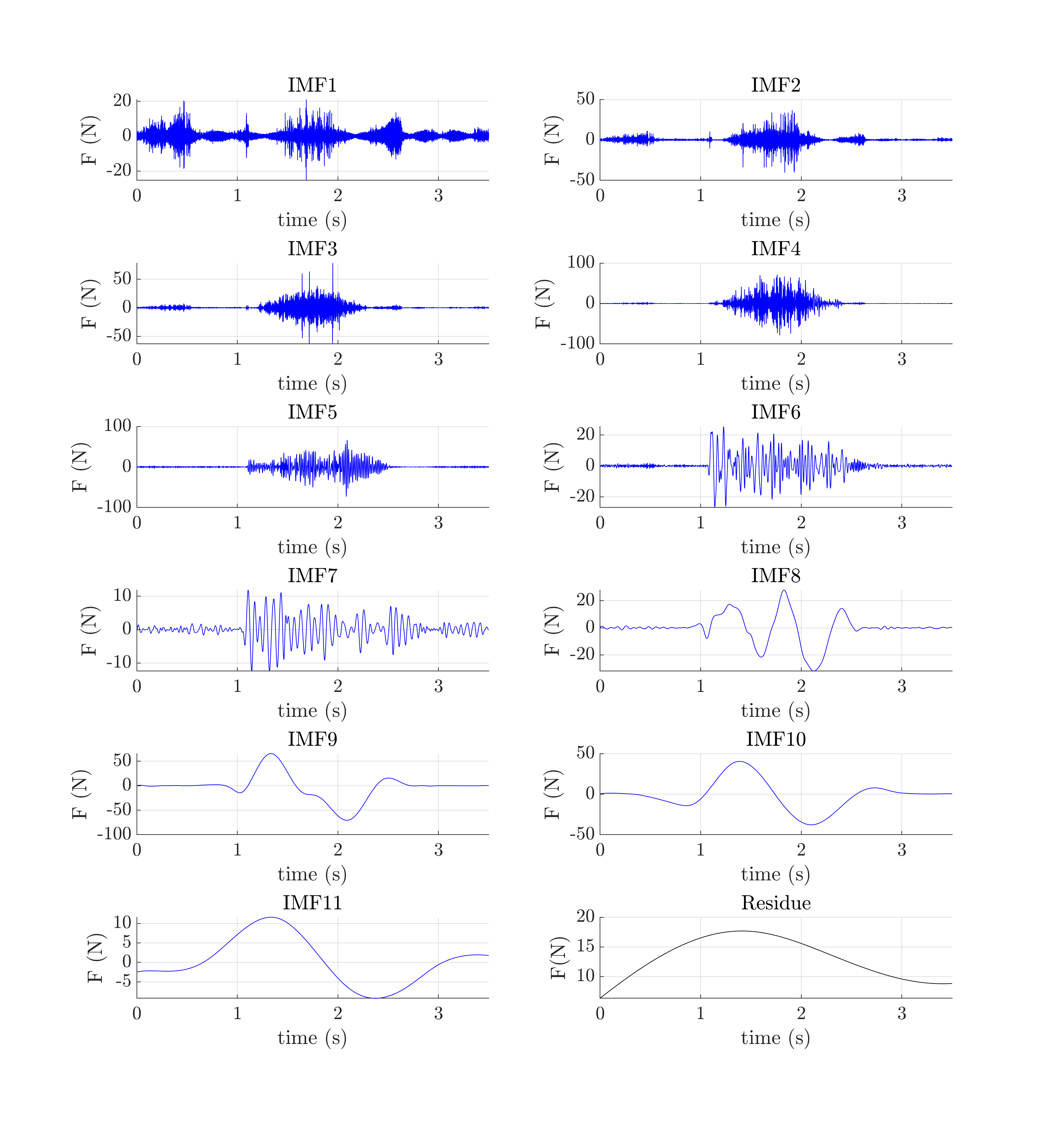}}
\caption{Time histories of (a) the theoretical and (b) measured force. 
The corresponding EEMDs are drawn in (c) and (d), respectively.}
\label{fig:EEMD_DryTheoryExp_EEMD_Comparisons}
\end{figure}

An estimate of the denoising output is the theoretical force 
computed from the acceleration, drawn in Figure 
\ref{fig:EEMD_DryTheoryExp_EEMD_Comparisons}(a), 
whereas the experimental force measured in one of the dry 
tests is shown in Figure \ref{fig:EEMD_DryTheoryExp_EEMD_Comparisons}(b). 
As suggested in Section \ref{guidelines_EEMD_denoising}, 
the EEMD of the theoretical and of the
experimental signals can be compared in
order to drive the selection of the EEMD denoising parameters. 
The resulting modes are provided in Figure 
\ref{fig:EEMD_DryTheoryExp_EEMD_Comparisons}(c) and (d) respectively. 
Both the decompositions consist of eleven IMFs and a residue.
Also in this case, due to the presence of the background
noise, it is not possible to establish a direct
comparison between the modes resulting from the theoretical and the
experimental decompositions. 
Nevertheless, it is possible to spot a good similarity of the last modes.
The shape of these mode suggests that a partial reconstruction 
including the residue, the last two modes and two further thresholded modes can provide 
a satisfactory denoising of the experimental inertial 
force.
In other words, the reconstruction is made by using equation 
(\ref{EEMD_reconstruction}) with $L=8$, 
$M=10$, and $N=11$. For the
thresholds it is assumed $T_i=2 \, \sigma_i$, where $\sigma_i$ is 
computed between t=1.7 s and t=1.75 s in each mode. 

\begin{figure}[htbp]
\centering
\includegraphics[width=0.85\textwidth]{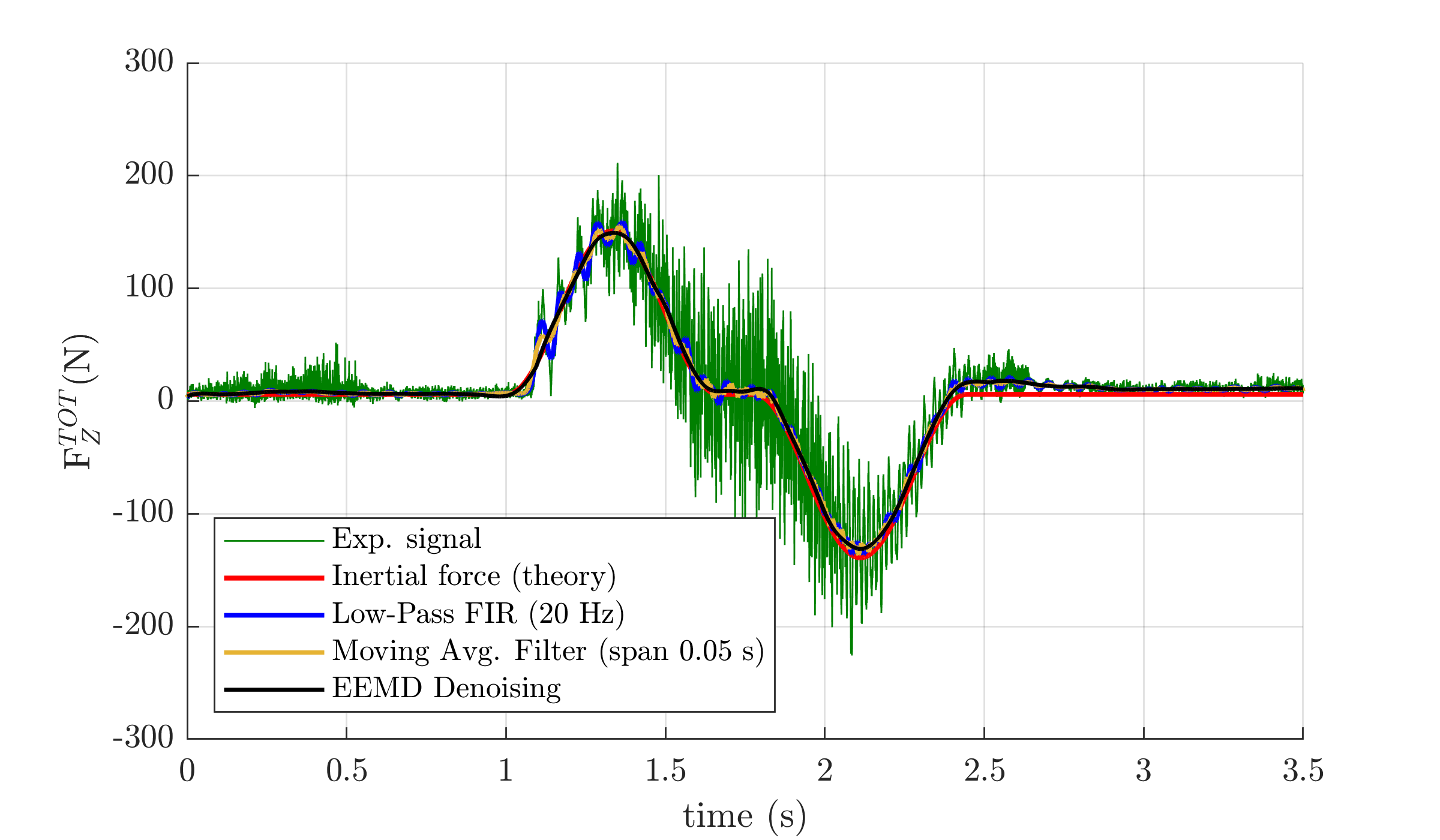}
\caption{Comparison between the different denoising techniques applied to 
the total force F$_z^{TOT}$ of the dry test: moving average filter (span 1000 samples i.e. 
0.05~s), low-pass FIR Hanning window filter (cut-off frequency 20 Hz) 
and the EEMD denoising techniques with the 
interval thresholding.}
\label{fig:EEMDDenoising_T2sigmaConstSpeed_Nkeep2_Nmore2_EEMD_Na01_Ne1000_DryTest_FzTOT_4T1092N_01}
\end{figure}

The denoised signal, the theoretical force, together with the outputs of the the classical 
moving average filter (with span 0.05~s), and of the low-pass FIR filter at 20 Hz are shown in Figure 
\ref{fig:EEMDDenoising_T2sigmaConstSpeed_Nkeep2_Nmore2_EEMD_Na01_Ne1000_DryTest_FzTOT_4T1092N_01}.
The comparison indicates that the EEMD 
denoising approach performs much better
than the other filtering methods, since it allows to remove 
the background noise while preserving quite faithfully the 
sharp variations of the signal. The EEMD-denoised signal 
also displays a good match with the theoretical estimate. 

\subsection{Water entry tests with horizontal velocity}
\begin{figure}[htbp]
\centering
\subfigure[Forward load cell $F_{zF}$]{\includegraphics[width=0.48\textwidth]{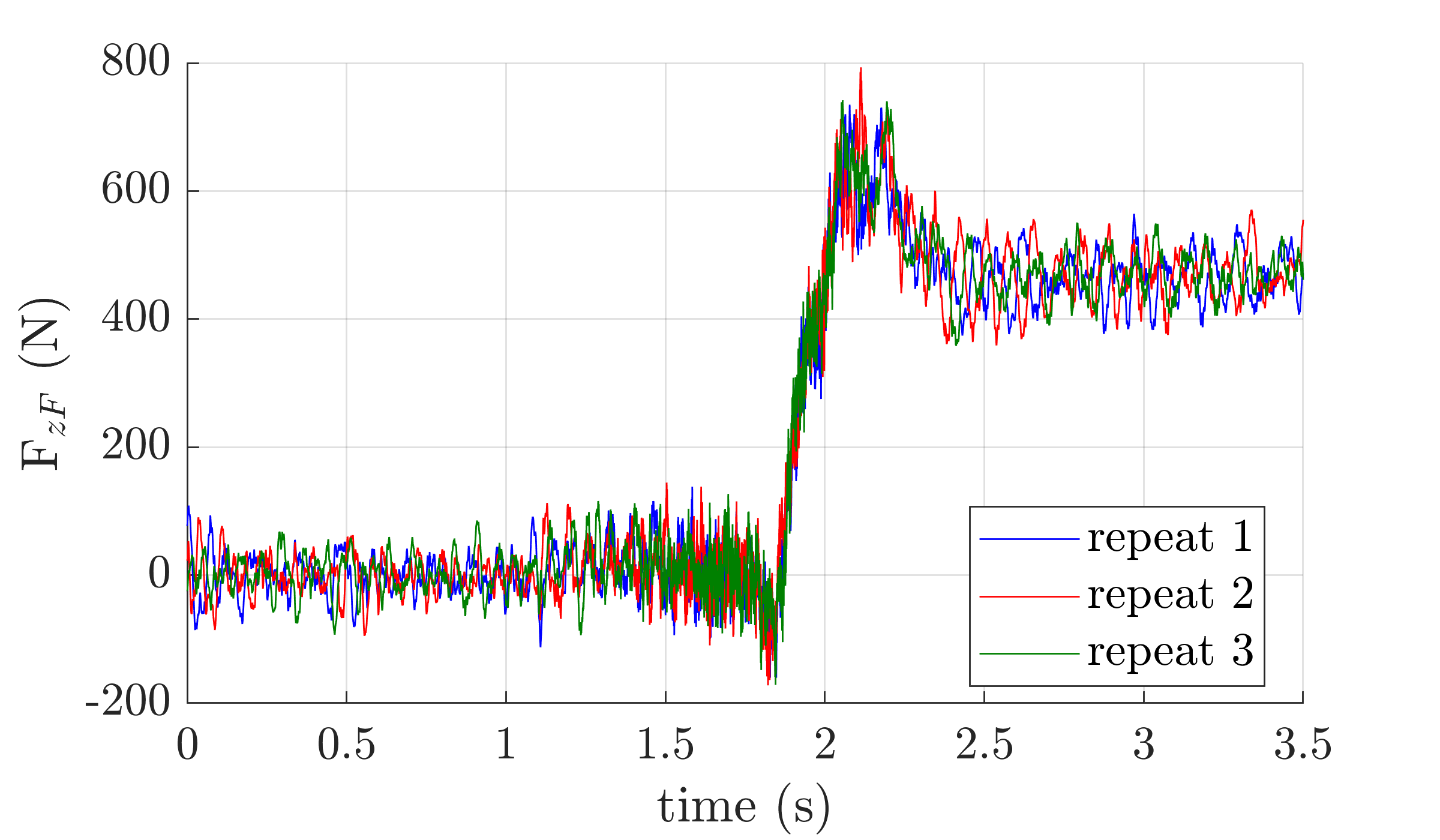}}
\quad
\subfigure[Rear load cell $F_{zR}$]{\includegraphics[width=0.48\textwidth]{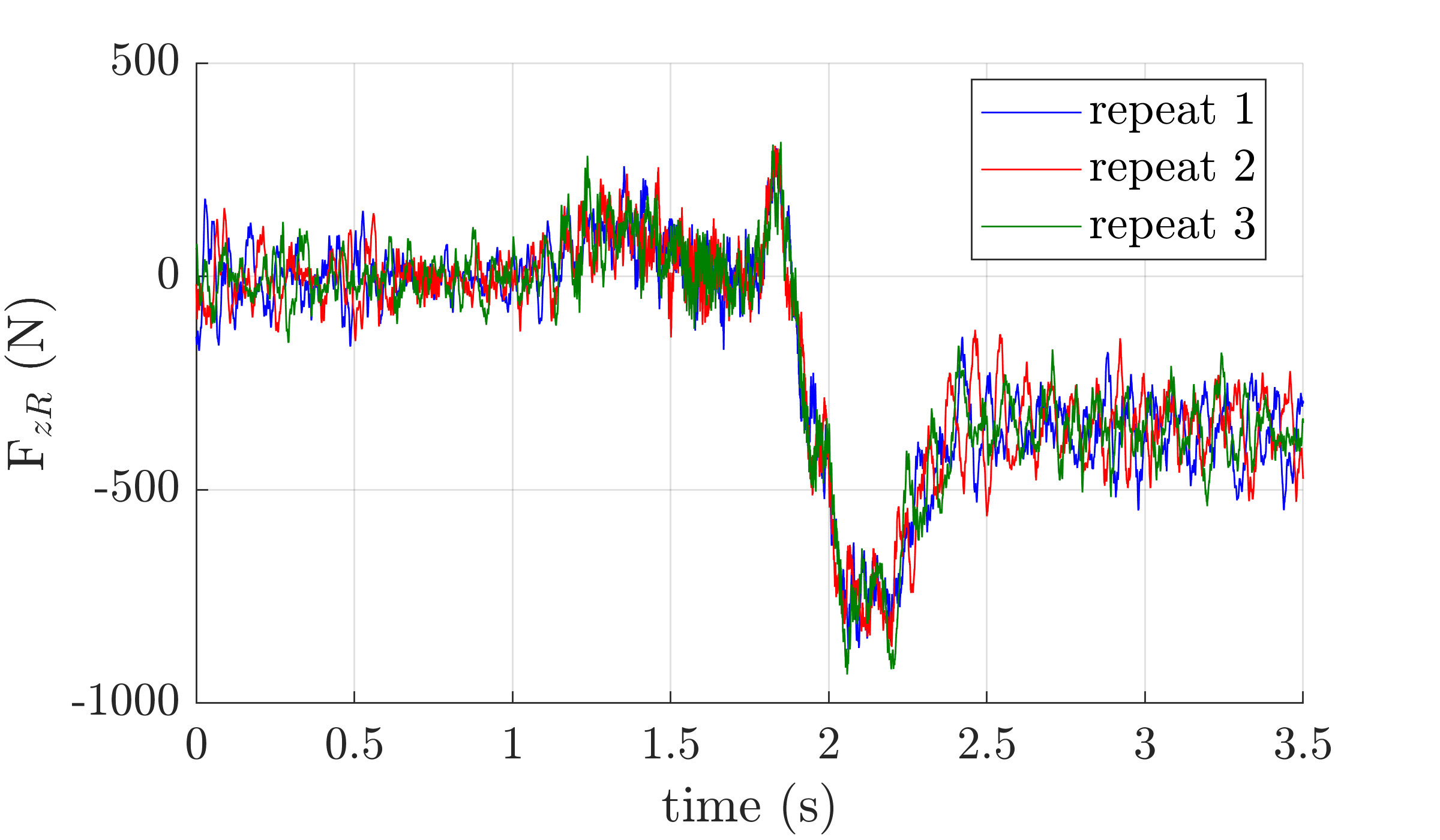}}
\caption{Vertical forces measured by the forward (a) and rear (b) load 
cells during the water entry test for three repeats.}
\label{fig:PlotRepeats_4T1422N_0X}
\end{figure}

Based on the above experience, the denoising method is 
applied to 
the time histories of the force recorded during the water entry
of the fuselage set at 6$^{\circ}$ and moving at a horizontal velocity of 12~m/s
and a vertical velocity at the impact of 0.45~m/s. 
For the purpose of the present study, only the forces measured by the 
load cells acting normal to the fuselage axis (see Figure
\ref{fig:FusRefRames}), i.e. $F_{zR}$ and $F_{zF}$, are considered. 
The raw data of the three repeats are depicted
in Figure \ref{fig:PlotRepeats_4T1422N_0X}. 
Despite the large noise, the data of the three repeats display a good
overlapping, thus denoting a satisfactory repeatability of the tests.
Aside from the background noise, there is a small time-shift of the curves,
which is caused by some differences in the position of the free surface as
a consequence of the residual waves in the tank.
The time histories of the $z$-component of the forces
$F_{zR}$ and $F_{zF}$ display a peak, positive at the rear and
negative at the forward cell, occurring about $t \approx ~1.778 $~s, which is
about the first contact with water.
After the peak, the force at the rear diminishes, attains a minimum and
grows up to reach a constant value in the last part of the test. A similar
behaviour, but opposite in sign, is observed for the forward cell.
It is worth noticing that the two forces do not balance at the end of the
test, meaning that there is a net force, which is the buoyancy 
(hydrostatic) component at that specific attitude. 
The inertial contribution is visible between $t=1$~s and $t=1.5$~s,
particularly at the rear. In the next phase, the effect
of the inertia is hidden by the action of the hydrodynamic loads. 
The background noise is present 
throughout the acquisition period, but it looks more intense starting 
from $t=1$~s, i.e. concurrent with the motion of the fuselage.

\begin{figure}[htbp]
\centering
\subfigure[Pure EMD]{\includegraphics[width=0.48\textwidth]{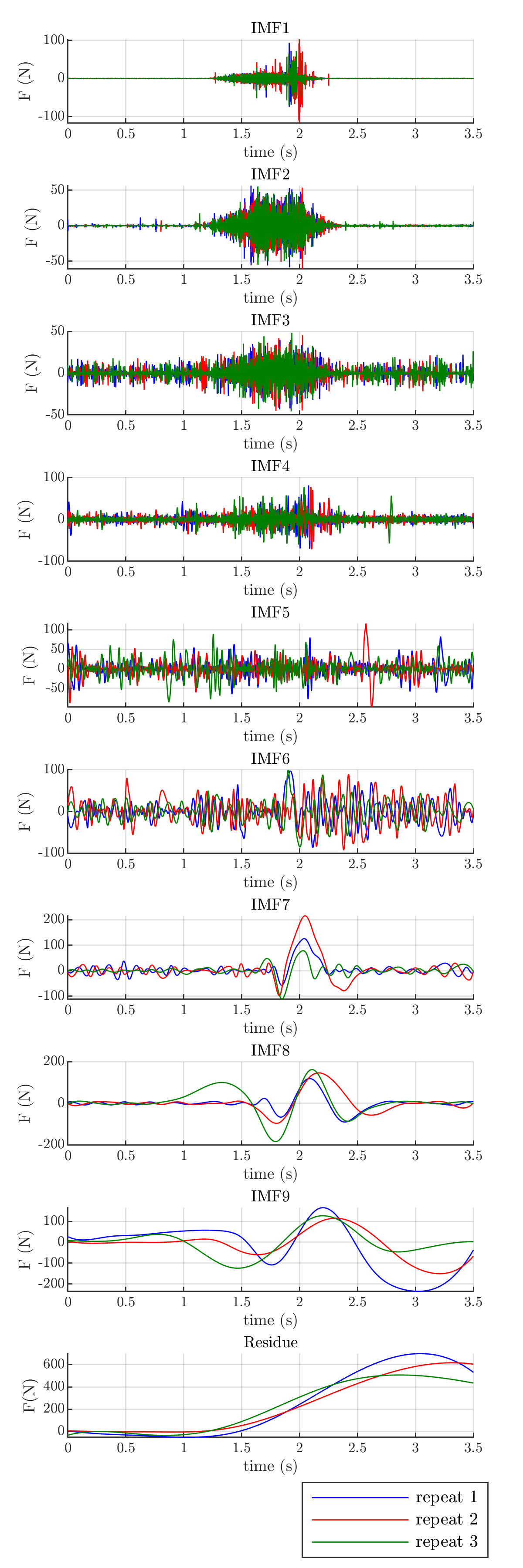}} \quad
\subfigure[EEMD]{\includegraphics[width=0.48\textwidth]{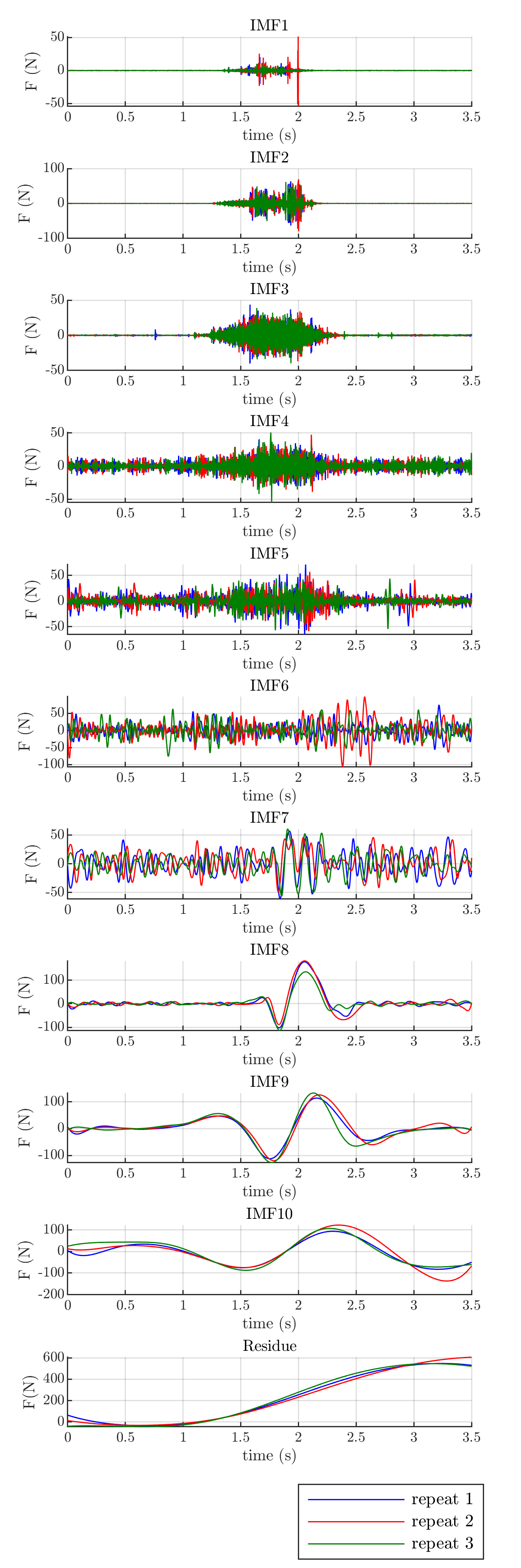}}
\caption{Comparison between the pure EMDs and the EEMDs 
performed on the three repeats of $F_{zF}$ during the water entry test.}
\label{fig:IMFs_FzF_4T1422N_0X}
\end{figure}

In order to reduce the background noise, the EMD and
EEMD are applied to the forces measured by the forward and rear cells.
The EEMD is computed by using $N_a=0.1$ and $N_e$=1000.
%
%
In Figure \ref{fig:IMFs_FzF_4T1422N_0X} the pure EMD and the EEMD
performed on the time histories of the three repeats of the forward force 
$F_{zF}$ are shown. Aside from the different number of modes, the 
advantage of the EEMD over the EMD is clearly
visible by the much closer overlapping of the most significant modes,
i.e. those above IMF$_7$, denoting a much lower sensitivity to the 
slight differences between the signal shapes in the three repeats.

\begin{figure}[htbp]
\centering
\includegraphics[width=0.85\textwidth]{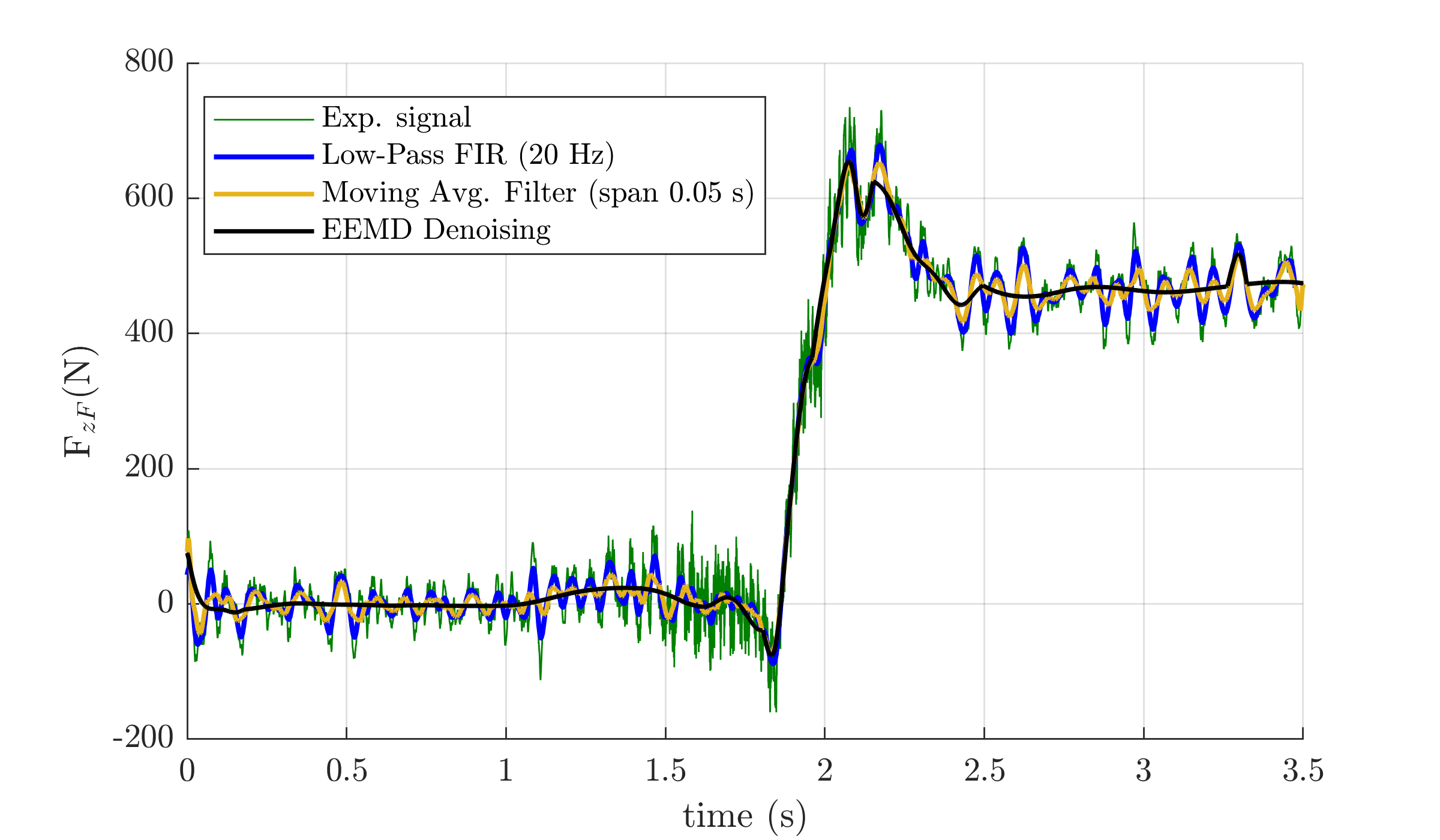}
\caption{Comparison between the different denoising techniques applied to 
the force $F_{zF}$: moving average filter (span 1000 samples i.e. 
0.05~s), low-pass FIR Hanning window filter (cut-off frequency 20 Hz) 
and the EEMD denoising techniques with
interval thresholding.}
\label{fig:Denoising_EEMD_Std01_Ne1000_Thresh35sigma_FzF_4T1422N_01_3modes2thresh}
\end{figure}

In order to retrieve the denoised signal, a partial
reconstruction is used, by including the residue, the last two modes 
(IMF$_{10}$ and IMF$_9$) in full and the previous two (IMF$_8$ and
IMF$_7$) after being treated with an interval thresholding, with a
mode-dependent threshold  $T_i = 3 \, \sigma_i$.
More formally, by using Equation (\ref{EEMD_reconstruction}), 
it is assumed that $L=7$ and $M=9$, with $N=10$. 
The standard deviation for the interval thresholding, $\sigma_i$,
is computed over the time interval $ t= 0.05 - 0.95 $~s. 
The reconstructed signal is shown in Figure
\ref{fig:Denoising_EEMD_Std01_Ne1000_Thresh35sigma_FzF_4T1422N_01_3modes2thresh}.
The results confirm that the denoising based on the EEMD is
rather efficient in removing the undesired oscillations without affecting
too much the sharpness of the signal. This is certainly an advantage
compared to classical filters.
It is difficult to establish whether the double peak occurring after 
the water impact has an actual physical origin or it is associated with the background noise. In order to 
understand its origin, numerical simulation of the water entry are ongoing. 
As explained in Section \ref{guidelines_EEMD_denoising}, such simulations should provide 
a reference force time history that can support the understanding of the experimental 
measurements and to tune the denoising method.

\begin{figure}[htbp]
\centering
\subfigure[Single 
EMD]{\includegraphics[width=0.48\textwidth]{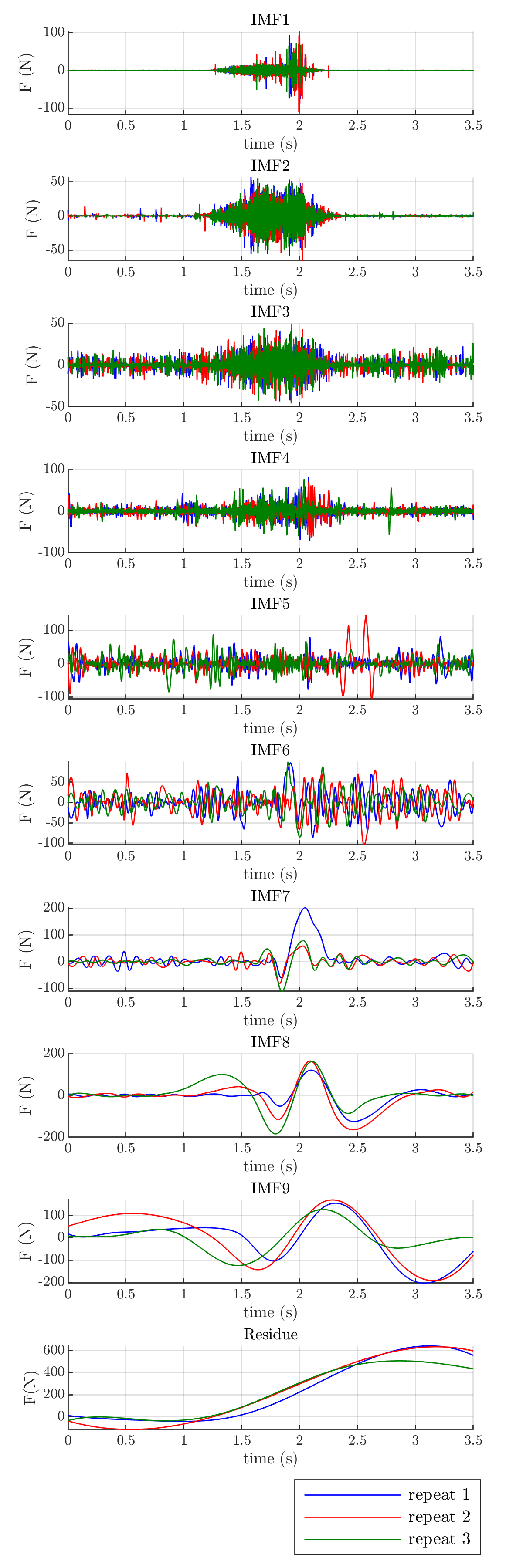}} \quad
\subfigure[EEMD]{\includegraphics[width=0.48\textwidth]{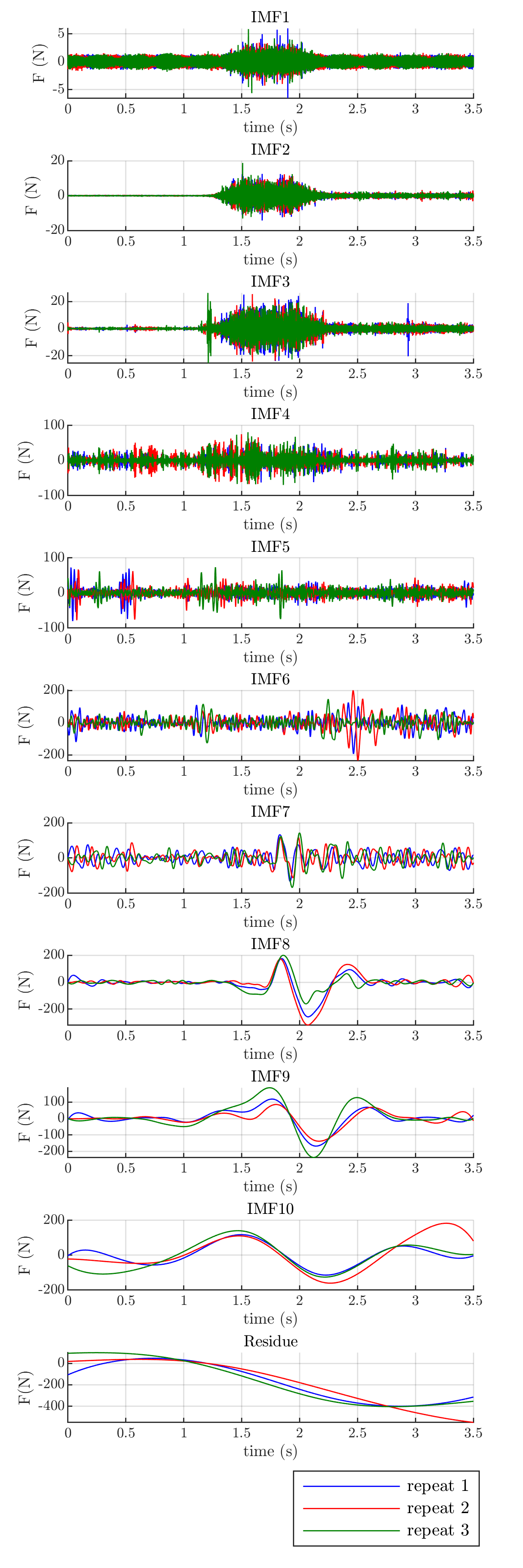}}
\caption{Comparison of the pure EMDs and the EEMDs of the $F_{zR}$ time 
histories for the three repeats of the water entry test.}
\label{fig:IMFs_FzR_4T1422N_0X}
\end{figure}

%
%
%
A similar analysis is performed for the force measured at the rear
$F_{zR}$ and results are provided in Figure \ref{fig:IMFs_FzR_4T1422N_0X}.
In this case both the pure EMD and the EEMD yields 
ten modes, but again the last EEMD modes and the residue exhibit a much
closer overlapping among the different repeats.

\begin{figure}[htbp]
\centering
\includegraphics[width=0.85\textwidth]{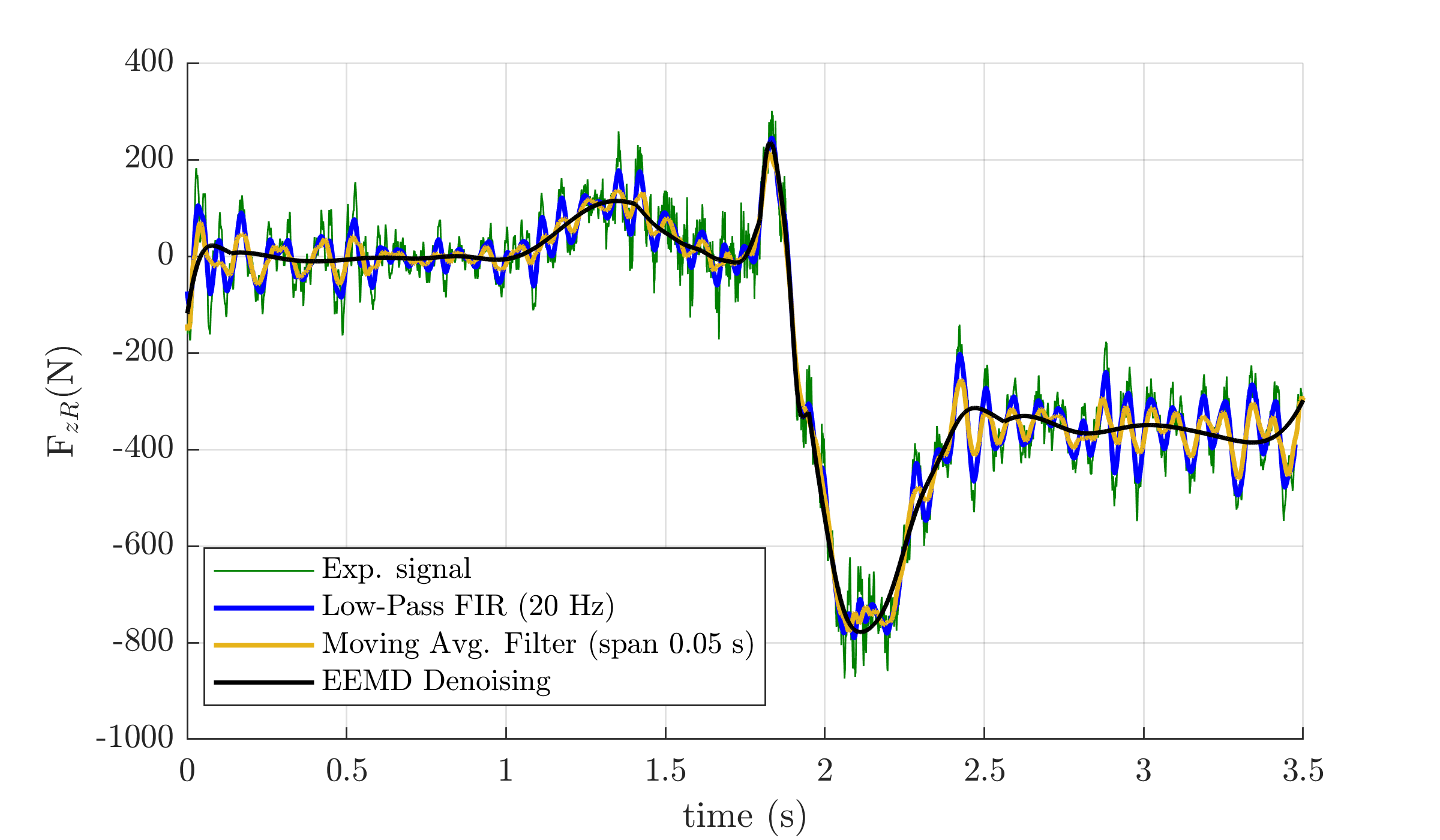}
\caption{Comparison between the different denoising techniques applied to 
the force $F_{zR}$: moving average filter (span 1000 samples i.e. 
0.05~s), low-pass FIR Hanning window filter (cut-off frequency 20 Hz) 
and the EEMD denoising techniques with 
interval thresholding.}
\label{fig:EEMDDenoising_T3sigmaConstSpeed_Nkeep2_Nmore2_EEMD_Na01_Ne1000_FzR_4T1422N_01}
\end{figure}

The denoised force is derived by using the same approach adopted for the
forward force. The results are shown in Figure
\ref{fig:EEMDDenoising_T3sigmaConstSpeed_Nkeep2_Nmore2_EEMD_Na01_Ne1000_FzR_4T1422N_01}.
As already observed, the denoising based on the EEMD
is more effective than the classical filters, even though some undesired 
oscillations are still present. The results for $F_{zR}$ are somewhat
cleaner than for $F_{zF}$.
\section{Conclusions}
%
In this paper a noise reduction strategy based on the Ensemble Empirical Mode
Decomposition (EEMD) has been developed and applied to transient signals affected
by a broad-band non-stationary noise. The signals are the
measured loads in scaled fuselage ditching tests in a towing
tank. The noise affecting the signals originates both from mechanical vibrations and from
electromagnetic interferences with the data acquisition system.

The denoising strategy has been first tested and tuned 
on a synthetic signal resembling the time histories of the
measured force, on which a consistent background noise is superimposed.
The denoising strategy is based on a partial reconstruction
including the residue, a few EEMD modes in full, typically the higher order ones,
and a few other modes to which an interval thresholding is applied. 
The latter modes, in fact, account for physically meaningful components to be preserved.
The proposed denoising strategy has been proved to be more efficient than 
classical low-pass FIR and moving average filters, since it
allows to reduce the noise without smoothing excessively the sharpness of the signal.
Hence, the strategy has been applied to two type of real force measurements. 
In the dry test case the denoising strategy has provided a signal
very close to the one expected from theory. In the ditching tests, 
the denoising approach has been found to outperform
the other filtering techniques, even though some residual oscillations still appear.

Owing to the features of the signal and of the noise, a
universally valid criterion to choose the thresholds and the modes
to be included in the reconstruction cannot be defined.
For a synthetic signal with a known superimposed noise, it is recommended
to compare the EEMD of the whole signal with the EEMD
of the signal only component, and to choose the denoising parameters
through a trial and error approach to achieve a satisfactory match between
the reconstructed signal and the signal only component.
In the case of real signals, the parameters of the EEMD denoising 
can be chosen similarly by using as a reference an expected denoised time history,
which can be estimated from theory or from numerical simulations, when possible.
Further improvements of the EEMD denoising techniques can be achieved,
for instance, by exploiting a time-windowing approach, 
e.g. \cite{weng2006ecg,kabir2012denoising}, or by developing
a method that selects the denoising parameters in an automatic way.

\section*{Funding}

This project has been partly funded from the European Union's Horizon 
2020 Research and Innovation Programme under Grant Agreement No. 724139 
(H2020-SARAH: increased SAfety \& Robust certification for ditching of 
Aircrafts \& Helicopters).


\bibliography{MSSP_paper_biblio}

\end{document}